\documentclass[useAMS,usenatbib]{mn2e}

\usepackage{lscape}
\usepackage{epsfig}
\usepackage{times}
 
\def\gtsim{\mathrel{\spose{\lower.5ex \hbox{$\mathchar"218$}}
     \raise.4ex\hbox{$\mathchar"13E$}}}
\def\ltsim{\mathrel{\spose{\lower.5ex\hbox{$\mathchar"218$}}
     \raise.4ex\hbox{$\mathchar"13C$}}}
\def\aFe{[$\alpha/{\rm Fe}$]}

\def\Hb{${\rm H}_{\beta}$}

\def\Mgb{{\rm Mg}\,$_b$}
\def\Fe{$\langle {\rm Fe}\rangle$}

\def\ZH{[$Z/{\rm H}$]}
\def\MgFe{[${\rm MgFe}$]$'$}

\def\Mgd{{\rm Mg}\,$_2$}

\def\Rdb{$r_{\rm{db}}$} 

\def\kms{$\rm km\;s^{-1}$}

\def\iraf{{\tt IRAF}}

\def\spose#1{\hbox to 0pt{#1\hss}}

\begin{document}

\title[Stellar Populations of Bulges in 14 Cluster Disc Galaxies]{Stellar Populations of Bulges in 14 Cluster Disc Galaxies\thanks{Based on observations made with ESO Telescopes at the La Silla observatory under programmes 70.B-0486 and 71.B-0202.}}

\author[L. Morelli et al.]{L.~Morelli$^{1,5}$, 
 E.~Pompei$^{2}$, A.~Pizzella$^{1}$, J.~M\'endez-Abreu$^{1,3}$, E.~M.~Corsini$^{1}$, L.~Coccato$^{4}$,
\newauthor R.~P.~Saglia$^{4}$, M.~Sarzi$^{6}$ and F.~Bertola$^{1}$\\
$^1$ Dipartimento di Astronomia, Universit\`a di Padova,
  vicolo dell'Osservatorio~3, I-35122 Padova, Italy.\\
$^2$ European Southern Observatory, 3107 Alonso de Cordova,
  Santiago, Chile.\\
$^3$ INAF-Osservatorio Astronomico di Padova,
  vicolo dell'Osservatorio 5, I-35122 Padova, Italy.\\
$^4$ Max-Planck Institut f\"ur extraterrestrische Physik,
  Giessenbachstrasse, D-85748 Garching, Germany.\\
$^5$ Department of Astronomy, Cat\'olica de Chile, Pontificia Universidad, 
 Vicuna Mackenna 4860, Casilla 306, Santiago 22, Chile\\
$^6$ Centre for Astrophysics Research, University of Hertfordshire,
College Lane, Hatfield, Herts AL10 9AB\\
}

%\date{Received..................; accepted...................}
%\date{{\it Draft version on \today}}
\date{Accepted 2008 June 11. Received 2008 May 13; in original form 2008 January 15}
%\pagerange{\pageref{firstpage}--\pageref{lastpage}} \pubyear{2002}

\maketitle

\begin{abstract}

Photometry and long-slit spectroscopy are presented for 14 S0 and
spiral galaxies of the Fornax, Eridanus and Pegasus cluster, and
NGC~7582 group.
The structural parameters of the galaxies are derived from the
$R-$band images by performing a two-dimensional photometric
decomposition of the surface-brightness distribution. This is assumed
to be the sum of the contribution of a bulge and disc component
characterized by elliptical and concentric isophotes with constant
(but possibly different) ellipticity and position angles.
The rotation curves and velocity dispersion profiles are measured from
the spectra obtained along the major axis of galaxies. The radial
profiles of the \Hb , Mg, and Fe line-strength indices are
presented too. Correlations between the central values of \Mgd , \Fe ,
\Hb , and $\sigma$ are found.
The age, metallicity and $\alpha/$Fe enhancement of the stellar
population in the center and at the radius where bulge and disc give
the same contribution to the total surface brightness are obtained
using stellar population models with variable element abundance ratios.
Three classes of bulges are identified. The youngest bulges ($\sim2$
Gyr) with ongoing star formation, intermediate-age bulges (4--8 Gyr)
have solar metallicity, and old bulges ($\sim10$ Gyr) have high
metallicity.
Most of the sample bulges display solar $\alpha/$Fe enhancement, no
gradient in age, and a negative gradient of metallicity.  The presence
of negative gradient in the metallicity radial profile favors a
scenario with bulge formation via dissipative collapse. This implies
strong inside-out formation that should give rise to a negative
gradient in the $\alpha$/Fe enhancement too. But, no gradient is
measured in the \aFe\/ radial profiles for all the galaxies, except
for NGC~1366.
In this galaxy there is a kinematically-decoupled component, which is
younger than the rest of host bulge. It possibly formed by enriched
material probably acquired via interaction or minor merging.
The bulge of NGC~1292 is the most reliable pseudobulge of our
sample. The properties of its stellar population are consistent with a
slow buildup within a scenario of secular evolution.

\end{abstract}


\section{Introduction}
\label{sec:introduction}

The relative importance of the dissipative collapse
\citep{eglbsa62,sandage90,gilwys98}, major and minor merging events
\citep{kauffmann96,coletal00,aguetal01}, and redistribution of disc material 
due to the presence of a bar or environmental effects
\citep{korken04} drives the variety of properties observed in
bulges.

The bulges of lenticulars and early-type spirals are similar to
low-luminosity elliptical galaxies.
Their photometric and kinematic properties satisfy the same
fundamental plane correlation (FP) found for ellipticals
\citep{bendetal92,bendetal93,bursetal97,aguetal05a}. The
surface-brightness radial profile of big bulges is well described by
the de Vaucouleurs law \citep{andretal95,caroetal98,molletal01} {\bf
  even if this could drastically change by taking into account the
  small-scale inner structures smoothed by the seeing in the
  ground-base observations \citep{balceetal03}.}
Some of them are rotationally-flattened oblate spheroids with little
or no anisotropy \citep{kormetal82,davill83,capetal06}. But, the
intrinsic shape of a large fraction of early-type bulges is triaxial,
as shown by the isophotal twisting \citep{lindblad56,zarlo86},
misalignment with respect to disc \citep{bertetal91,mendetal07}, and
non-circular gas motions
\citep{bertetal89,gerhetal89,berman01,corsetal03,coccetal04}.
The bulk of their stellar population formed between redshift 3 and 5
($\simeq$12 Gyr) in a short time-scale
\citep{bernetal98,mehletal03,thometal05}. The enrichment of
interstellar medium is strongly related to the time delay between SNII
and SNIa, which contributed most of the $\alpha$ elements and Iron,
respectively \citep{wofago92,thmabe03}.

On the contrary, the bulges of late-type spiral galaxies are
reminiscent of discs.
They are flat components \citep{fatpel03} with exponential
surface-brightness radial profiles \citep{AndSan94,dejong96,macartetal03}
and rotate as fast as discs \citep{korm93,kormetal02}.
Moreover, the stellar population in late-type bulges is younger than
in early-type bulges \citep{tragetal99,goud99,thda06}. They appear to
have lower metallicity \citep{gandetal07} and lower $\alpha/$Fe
enhancement with respect to early type galaxies
\citep{procetal02,peleetal07,afasil05}. 

In the current paradigm, early-type bulges were formed by rapid
collapse and merging events while late-type bulges have been slowly
assembled by internal and environmental secular processes
\citep{korken04}. But many questions are still open.  For instance,
the monolithic collapse can not explain the presence in bulges of
kinematically-decoupled components
\citep{pizzetal02,krajaf03,emsetal04,mcdetal06}.
Moreover, the environment plays a role in defining the properties of
galaxies \citep[e.g.,][]{dresler80,cozietal01,clemetal06,brouetal07}. Recent studies of
early-type galaxies in different environments
\citep{beui02,thometal05,thda06} have shown that age, metallicity, and
$\alpha/$Fe enhancement are more correlated with the total mass of the
galaxy than local environment.

To investigate the formation and evolution of the bulges, there are
two possible approaches: going backward in redshift and look to
evolution of galaxies through cosmic times or analyze in detail nearby
galaxies to understand the properties of their stellar population in
terms of dominant mechanism at the epochs of star formation and mass
assembly. In this work, we present a photometric and spectroscopic
study of the bulge dominated region of a sample of spiral galaxies in
the Fornax and Pegasus clusters. Our aim is to estimate the age and
metallicity of the stellar population and the efficiency and timescale
of the last episode of star formation to disentangle between early
rapid assembly and late slow growing.

The galaxy sample is presented in Sect. \ref{sec:sample}. The
photometric observations are described in
Sect. \ref{sec:observation_photometry}. The structural parameters of
the bulge and disc of the sample galaxies are derived by analyzing
their two-dimensional surface brightness distribution in
Sect. \ref{sec:decomposition}. The spectroscopic observations are
described in Sect. \ref{sec:observation_spectroscopy}.  The stellar
kinematics and line-strength indices are measured from long-slit
spectra in Sect. \ref{sec:kinematics}. The central values of the
line-strength indices are derived in Sect. \ref{sec:linestreng_cent}.
They are used to estimate the age, metallicity, and
$\alpha/$Fe-enhancement of the stellar population of the bulge in
Sect. \ref{agemet_cent} . Their gradients in the bulge dominated
region are discussed in Sect. \ref{sec:agemetalpha_grad}. The
identification of pseudobulges hosted by sample galaxies is performed
in Sect. \ref{slowfastrotator}. Finally, conclusions are given in
Sect. \ref{conclusion}.

\section{Sample selection}
\label{sec:sample}

The main goal of this paper is to study the stellar populations in
bulges. In order to simplify the interpretation of the results we
selected a sample of disc galaxies, which do not show any
morphological signature of having undergone a recent interaction
event. All the observed galaxies are classified as unbarred or weakly
barred galaxies by \citet[hereafter RC3]{rc3}.  They are bright
($B_T\leq13.5$, RC3) and nearby ($V_{\rm CMB} < 4500$ \kms , RC3)
lenticulars and spirals with a low-to-intermediate inclination ($i
\leq 65^\circ$, RC3). 12 of them were identified as member of either
the Fornax, Eridanus and Pegasus cluster and 2 are member of the NGC
7582 group \citep{ferguson89,fouqetal92,garcia93,nishetal00}.  
The final sample is formed by 12 disc cluster and 2 group galaxies.

For every galaxy we provide the local galaxy number density using the
method of distance to the $5^{\rm th}$ nearest neighbour
\citep{baloetal04}. The projected galaxy density can be defined by
$\Sigma_5 = 5/(\pi d^2_5)$, where $d_5$ is the distance to the
$5^{th}$ nearest neighbour. To this aim we adopted the catalog of
galaxies available in the Hyperleda database \citep{patuetal03}. We
considered only galaxies brighter than $M_B>-16$ with systemic
velocity in a range between $\pm 1000$ \kms \ with respect to the
velocity of the sample galaxy (to avoid the background/foreground
contamination). The sample galaxies reside in a high-density
environment. In fact, their $\Sigma_5$ is much higher than the typical
value of the field galaxies \citep[$\Sigma_5 =0.05 $][]{baloetal04}.

The galaxy number density along with an overview of the basic
properties of the sample galaxies is given in Table \ref{tab:sample}.

%%%%%%%%%%%%%%%%%%%%%%%%%%%%%%%%%%%%%%%%%%%%%%%%%%%%%%%%%%%%%%%%%%%%%%%%%%%%%%%%
\begin{table*}
\caption{Properties of the sample galaxies. The columns show the
  following: (2) the morphological classification from RC3; (3)
  numerical morphological type from RC3. The values marked with * were
  derived from the galaxy $B/T$ ratio following the morphology-$B/T$
  relation by \citet{grah01};
  (4) major-axis position angle adopted for
  spectroscopic observations; (5) apparent isophotal diameters
  measured at a surface-brightness level of $\mu_B = 25$ mag
  arcsec$^{-2}$ from RC3; (6) total observed blue magnitude from RC3,
  except for ESO 358-50 from Lyon Extragalactic Database (LEDA); (7)
  radial velocity with respect to the CMB radiation from RC3; (8)
  distance obtained as luminosity-weighted mean radial velocity
  \citep{garcia93} divided by $H_0 = 75$ km s$^{-1}$ Mpc$^{-1}$; (9)
  absolute total blue magnitude from $B_T$ corrected for inclination
  and extinction as in LEDA and adopting $D$; (10) membership to
  Fornax (F) Eridanus (E) Pegasus (P) cluster or NGC~7582 group
  according to \citep{ferguson89,fouqetal92,garcia93,nishetal00};
  (11) projected galaxy density defined as in \citet{baloetal04}}
\begin{center}
\begin{small}
\begin{tabular}{llcr ccr cccc}
\hline
\noalign{\smallskip}
\multicolumn{1}{c}{Galaxy} &
\multicolumn{1}{c}{Type} &
\multicolumn{1}{c}{T} &
\multicolumn{1}{c}{PA} &
\multicolumn{1}{c}{$D_{25}\times d_{25}$} &
\multicolumn{1}{c}{$B_T$} &
\multicolumn{1}{c}{$V_{\rm CMB}$} &
\multicolumn{1}{c}{$D$} &
\multicolumn{1}{c}{$M_{B_T}^0$} &
\multicolumn{1}{c}{Cluster} &
\multicolumn{1}{c}{Density} \\ 
\noalign{\smallskip}
\multicolumn{1}{c}{} &
\multicolumn{1}{c}{(RC3)} &
\multicolumn{1}{c}{(RC3)} &
\multicolumn{1}{c}{($^\circ$)} &
\multicolumn{1}{c}{(arcmin)} &
\multicolumn{1}{c}{(mag)} &
\multicolumn{1}{c}{(\kms)} &
\multicolumn{1}{c}{(Mpc)} &
\multicolumn{1}{c}{(mag)} &
\multicolumn{1}{c}{} &
\multicolumn{1}{c}{$\Sigma_5$} \\
\noalign{\smallskip}
\multicolumn{1}{c}{(1)} &
\multicolumn{1}{c}{(2)} &
\multicolumn{1}{c}{(3)} &
\multicolumn{1}{c}{(4)} &
\multicolumn{1}{c}{(5)} &
\multicolumn{1}{c}{(6)} &
\multicolumn{1}{c}{(7)} &
\multicolumn{1}{c}{(8)} &
\multicolumn{1}{c}{(9)} &
\multicolumn{1}{c}{(10)}&
\multicolumn{1}{c}{(11)} \\
\noalign{\smallskip}
\hline
\noalign{\smallskip}  
ESO 358-50  & SA0$^0$?      &-2.0    &173 & $1.7\times0.7$ & 13.87 & 1154 & 17.0 & $-17.34$ & F     & 11.0 \\
ESO 548-44  & SA(r)0$^+$:   &-1.5    & 60 & $1.2\times0.5$ & 14.53 & 1561 & 22.6 & $-17.66$ & E     & 19.3 \\
IC 1993     &(R$'$)SAB(rs)b & 3.0    & 57 & $2.5\times2.1$ & 12.43 &  877 & 17.0 & $-18.81$ & F     & 6.2  \\
IC 5267     & SA(s)0/a      & 0.0    &140 & $5.2\times3.9$ & 11.43 & 1480 & 21.5 & $-20.38$ & N 7582& 2.2  \\
IC 5309     & Sb            & 3.0    & 23 & $1.3\times0.6$ & 14.52 & 3840 & 50.2 & $-19.64$ & P     & 7.6  \\
NGC 1292    & SAb(s)        & 5.0    &  7 & $3.0\times1.3$ & 11.29 & 1227 & 17.0 & $-18.90$ & F     & 4.0  \\
NGC 1351    & SA$^-$pec:    &-3.0    &140 & $2.8\times1.7$ & 12.46 & 1407 & 17.0 & $-18.67$ & F     &31.3  \\
NGC 1366    & S0$^0$        &-2.0    &  2 & $2.1\times0.9$ & 11.97 & 1182 & 17.0 & $-18.30$ & F     &10.1  \\
NGC 1425    & SA(s)b        & 3.0    &129 & $5.8\times2.6$ & 11.29 & 1402 & 17.0 & $-20.25$ & F     & 2.1  \\
NGC 7515    & S?            & 3.0$^*$& 15 & $1.7\times1.6$ & 13.16 & 4117 & 50.2 & $-20.44$ & P     & 2.1  \\
NGC 7531    & SAB(r)bc      & 4.0    & 15 & $4.5\times1.8$ & 12.04 & 1361 & 21.5 & $-19.25$ & N 7582& 4.8  \\
NGC 7557    & S?            & 4.0$^*$&163 & $0.6\times0.6$ & 15.15 & 3366 & 50.2 & $-18.74$ & P     & 2.4  \\
NGC 7631    & SA(r)b:       & 3.0    & 79 & $1.8\times0.7$ & 13.93 & 3396 & 50.2 & $-20.38$ & P     &19.1  \\
NGC 7643    & S?            & 6.0$^*$& 45 & $1.4\times0.7$ & 14.12 & 3520 & 50.2 & $-19.88$ & P     &1.2   \\
\noalign{\smallskip}
\hline
\noalign{\medskip}
\end{tabular}
\end{small}
\label{tab:sample}
\end{center}
\end{table*}
%%%%%%%%%%%%%%%%%%%%%%%%%%%%%%%%%%%%%%%%%%%%%%%%%%%%%%%%%%%%%%%%%%%%%%%%%%%%%%%%

\section{Surface photometry}
\label{sec:photometry}

\subsection{Observations and data reduction}
\label{sec:observation_photometry}

The photometric observations of the sample galaxies were carried out
in two runs at the European Southern Observatory (ESO) in La Silla
(Chile) on December 2002, 9-12 (run 1), September 2003, 26-28 (run 2).

We imaged the galaxies at the ESO 3.6-m Telescope with the ESO Faint
Object Spectrograph and Camera 2 (EFOSC2). The detector was the No.~40
Loral/Lesser CCD with $2048\,\times\,2048$ pixels of $15\,\times\,15$
$\rm \mu m^2$. A $2\times2$ pixel binning was adopted giving an
effective scale of $0.314$ arcsec pixel$^{-1}$ with a field of view of
$5.3\times8.6$ arcmin$^2$. The gain and readout noise were set to
1.3 e$^-$ ADU$^{-1}$ and 9 e$^-$, respectively.

We used the No. 642 Bessel $R$-band filter centred at 6431 \AA\ with
a FWHM of 1654 \AA . For each galaxy we took $2\times60-$s images
with a offset of few pixels to be able to clean cosmic rays and bad
pixels.
Every night we observed several fields of standard stars at different
air-masses to be used for the flux calibration. For each field we took
different exposures ranging from 5 to 15 s to have good
signal-to-noise ratio ($S/N$) and well sampled PSF for all the
standard stars.
The typical value of the seeing FWHM during the galaxy exposures was
$1.0$ arcsec as measured by fitting a two-dimensional Gaussian to the
field stars.

All images were reduced using standard \iraf\footnote{\iraf\ is
  distributed by NOAO, which is operated by AURA Inc., under contract
  with the National Science Foundation.} tasks. We first subtracted a
bias frame consisting of ten exposures for each night. The images were
flat-fielded using sky flats taken at the beginning and/or end of each
observing night. The sky background level was removed by fitting
  a Legendre polynomial (with a degree free to range between 0 and 2)
  to the regions free of sources in the images. Special care was taken
  during sky subtraction to reach the outermost parts of the
  objects. No scattered light was observed in the images, and the
  adopted polynomial degree was either 0 or 1. Cosmic rays and bad
pixels were removed by combining the different exposures using field
stars as a reference and adopting a sigma clipping rejection
algorithm. Residual cosmic rays and bad pixels were corrected by
manually editing the combined image.

The photometric calibration constant includes only the correction for
atmospheric extinction, which is taken from the differential aerosol
extinction for ESO \citep{burketal95}.  No color term has been
considered and no attempt was made to correct for internal and
Galactic extinction. Fig. \ref{fig:decomposition} shows the calibrated
$R$-band images of the sample galaxies.

Isophote-fitting with ellipses, after masking foreground stars and
residual bad columns, was carried out using the \iraf\ task {\tt
ELLIPSE}. In all cases, we first fit ellipses allowing their centres
to vary. Within the errors, no variation in the ellipse centres was
found for all the galaxies studied in this paper. The final ellipse
fits were done at fixed ellipse centres. The ellipse-averaged profiles
of surface brightness, position angle and ellipticity are shown in
Fig. \ref{fig:decomposition} for all the sample galaxies.

\subsection{Photometric decomposition}    
\label{sec:decomposition}
    
Let $(x,y,z)$ be Cartesian coordinates with the origin corresponding
to the position of the galaxy surface-brightness peak, the $x-$axis
parallel to direction of right ascension and pointing westward, the
$y-$axis parallel to direction of declination and pointing northward,
and the $z-$axis along the line-of-sight and pointing toward the
observer. The plane of the sky is confined to the $(x,y)$ plane.

We assumed the galaxy surface-brightness distribution to be the sum of
the contributions of a bulge and a disc component.
Bulge isophotes are ellipses centred on $(x_0,y_0)$, with constant
position angle PA$_{\rm b}$ and constant axial ratio $q_{\rm b}$. Disc
isophotes are ellipses centred on $(x_0,y_0)$, with constant position
angle PA$_{\rm d}$ and constant axial ratio $q_{\rm d}$, implying that
the galaxy inclination is $i = \arccos q_{\rm d}$.
We adopted the S\'ersic law \citep{sersic68} to describe the surface
brightness of the bulge component
\begin{eqnarray} 
I_{\rm b}(x,y)=I_{\rm e}
10^{-b_{n}\left[\left(r_{\rm b}/r_{\rm e}\right)^{1/n}-1\right]},
\label{eqn:bulge_surfbright}
\end{eqnarray}
\noindent 
where $r_{\rm e}$, $I_{\rm e}$ and $n$ are respectively the effective
radius, surface brightness at $r_{\rm e}$, and a shape parameter
describing the curvature of the profile. The value of $b_n$ is coupled
to $n$ so that half of the total flux is always within $r_{\rm e}$ and
can be approximated as $b_{n}=0.868n-0.142$ \citep{caonetal93}. The
radius $r_{\rm b}$ is given by
\begin{eqnarray}
r_{\rm b} & = &\left[(-(x-x_0)\sin{{\rm PA}_{\rm b}} + 
(y-y_0) \cos{{\rm PA}_{\rm b}})^2 + \right. \nonumber \\ 
    &   &\left.(-(x-x_0)\cos{{\rm PA}_{\rm b}} - 
(y-y_0) \sin{{\rm PA}_{\rm b}})^2/q_{\rm b}^2\right]^{1/2} 
\label{eqn:bulge_radius}
\end{eqnarray}
 
For the surface brightness distribution of the disc component we
assumed the exponential law \citep{free70}
\begin{equation}
I_{\rm d}(x,y) = I_0\,e^{-r_{\rm d}/h},
\label{eqn:disc_surfbright}
\end{equation}
where $h$ and $I_0$ are the scale length and central surface
brightness of the disc, respectively. The radius $r_{\rm d}$ is given
by
\begin{eqnarray}
r_{\rm d} & = &\left[(-(x-x_0) \sin{{\rm PA}_{\rm d}} + 
(y-y_0) \cos{{\rm PA}_{\rm d}})^2 + \right. \nonumber \\ 
    &   &\left.(-(x-x_0) \cos{{\rm PA}_{\rm d}} - 
(y-y_0) \sin{{\rm PA}_{\rm d}})^2/q_{\rm d}^2\right]^{1/2} 
\label{eqn:disc_radius}
\end{eqnarray}

To derive the photometric parameters of the bulge ($I_{\rm e}$,
$r_{\rm e}$, $n$, PA$_{\rm b}$, and $q_{\rm b}$) and disc ($I_0$, $h$,
PA$_{\rm d}$, and $q_{\rm d}$) and the position of the galaxy centre
$(x_0,y_0)$ we fitted iteratively a model of the surface brightness
$I_{\rm m}(x,y) = I_{\rm b}(x,y) + I_{\rm d}(x,y)$ to the pixels of
the galaxy image using a non-linear least-squares minimization based
on the robust Levenberg-Marquardt method by \citet{moreetal80}. We
adopted the technique for photometric decomposition developed in
GASP2D by \citet{mendetal07}. The actual computation has been done
using the {\tt MPFIT}\footnote{The updated version of this code is
  available on http://cow.physics.wisc.edu/~craigm/idl/idl.html}
algorithm implemented by C. B. Markwardt under the {\tt
  IDL}\footnote{Interactive Data Language} environment.
Each image pixel has been weighted according to the variance of its
total observed photon counts due to the contribution of both the galaxy
and sky, and determined assuming photon noise limitation and taking
into account for the detector read-out noise.
The seeing effects were taken into account by convolving the model
image with a circular Gaussian PSF with the FWHM measured from the
stars in the galaxy image. The convolution was performed as a product
in Fourier domain before the least-squares minimization.

%
%%%%%%%%%%%%%%%%%%%%%%%%%%%%%%%%%%%%%%%%%%%%%%%%%%%%%%%%%%%%%%%%%%%%%%%%%%%%%%%%
%% Figure 1
\begin{figure*}
\centering
\includegraphics[angle=90.0,width=0.411\textwidth]{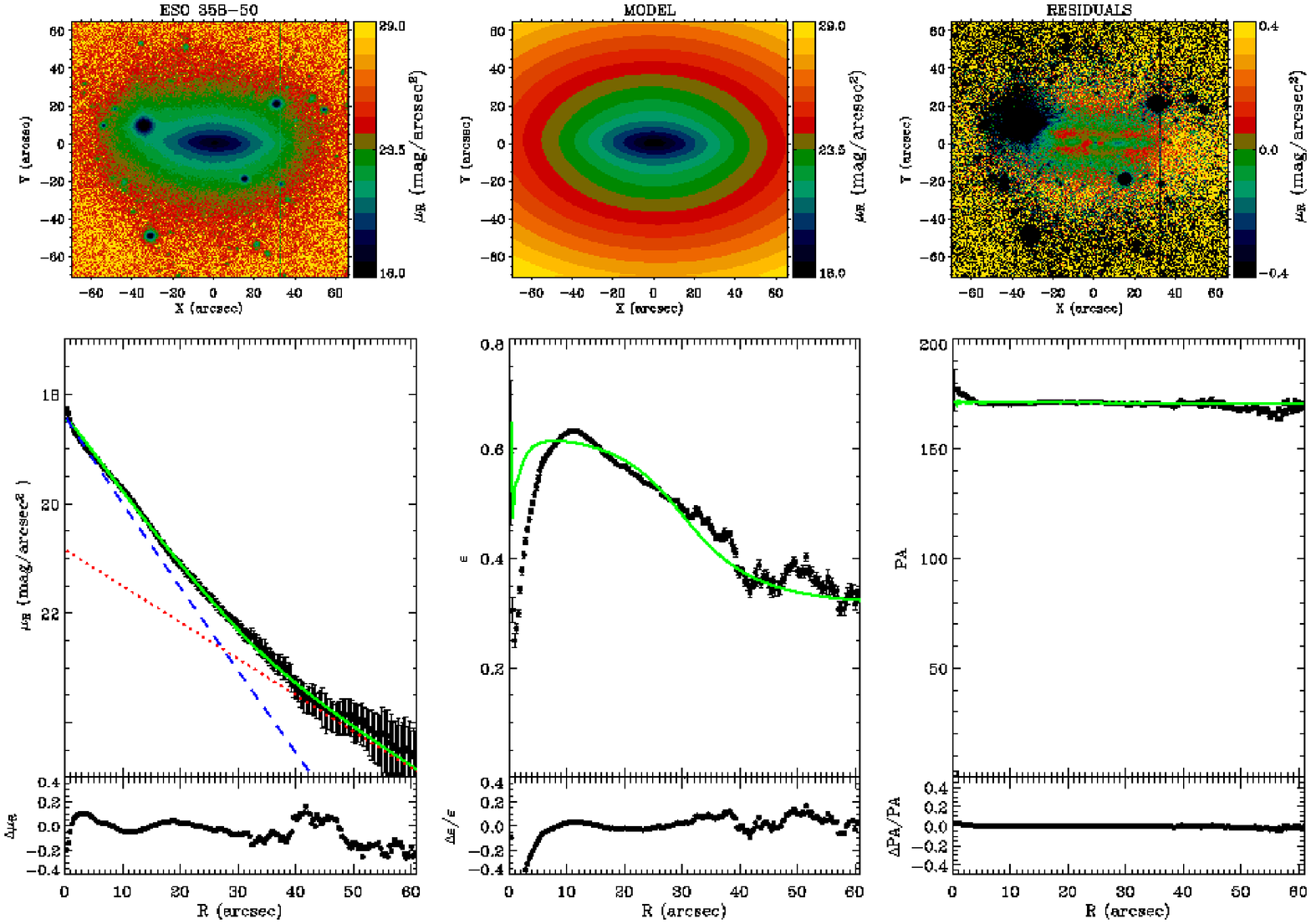}
\includegraphics[angle=90.0,width=0.411\textwidth]{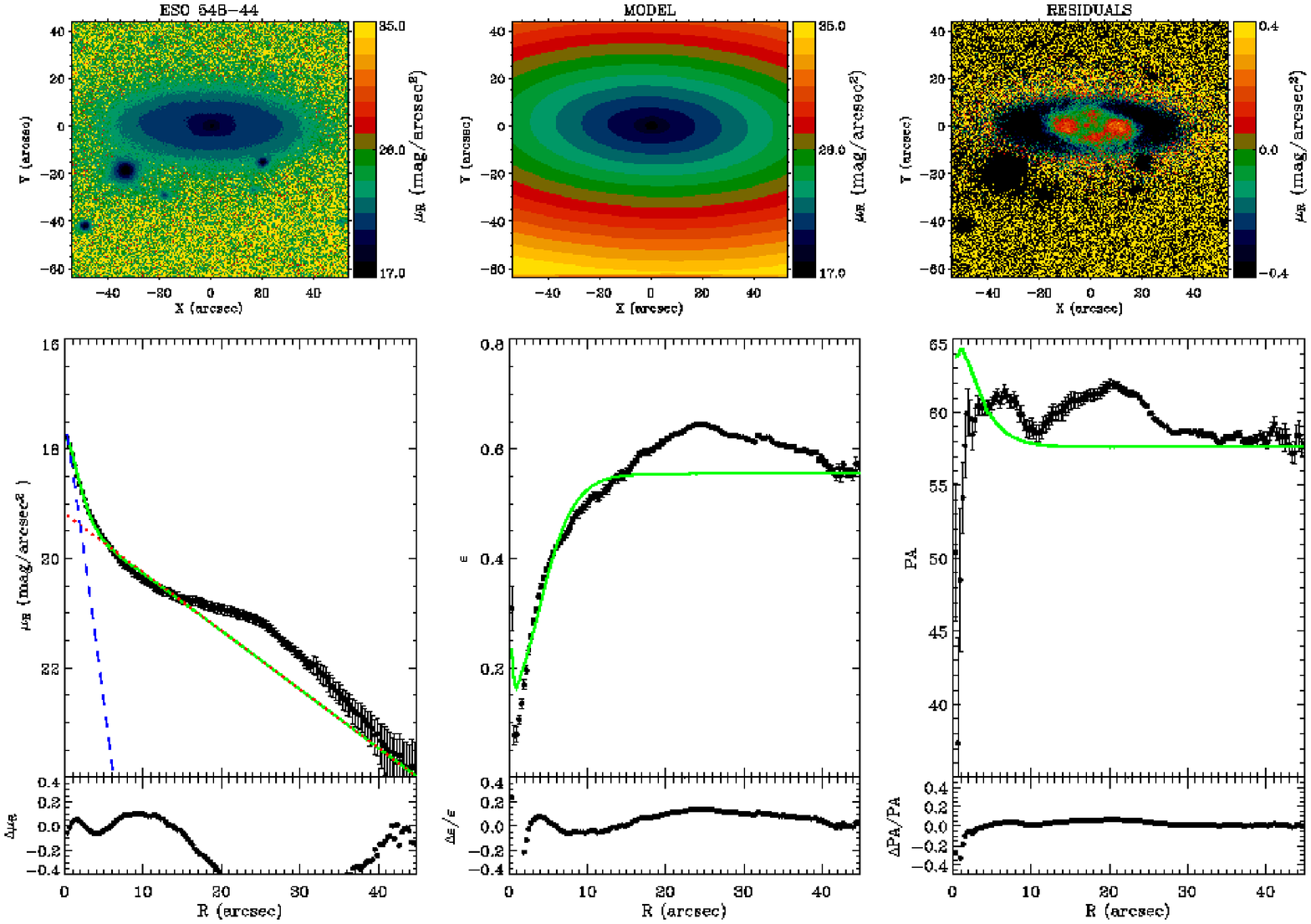}
\includegraphics[angle=90.0,width=0.411\textwidth]{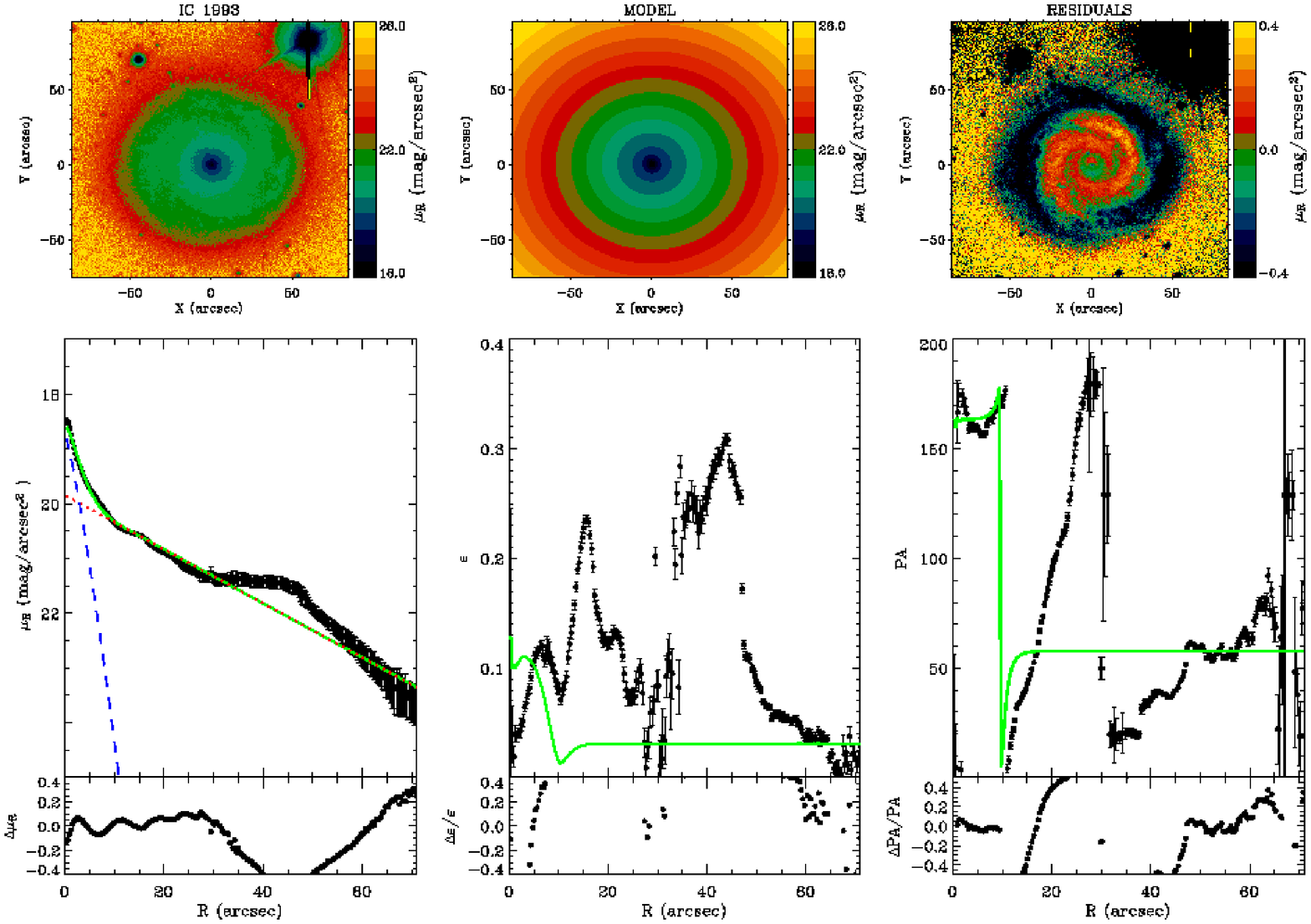}
\includegraphics[angle=90.0,width=0.411\textwidth]{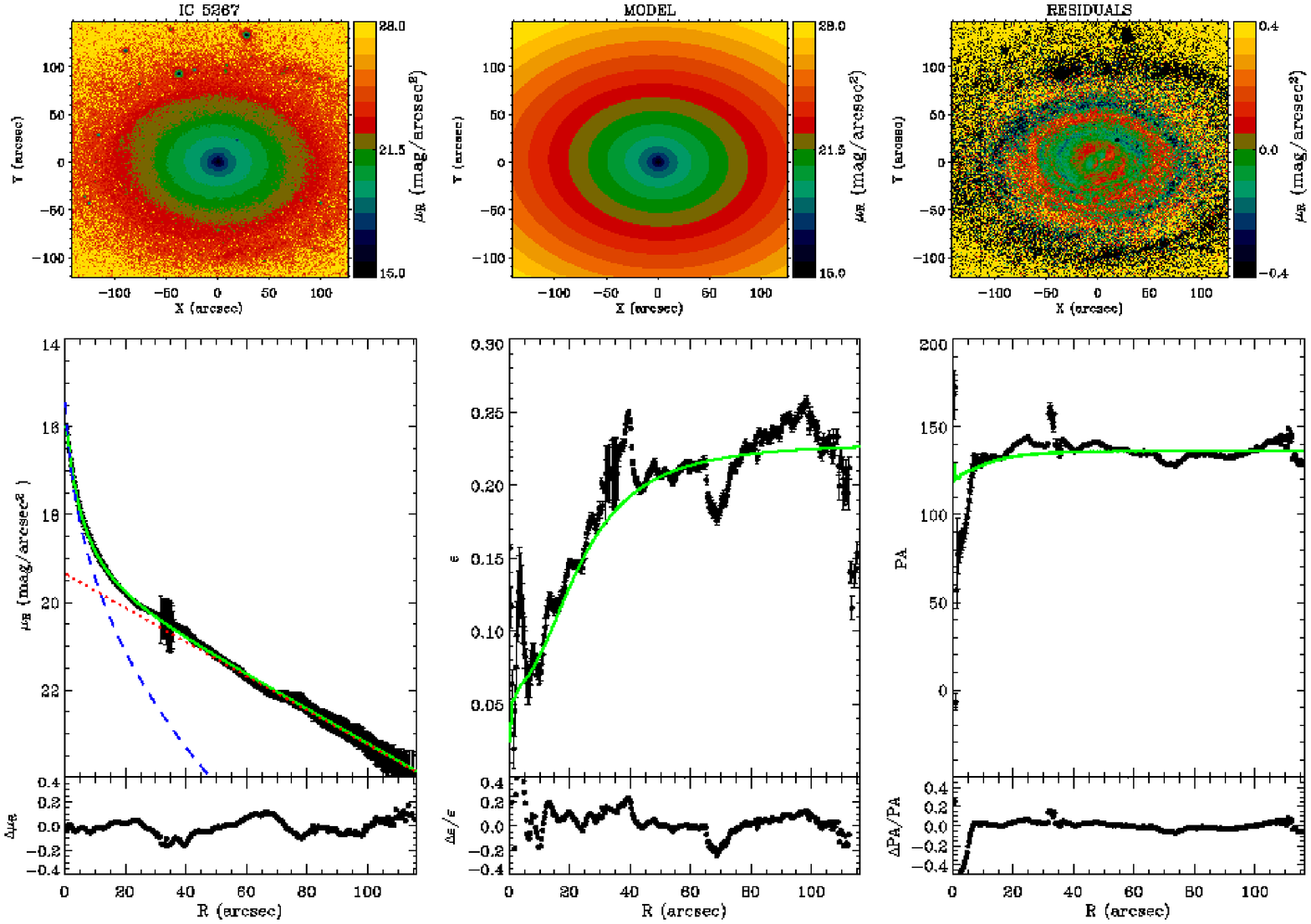}
\includegraphics[angle=90.0,width=0.411\textwidth]{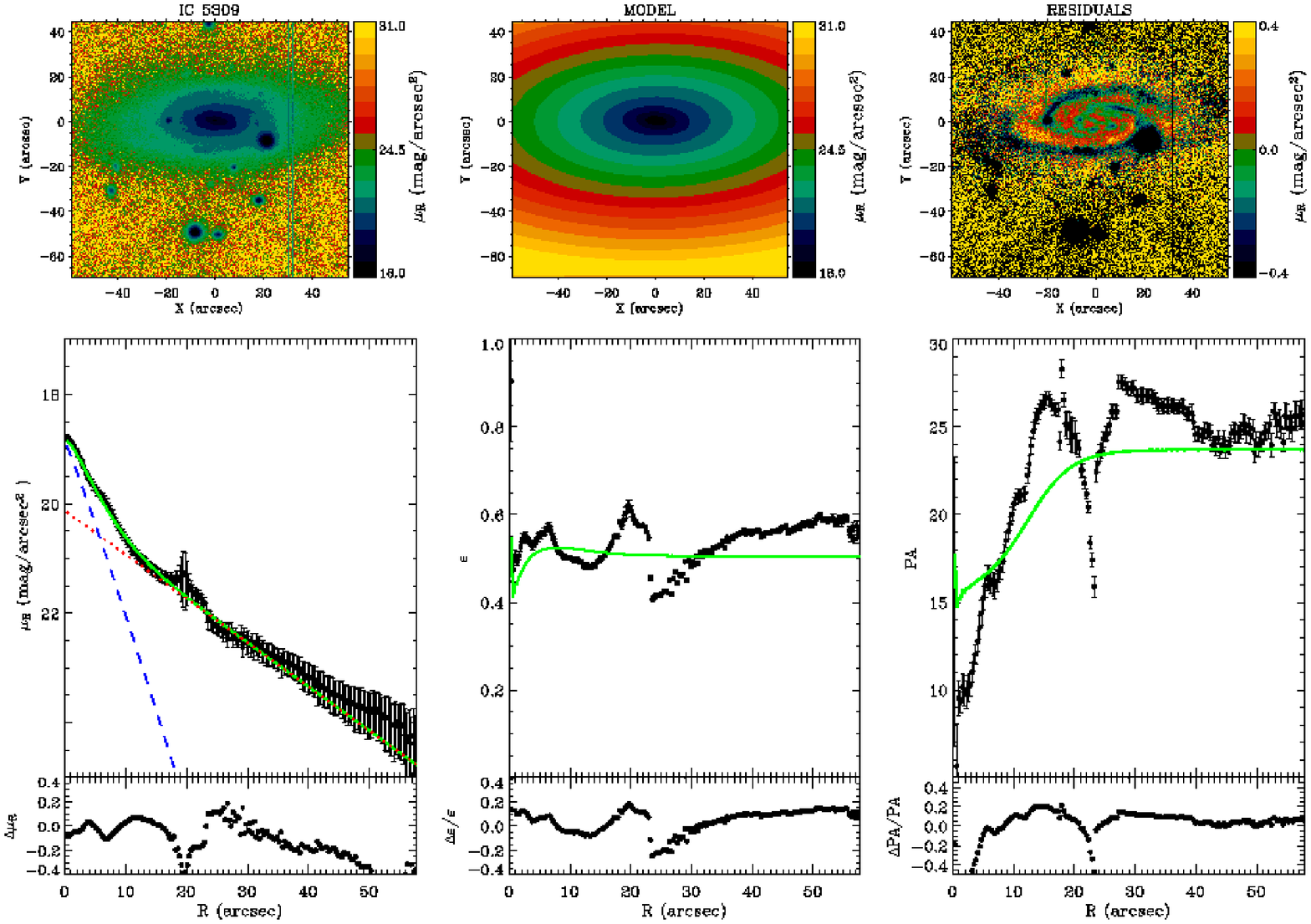}
\includegraphics[angle=90.0,width=0.411\textwidth]{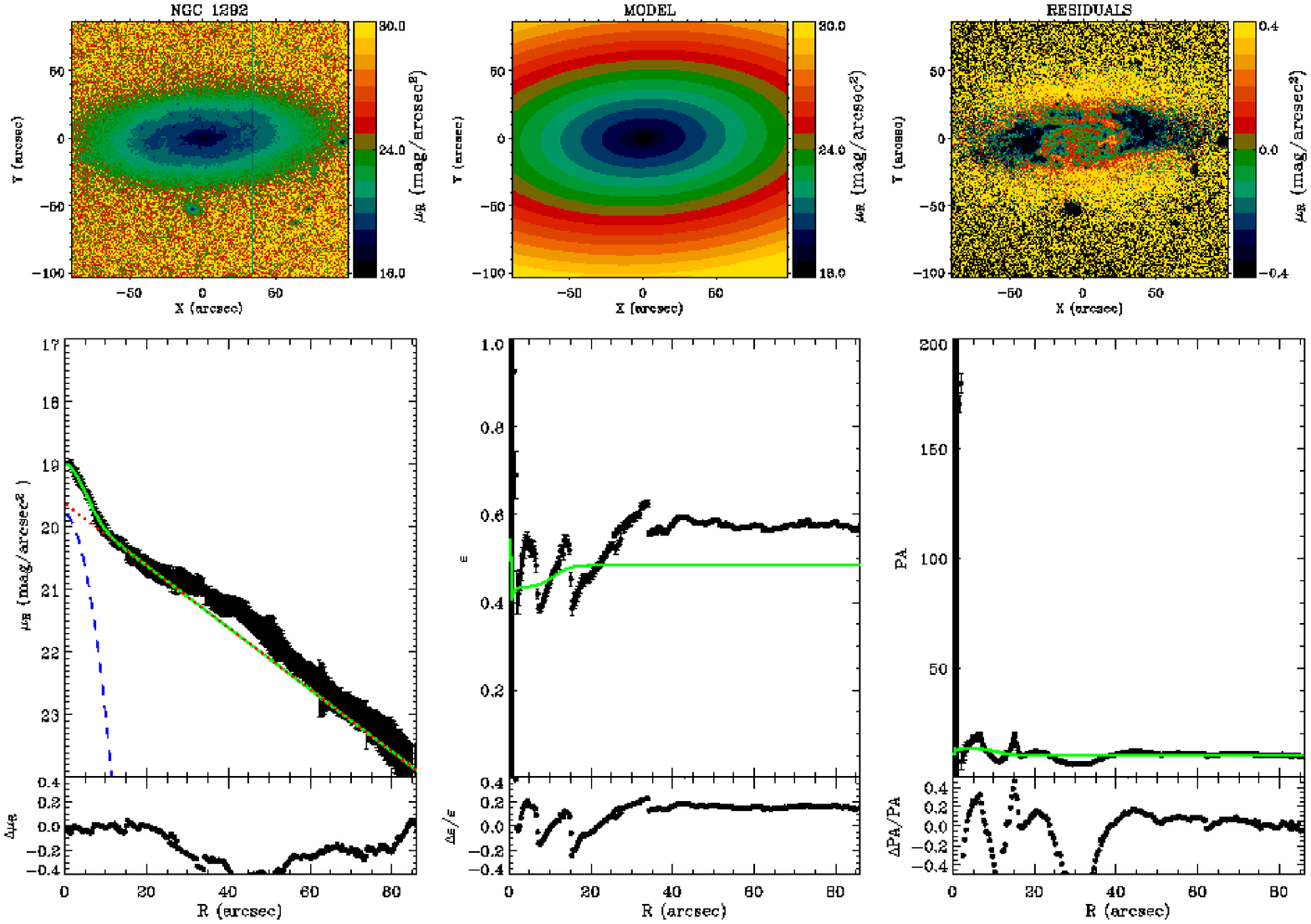}
\includegraphics[angle=90.0,width=0.411\textwidth]{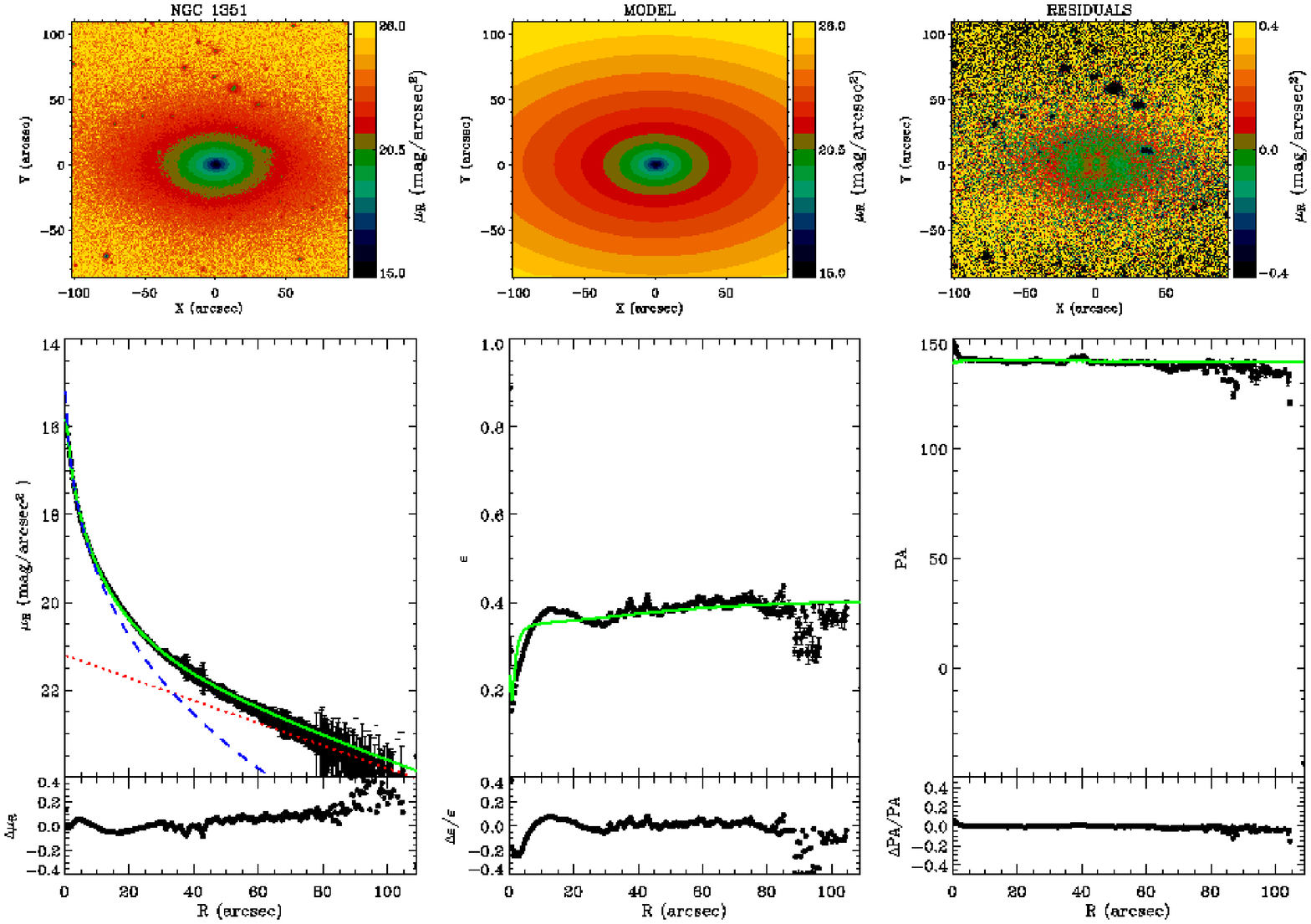}
\includegraphics[angle=90.0,width=0.411\textwidth]{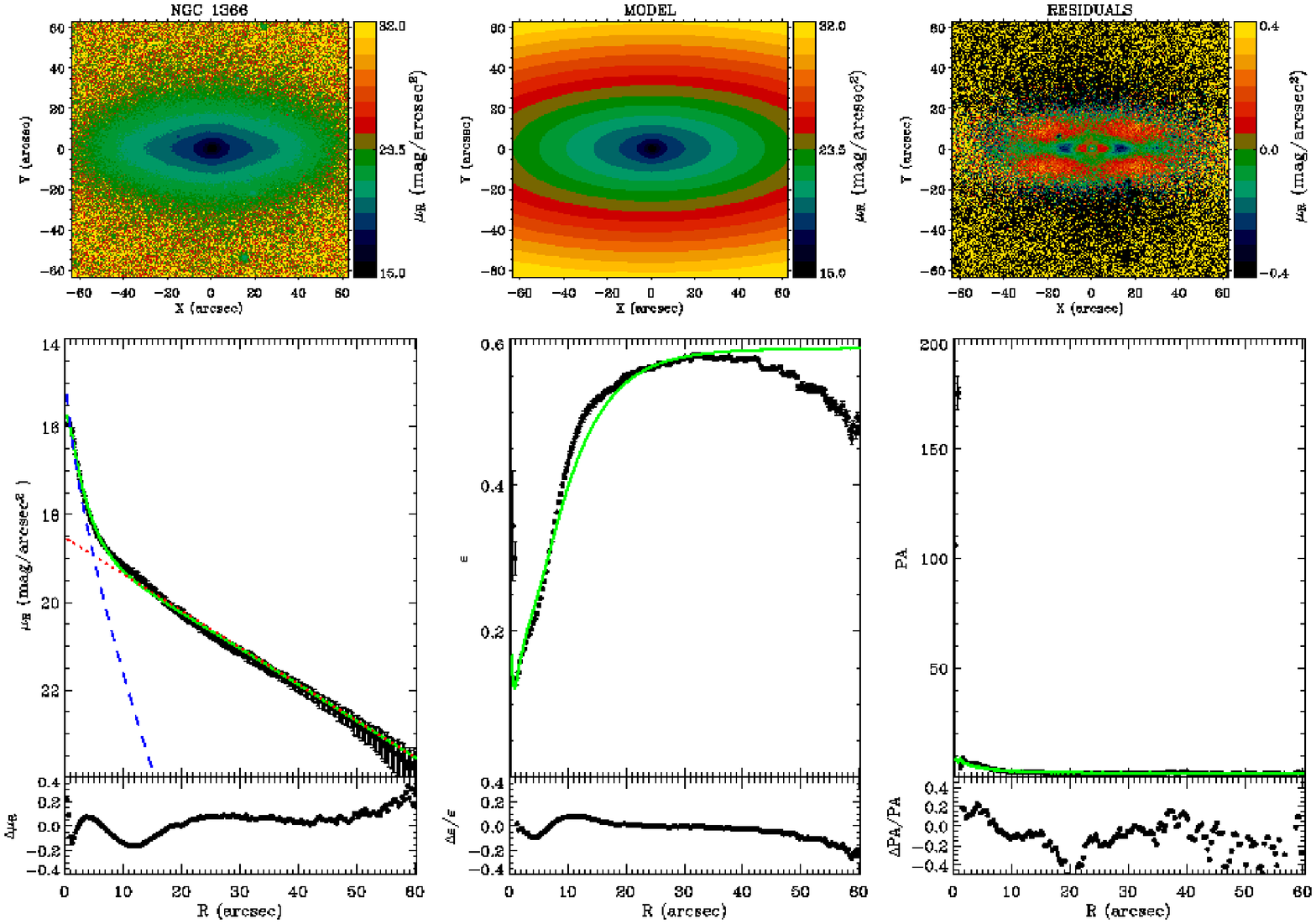}\\
\caption{Two-dimensional photometric decomposition of the sample
  galaxies.  Upper panels (from left to right): Map of the observed,
  modeled and residual (observed-modeled) surface-brightness
  distribution of the galaxy. The surface-brightness range of each
  image is indicated at the right of the panel. All images were
  rotated to have the galaxy major axis parallel to rows. In each
  panel the spatial coordinates with respect to the galaxy centre are
  given in arcsec. Lower panels (from left to right): Ellipse-averaged
  radial profile of surface-brightness, position angle, and
  ellipticity measured in the observed (dots with error-bars) and
  modeled image (solid line).  The dashed and dotted lines represent
  the intrinsic surface-brightness contribution of the bulge and disc,
  respectively. The difference between the ellipse-averaged radial
  profiles extracted from the observed and modeled images is also
  shown.
\label{fig:decomposition}}
\end{figure*}
%\end{scriptsize}
%\newpage
\begin{figure*}
\centering
\includegraphics[angle=90.0,width=0.411\textwidth]{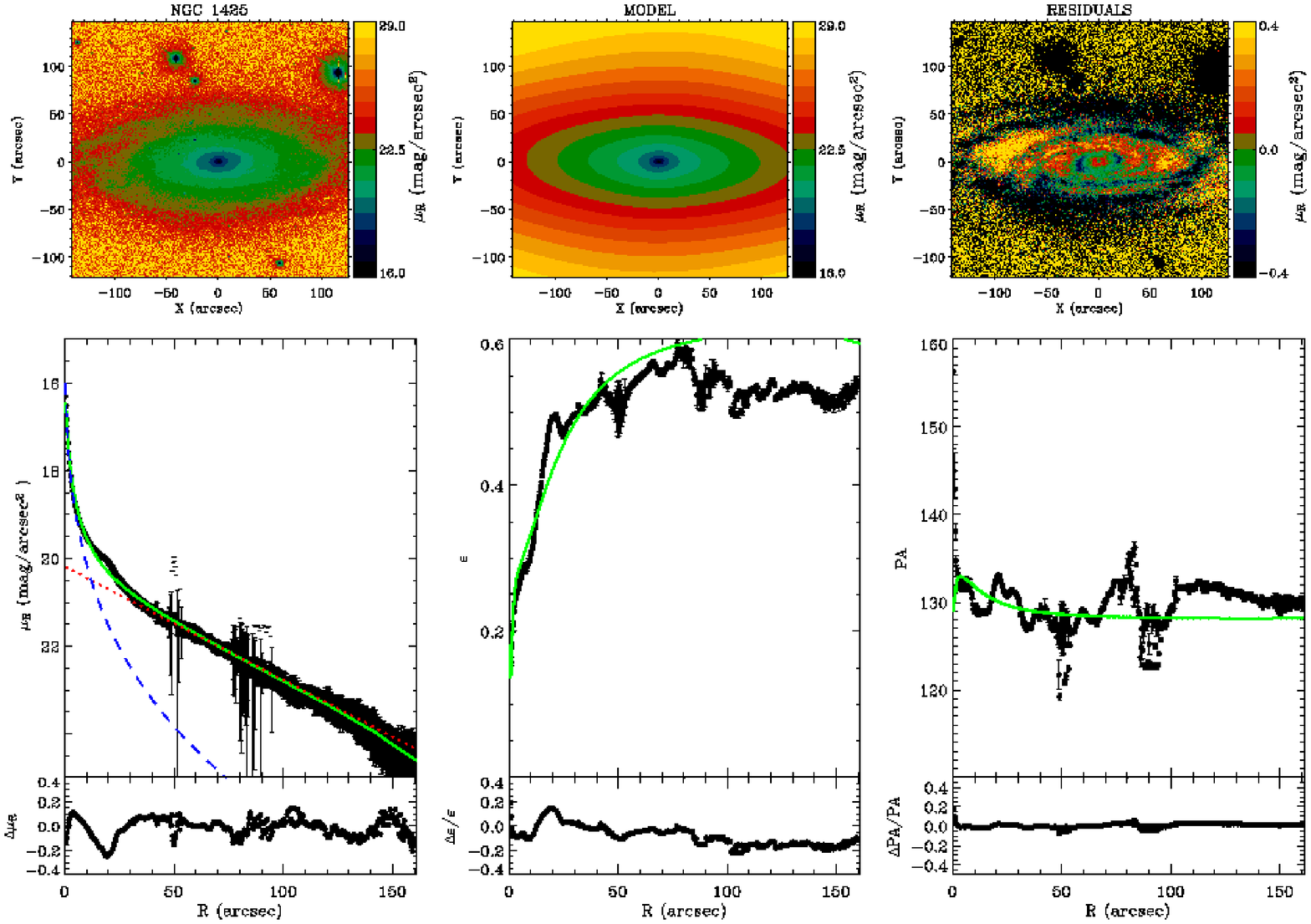}
\includegraphics[angle=90.0,width=0.411\textwidth]{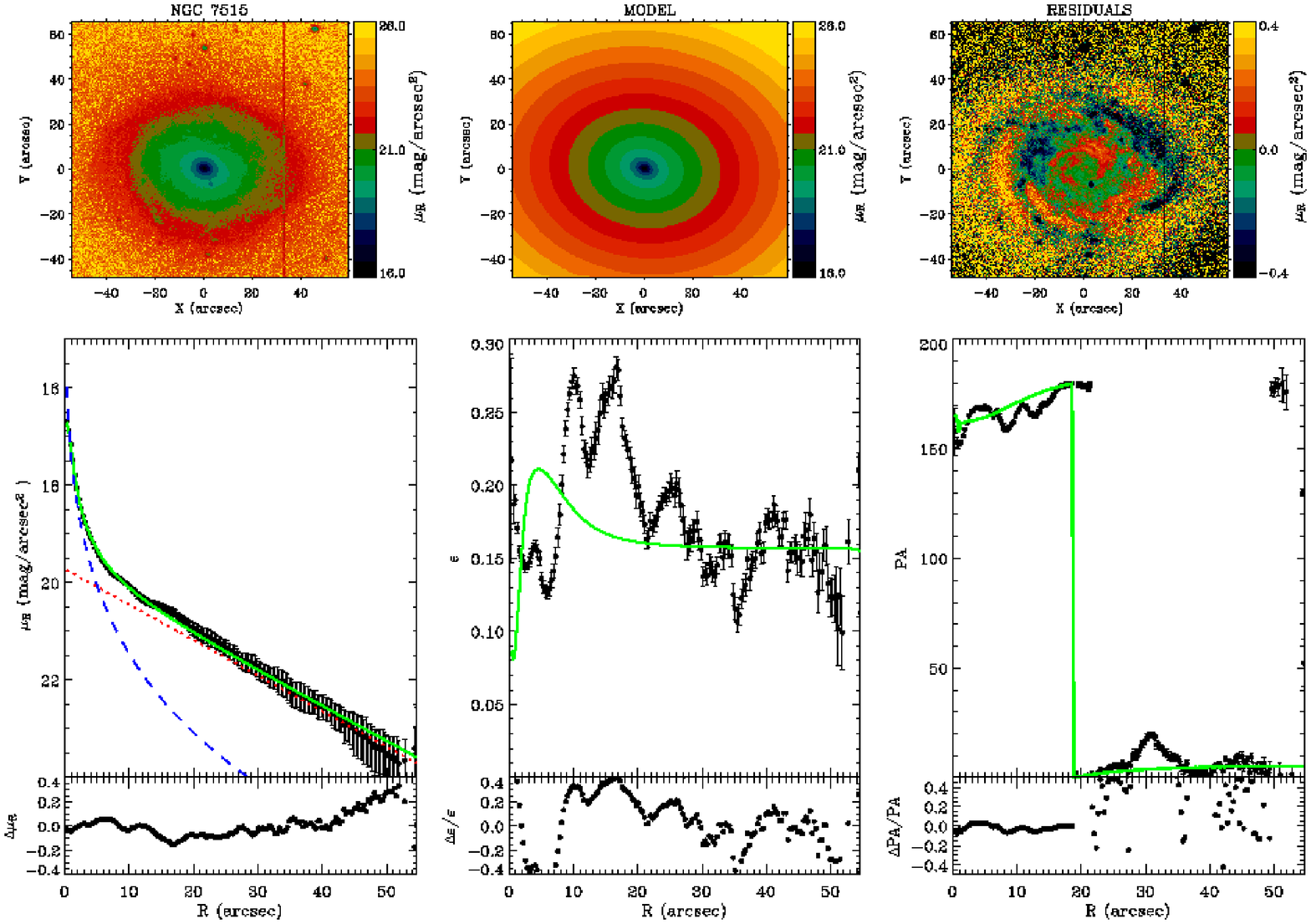}
\includegraphics[angle=90.0,width=0.411\textwidth]{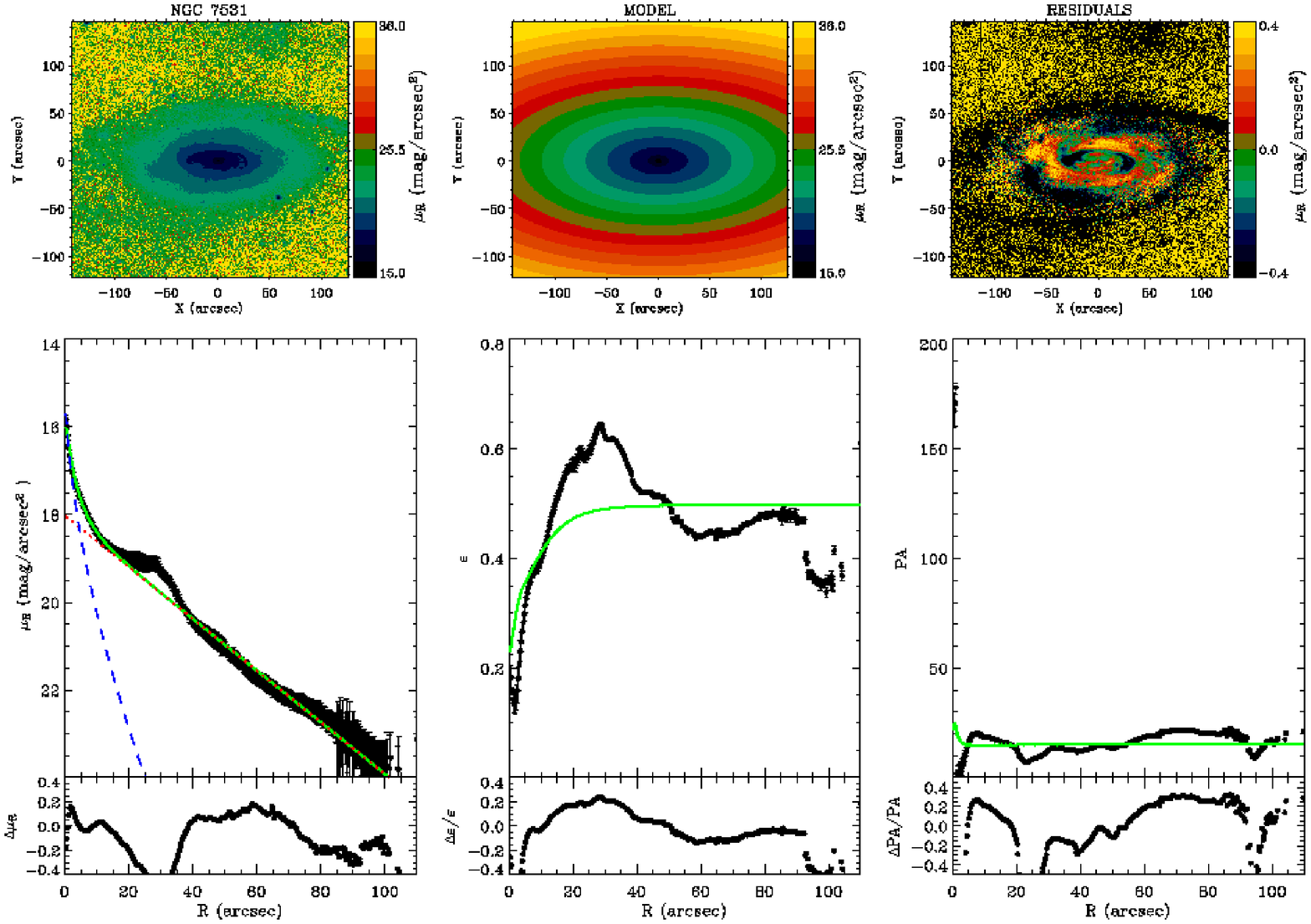}
\includegraphics[angle=90.0,width=0.411\textwidth]{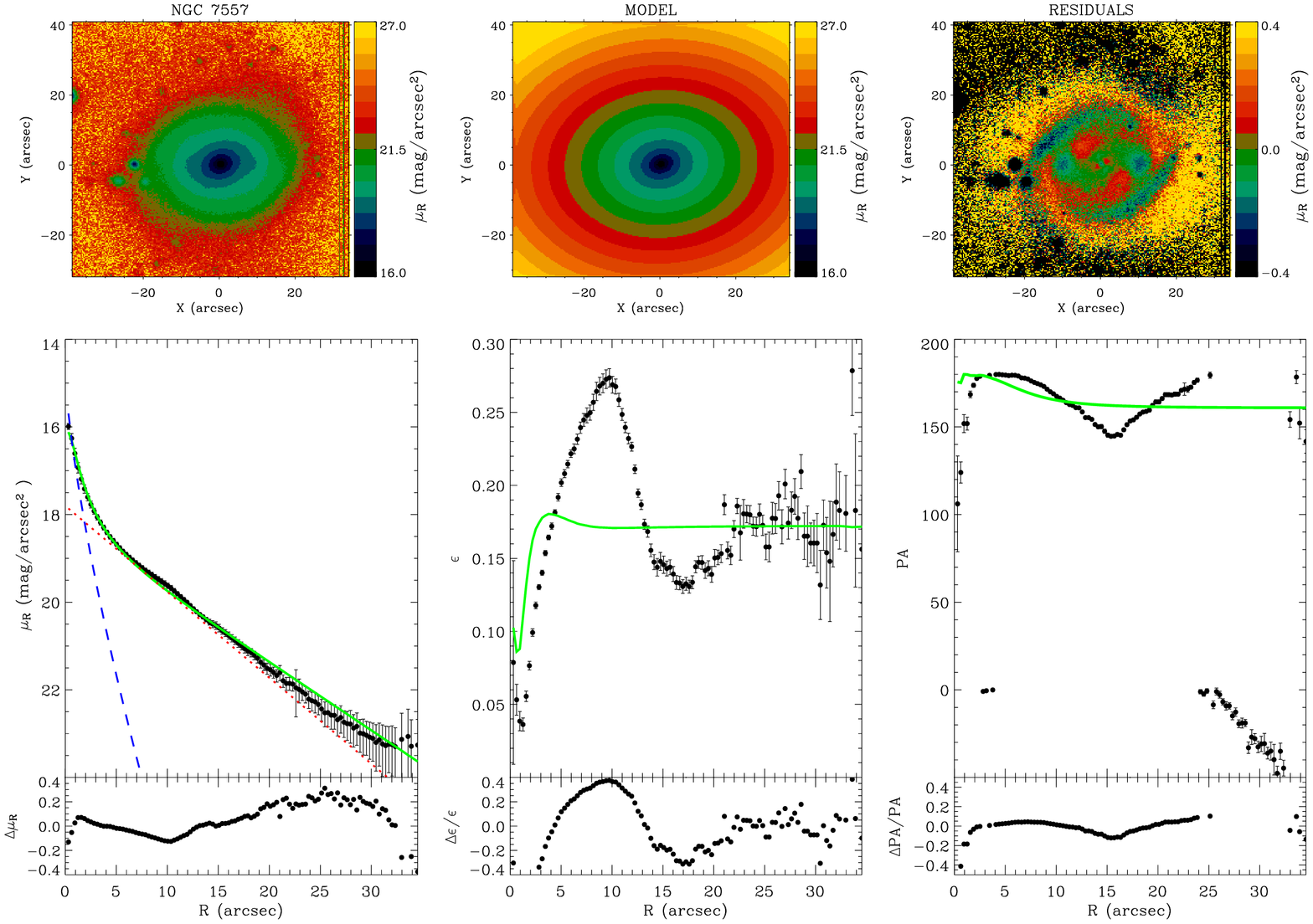}
\includegraphics[angle=90.0,width=0.411\textwidth]{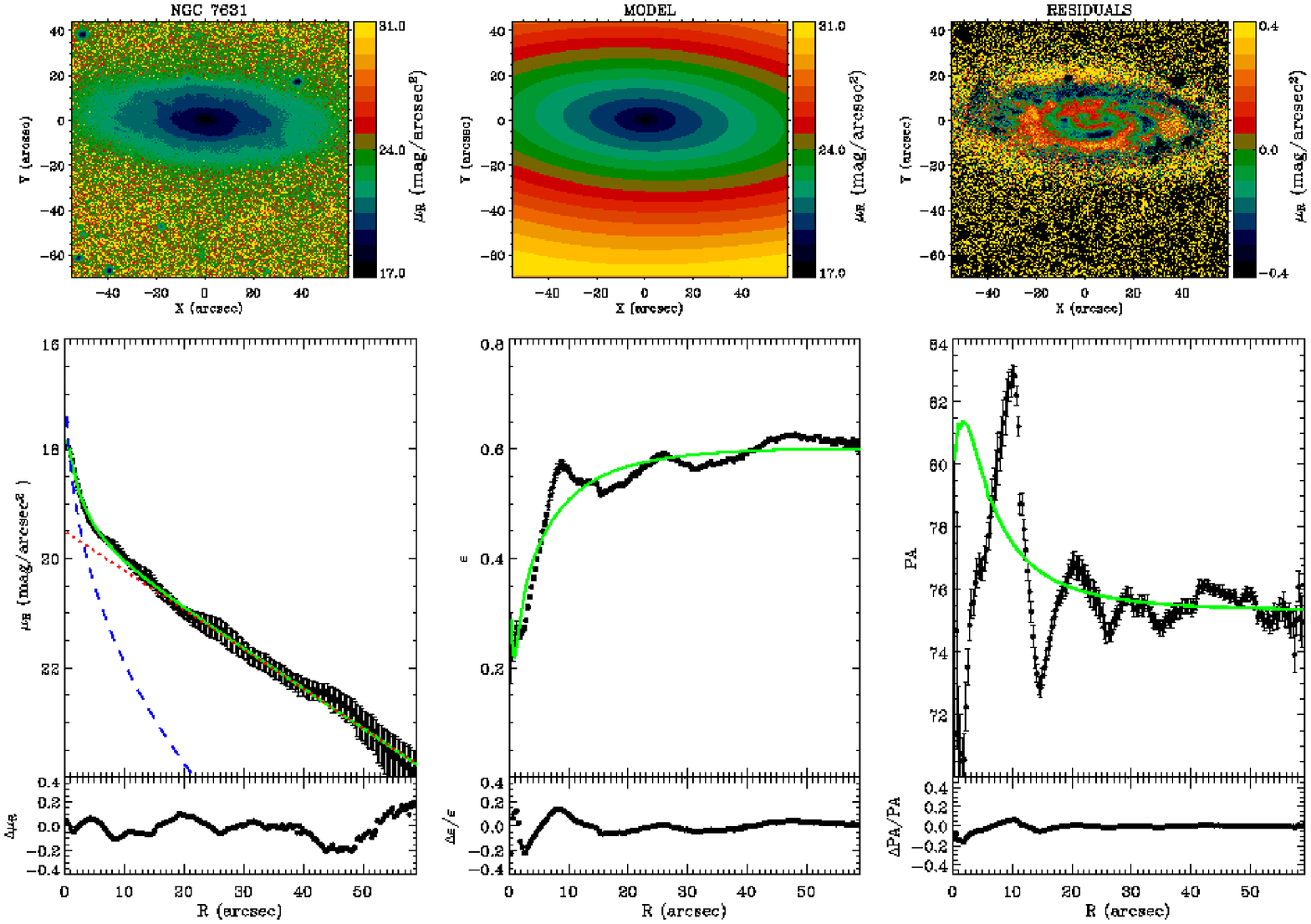}
\includegraphics[angle=90.0,width=0.411\textwidth]{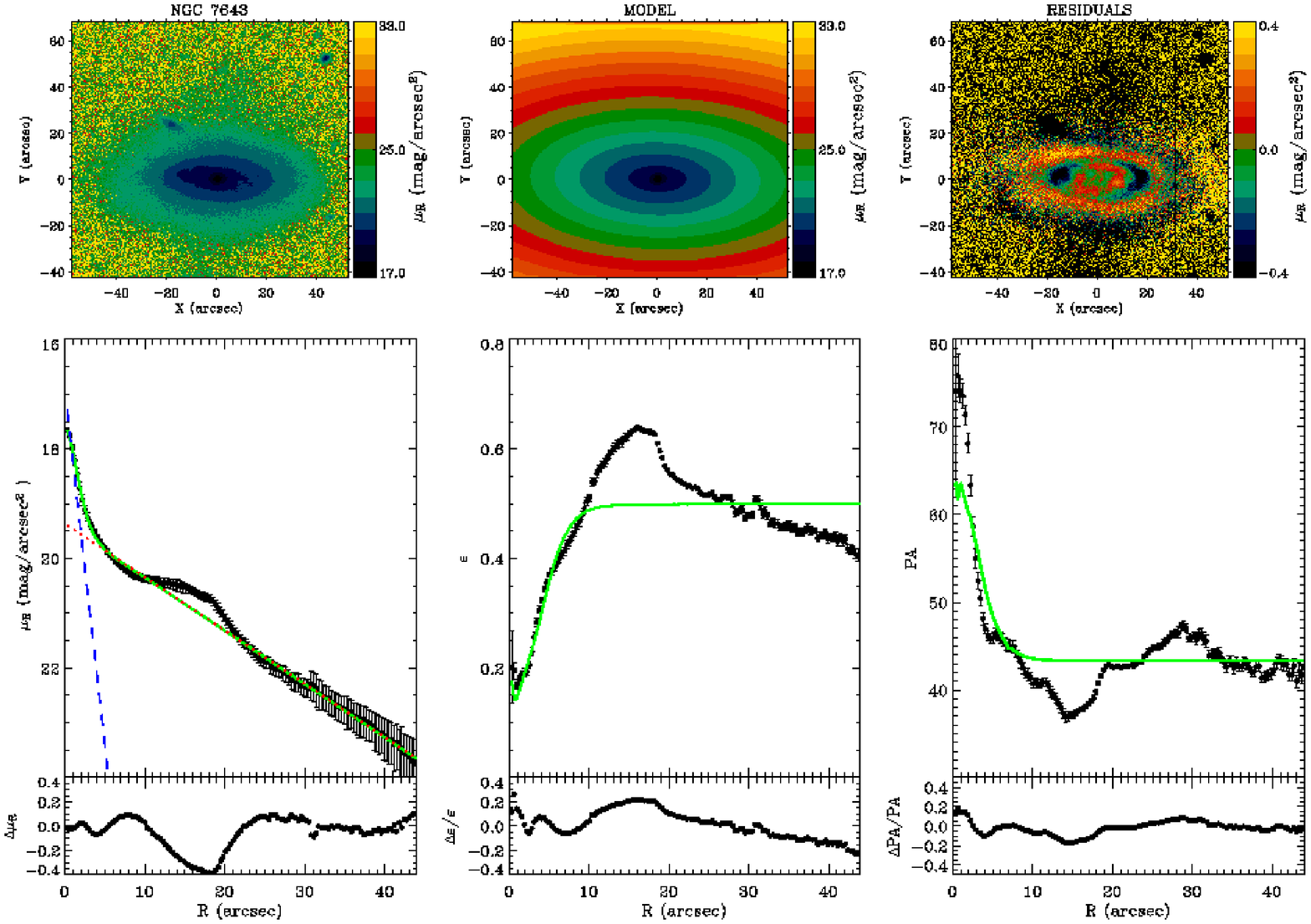}\\
\contcaption{}
\end{figure*}
%\newpage
%%%%%%%%%%%%%%%%%%%%%%%%%%%%%%%%%%%%%%%%%%%%%%%%%%%%%%%%%%%%%%%%%%%%%%%%%%%%%%%%
%

To derive the free parameters of the model surface-brightness
distribution, we adopted as initial trial for least-squares
minimization the values obtained by performing a standard photometric
decomposition with a parametric technique similar to that adopted by
\citet{Kormendy77}. The ellipse-averaged surface-brightness profile of
the galaxy has been fitted in two steps by considering first
separately and then simultaneously the light contributions of the bulge
and disc.

We began by fitting an exponential law to the galaxy
surface-brightness profile at large radii, where it is dominated by
the light contribution of the disc. In this way we derived the values
of $I_0$ and $h$. Then we extrapolated the exponential profile to
small radii and subtracted it from the galaxy surface-brightness
profile. This gave us a first estimate of the light distribution of
the bulge. We fitted to this a Sers\'ic profile by assuming the bulge
shape parameter to be $n=0.5,1,1.5,$\ldots$,6$ and deriving the
corresponding values of $I_{\rm e}$ and $r_{\rm e}$. Finally, for each
set of $I_{\rm e}$, $r_{\rm e}$, $n$, $I_0$, and $h$ we fitted
simultaneously both bulge and disc to the galaxy surface-brightness
profile and we assumed those giving the best fit as the trial values
for the two-dimensional fit, where all the parameters are allowed to
vary.

 The trial values of PA$_{\rm b}$ and $q_{\rm b}$ were obtained by
 interpolating at $r_{\rm e}$ the ellipse-averaged position-angle and
 ellipticity profiles, respectively. We assumed the coordinates of
 the image pixel where the maximum galaxy surface brightness is
 measured as the trial values of the coordinates $(x_0,y_0)$ of the galaxy
 centre.
The parameters derived for the structural components of the sample
galaxies are collected in Table \ref{tab:parameters}.  The result of
the photometric decomposition of the surface brightness distribution
of the sample galaxies is shown in Fig. \ref{fig:decomposition}.

The errors given in Table \ref{tab:phot_para} were obtained through a series of Monte
Carlo simulations. Due to the formal errors obtained from the $\chi^2$
minimization method are usually not representative of the real errors
in the structural parameters. We have carried out extensive
simulations on artificial galaxies in order to give a reliable
estimation of these errors.

We generated a set of 2000 images of galaxies with a S\'ersic bulge
and an exponential disc. The structural parameters of the artificial
galaxies were randomly chosen among the following ranges

\begin{equation}
0.5 \leq r_e \leq 3~\rm{kpc} \qquad  0.5 \leq q_b \leq 0.9 \qquad 0.5 \leq
n \leq 6
\end{equation}
\noindent
for the bulges, and

\begin{equation}
1 \leq h \leq 6~\rm{kpc} \qquad  0.5 \leq q_b \leq 0.9
\end{equation}
\noindent
for the discs. The artificial galaxies also satisfy the following
conditions

\begin{equation}
q_d \leq q_b \qquad  10 \leq m_R \leq 14~\rm{mag}
\end{equation}
where $m_R$ is the total $R$-band magnitude of the galaxy.  The
simulated galaxies were assumed to be at a distance of $\sim$ 29 Mpc,
which correspond to a scale of $\sim$ 141 pc $\rm arcsec^{-1}$. The pixel
scale used was 0.314 arcsec pixel$^{-1}$ and the CCD gain and RON were
0.7 e$^-$ ADU$^{-1}$ and 8 e$^-$ to mimic the instrumental setup of
the observed galaxies. Finally, a background level and a photon noise
were added to the resulting images to yield a signal-to-noise ratio
similar to that of the observed images.

To estimate the errors associated to the decomposition, the code
GASP 2D was applied to the artificial images as if they were
real. Then, the relative errors (1- v$_i$/v$_o$) were estimated by
comparing the input (v$_i$) and output (v$_o$) values. To
assign to every single galaxy the corresponding error for every
structural parameter, we divided our catalogue of artificial galaxies
in bins of 0.5 magnitudes, we assumed that the errors were normally
distributed, with mean and standard deviation corresponding to the
systematic and typical error respectively. Then we placed our observed
galaxy in its magnitude bin and assigned to every parameter the
corresponding error.

\section{Long-slit spectroscopy}
\label{sec:spectroscopy}

\subsection{Observations and data reduction}
\label{sec:observation_spectroscopy}

The spectroscopic observations of the sample galaxies were carried out
in three runs at ESO in La Silla (Chile) on December 2002, 9-12 (run
1), September 2003, 26-28 (run 2), and January 2005, 25 (run 3).  

In run 1 and 2 the 3.6-m ESO Telescope mounted the EFOSC2
spectrograph. The grism No.~9 with 600 $\rm grooves\,mm^{-1}$ was used
in combination with the 1.0 arcsec $\times$ 5.0 arcmin slit and the
No.~40 Loral/Lesser CCD with $2048\,\times\,2048$ pixels of
$15\,\times\,15$ $\rm \mu m^2$. A $2\times2$ pixel binning was
adopted.  The wavelength range between 4700 and 6770 \AA\ was covered
with a reciprocal dispersion of 1.98 \AA\ pixel$^{-1}$ after pixel
binning. This guarantees an adequate oversampling of the instrumental
broadening function. Indeed, the instrumental dispersion, obtained by
measuring the width of emission lines of a comparison spectrum after
the wavelength calibration, was $5.10$ \AA\ (FWHM). This corresponds to
$\sigma_{\rm inst}\sim110$ \kms\ at 5735 \AA . The spatial scale was
$0.314$ arcsec pixel$^{-1}$ after pixel binning.

In run 3 the New Technology Telescope mounted the ESO 
 Multi-Mode Instrument (EMMI) in red medium-dispersion
spectroscopic (REMD) mode. It used the grating No.~6 with 1200 $\rm
grooves\,mm^{-1}$ with a 1.0 arcsec $\times$ 5.5 arcmin slit. The
detector was a mosaic of the No.~62 and No.~63 MIT/LL CCDs. Each CCD
has $2048\,\times\,4096$ pixels of $15\,\times\,15$ $\rm \mu m^2$.  We
adopted a $2\times2$ pixel binning. It yielded a wavelength coverage
between about 4850 \AA\ and 5490 \AA\ with a reciprocal dispersion of
0.40 $\rm \AA\,pixel^{-1}$ after pixel binning.  The instrumental
dispersion is $1.00$ \AA\ (FWHM) corresponding to $\sigma_{\rm
inst}\sim25$ \kms\ at 5170 \AA . The spatial scale was $0.332$ arcsec
pixel$^{-1}$ after pixel binning.

%%%%%%%%%%%%%%%%%%%%%%%%%%%%%%%%%%%%%%%%%%%%%%%%%%%%%%%%%%%%%%%%%%%%%%%%%%%%%%%%
\begin{landscape}
\begin{table}
\caption{The photometric parameters of bulge and disc in the sample
  galaxies.  The columns show the following: 
(2) the effective surface brightness of the bulge; 
(3) effective radius of the bulge; 
(4) shape parameter of the bulge; 
(5) axial ratio of the bulge isophotes; 
(6) position angle of the bulge major-axis; 
(7) central surface brightness of the disc; 
(8) scale length of the disc; 
(9) axial ratio of the disc isophotes; 
(10) position angle of the disc major-axis; 
(11) bulge-to-total luminosity ratio; 
(12) radius where the bulge and the disc surface brightness are the
  same;
(13) total absolute magnitude of the bulge.
}
\begin{scriptsize}
\label{tab:parameters} 
\begin{tabular}{lrrrrrrrrrrrr}
\hline
\noalign{\smallskip}
\multicolumn{1}{c}{Galaxy} &
\multicolumn{1}{c}{$\mu_{\rm e}$} &
\multicolumn{1}{c}{$r_{\rm e}$} &
\multicolumn{1}{c}{$n$} &
\multicolumn{1}{c}{$q_{\rm b}$} &
\multicolumn{1}{c}{PA$_{\rm b}$} &
\multicolumn{1}{c}{$\mu_0$} &
\multicolumn{1}{c}{$h$} &
\multicolumn{1}{c}{$q_{\rm d}$} &
\multicolumn{1}{c}{PA$_{\rm d}$} &
\multicolumn{1}{c}{$B/T$} &
\multicolumn{1}{c}{\Rdb} &
\multicolumn{1}{c}{$m_R$} \\
\multicolumn{1}{c}{ } &
\multicolumn{1}{c}{(mag/arcsec$^2$)} &
\multicolumn{1}{c}{(arcsec)} &
\multicolumn{1}{c}{} &
\multicolumn{1}{c}{} &
\multicolumn{1}{c}{($^{\circ}$)} &
\multicolumn{1}{c}{(mag/arcsec$^2$)} &
\multicolumn{1}{c}{(arcsec)} &
\multicolumn{1}{c}{} &
\multicolumn{1}{c}{($^{\circ}$)} &
\multicolumn{1}{c}{} &
\multicolumn{1}{c}{(arcsec)} &
\multicolumn{1}{c}{(mag)} \\
\multicolumn{1}{c}{(1)} &
\multicolumn{1}{c}{(2)} &
\multicolumn{1}{c}{(3)} &
\multicolumn{1}{c}{(4)} &
\multicolumn{1}{c}{(5)} &
\multicolumn{1}{c}{(6)} &
\multicolumn{1}{c}{(7)} &
\multicolumn{1}{c}{(8)} &
\multicolumn{1}{c}{(9)} &
\multicolumn{1}{c}{(10)} &
\multicolumn{1}{c}{(11)} &
\multicolumn{1}{c}{(12)} &
\multicolumn{1}{c}{(13)} \\
\noalign{\smallskip}
\hline
\noalign{\smallskip}
ESO 358-50  & $20.00 \pm0.05$ & $11.63 \pm 0.48$  & $1.03 \pm 0.03$ & $0.36 \pm 0.01$ & $171.25 \pm 2.85$  & $20.57 \pm0.05$ & $16.12 \pm 0.67$ & $0.68 \pm 0.02$ & $170.62 \pm 2.84$  & $0.63 \pm 0.10$  & $27.0 \pm 1.1$  & $-17.83 \pm 0.22$\\
ESO 548-44  & $19.59 \pm0.05$ & $1.72  \pm 0.07$  & $0.94 \pm 0.02$ & $0.79 \pm 0.02$ & $ 65.72 \pm 3.19$  & $19.62 \pm0.05$ & $10.02 \pm 0.42$ & $0.44 \pm 0.01$ & $ 57.64 \pm 2.84$  & $0.05 \pm 0.01$  & $ 2.2 \pm 0.5$  & $-16.23 \pm 0.22$\\
IC 1993     & $19.21 \pm0.04$ & $3.22  \pm 0.09$  & $0.82 \pm 0.01$ & $0.87 \pm 0.02$ & $163.01 \pm 0.41$  & $18.91 \pm0.03$ & $21.70 \pm 0.55$ & $0.97 \pm 0.02$ & $ 57.87 \pm 2.34$  & $0.03 \pm 0.01$  & $ 4.8 \pm 0.6$  & $-15.93 \pm 0.21$\\
IC 5267     & $18.76 \pm0.02$ & $9.01  \pm 0.17$  & $2.56 \pm 0.06$ & $0.94 \pm 0.02$ & $123.51 \pm 2.05$  & $18.86 \pm0.02$ & $27.46 \pm 0.51$ & $0.77 \pm 0.01$ & $136.63 \pm 2.36$  & $0.25 \pm 0.04$  & $14.0 \pm 0.7$  & $-20.30 \pm 0.20$\\
IC 5309     & $20.80 \pm0.06$ & $5.41  \pm 0.24$  & $0.93 \pm 0.02$ & $0.44 \pm 0.01$ & $ 15.28 \pm 2.62$  & $20.40 \pm0.06$ & $13.34 \pm 0.68$ & $0.49 \pm 0.01$ & $ 23.71 \pm 3.01$  & $0.17 \pm 0.02$  & $ 7.0 \pm 0.9$  & $-18.44 \pm 0.23$\\
NGC 1292    & $20.77 \pm0.04$ & $4.71  \pm 0.14$  & $0.52 \pm 0.01$ & $0.57 \pm 0.01$ & $ 14.60 \pm 2.67$  & $19.82 \pm0.03$ & $21.62 \pm 0.59$ & $0.51 \pm 0.01$ & $ 10.06 \pm 2.80$  & $0.03 \pm 0.01$  & $ ...        $  & $-15.77 \pm 0.21$\\
NGC 1351    & $19.87 \pm0.03$ & $13.14 \pm 0.33$  & $3.39 \pm 0.07$ & $0.64 \pm 0.01$ & $140.53 \pm 2.60$  & $21.20 \pm0.02$ & $41.51 \pm 0.87$ & $0.59 \pm 0.01$ & $139.61 \pm 2.44$  & $0.53 \pm 0.12$  & $32.5 \pm 1.2$  & $-19.71 \pm 0.20$\\
NGC 1366    & $18.00 \pm0.04$ & $2.59  \pm 0.07$  & $1.50 \pm 0.03$ & $0.80 \pm 0.01$ & $  5.38 \pm 2.40$  & $19.07 \pm0.03$ & $12.81 \pm 0.32$ & $0.41 \pm 0.01$ & $  2.08 \pm 2.32$  & $0.20 \pm 0.06$  & $ 5.0 \pm 0.4$  & $-18.45 \pm 0.21$\\
NGC 1425    & $21.66 \pm0.03$ & $15.52 \pm 0.40$  & $3.70 \pm 0.07$ & $0.71 \pm 0.01$ & $133.42 \pm 2.71$  & $20.83 \pm0.02$ & $41.37 \pm 0.87$ & $0.34 \pm 0.01$ & $127.85 \pm 2.42$  & $0.18 \pm 0.06$  & $15.0 \pm 1.3$  & $-19.09 \pm 0.20$\\
NGC 7515    & $20.88 \pm0.04$ & $8.14  \pm 0.24$  & $5.09 \pm 0.10$ & $0.76 \pm 0.01$ & $162.23 \pm 1.56$  & $19.55 \pm0.03$ & $14.68 \pm 0.40$ & $0.84 \pm 0.02$ & $  8.20 \pm 2.50$  & $0.27 \pm 0.07$  & $ 6.0 \pm 0.6$  & $-20.25 \pm 0.20$\\
NGC 7531    & $18.24 \pm0.02$ & $4.58  \pm 0.08$  & $1.70 \pm 0.04$ & $0.66 \pm 0.01$ & $ 14.02 \pm 2.49$  & $17.92 \pm0.02$ & $18.15 \pm 0.34$ & $0.50 \pm 0.01$ & $ 15.00 \pm 2.45$  & $0.10 \pm 0.02$  & $ 5.0 \pm 0.4$  & $-18.53 \pm 0.20$\\
NGC 7557    & $17.06 \pm0.05$ & $1.31  \pm 0.05$  & $1.38 \pm 0.03$ & $0.82 \pm 0.01$ & $180.46 \pm 3.09$  & $17.45 \pm0.04$ & $ 5.46 \pm 0.18$ & $0.87 \pm 0.02$ & $172.667\pm 3.16$  & $0.16 \pm 0.04$  & $ 2.1 \pm 0.9$  & $-19.34 \pm 0.21$\\
NGC 7631    & $21.34 \pm0.05$ & $7.01  \pm 0.26$  & $2.68 \pm 0.06$ & $0.63 \pm 0.01$ & $ 83.40 \pm 2.55$  & $19.67 \pm0.04$ & $14.90 \pm 0.50$ & $0.38 \pm 0.01$ & $ 75.06 \pm 2.55$  & $0.12 \pm 0.04$  & $ 3.0 \pm 0.8$  & $-19.27 \pm 0.21$\\
NGC 7643    & $19.08 \pm0.05$ & $1.37  \pm 0.05$  & $1.01 \pm 0.03$ & $0.75 \pm 0.02$ & $ 63.30 \pm 3.50$  & $19.73 \pm0.05$ & $10.97 \pm 0.46$ & $0.50 \pm 0.01$ & $ 43.32 \pm 2.86$  & $0.05 \pm 0.01$  & $ 2.5 \pm 0.5$  & $-17.89 \pm 0.22$\\

\noalign{\smallskip}
\hline
\noalign{\bigskip}
\label{tab:phot_para}
\end{tabular}
\end{scriptsize}
\end{table}
\end{landscape}
%%%%%%%%%%%%%%%%%%%%%%%%%%%%%%%%%%%%%%%%%%%%%%%%%%%%%%%%%%%%%%%%%%%%%%%%%%%%%%%%
\noindent

We obtained $2\times45$-minutes spectra for all the sample galaxies in
run 1 and 2. 
In run 3 we obtained new 30-minutes spectra of ESO 358-50, ESO 548-44,
NGC 1292, and IC 1993 which turned out to have a central velocity
dispersion lower than 100 \kms\.
At the beginning of each exposure, the slit was positioned on the
galaxy nucleus using the guiding camera.  Then it was aligned along
the galaxy major axis, according to the position angle given in Table
\ref{tab:sample}.
During the three observing runs, we took spectra of several giant stars
which were selected from \citet{wortetal94} to be used as templates in
measuring the stellar kinematics and line strength indices. In run 1
and 2 we observed HR~296 (K0III-IV), HR~489 (K3III), HR~2429 (K1III),
HR~2503 (K4III), HR 2701 (K0III), HR~2970 (K0III), HR~3145 (K2III),
HR~3418 (K2III), HR~7149 (K2III), HR~7317 (K3III), and HR~7430
(G9III). In run 3 we observed HR~294 (K0III), HR~510 (G8III), HR~1318
(K3III), and HR~2035 (G8III). Additionally, we observed three
spectrophotometrical standard stars in order to flux-calibrate the
galaxy and line strength standard stars before the line
indices were measured. A spectrum of the comparison helium-argon arc
lamp was taken before and/or after each target exposure to allow an
accurate wavelength calibration.
The value of the seeing FWHM during the galaxy exposures ranged
between 0.5 and 1.3 arcsec as measured by fitting a two-dimensional
Gaussian to the guide star.

All the spectra were bias subtracted, flat-field corrected, cleaned of
cosmic rays, and wavelength calibrated using standard \iraf\ routines.
The bias level was determined from the bias frames obtained during the
observing nights to check the CCD status. The flat-field correction was
performed by means of both quartz lamp and twilight sky spectra (which
were normalized and divided into all the spectra) to correct for
pixel-to-pixel sensitivity variations and large-scale illumination
patterns due to slit vignetting. Cosmic rays were identified by
comparing the counts in each pixel with the local mean and standard
deviation (as obtained from Poisson statistics by taking into account
the gain and read-out noise of the detector) and then corrected by
interpolating over. The residual cosmic rays were corrected by
manually editing the spectra.
Each spectrum was rebinned using the wavelength solution obtained from
the corresponding arc-lamp spectrum.  We checked that the wavelength
rebinning had been done properly by measuring the difference between
the measured and predicted wavelengths for the brightest night-sky
emission lines in the observed spectral range \citep{ostetal96}. The
resulting accuracy in the wavelength calibration is better than 5
\kms.
All the spectra were corrected for CCD misalignment following Bender,
Saglia \& Gerhard (1994, BSG94). The spectra obtained for
the same galaxy in the same run were co-added using the center of the
stellar continuum as reference. This allowed to improve the $S/N$ of the
resulting two-dimensional spectrum. In such a spectrum, the
contribution of the sky was determined by interpolating a one-degree
polynomium along the outermost 20 arcsec at the two edges of the slit,
where the galaxy light was negligible, and then subtracted. A sky
subtraction better than $1\%$ was achieved.  A one-dimensional
spectrum was obtained for each kinematical template star as well as
for each flux standard star. The spectra of the kinematical templates
were deredshifted to laboratory wavelengths.

\subsection{Measuring stellar kinematics and line-strength indices}
\label{sec:kinematics}

We measured the stellar kinematics from the galaxy absorption features
present in the wavelength range and centered on the Mg line triplet
($\lambda\lambda$ 5164, 5173, 5184 \AA) by applying the Fourier
Correlation Quotient method (Bender 1990) as done by BSG94.
The spectra were rebinned along the dispersion direction to a natural
logarithmic scale, and along the spatial direction to obtain a
$S/N\geq40$ per resolution element.  For run 3 it was necessary to
average the whole spectra obtaining a one-dimensional spectrum in
order to achieve the desired $S/N$.  In few spectra of run 1 and 2 the
$S/N$ decreases to 10 at the outermost radii.
To measure the stellar kinematics of the sample galaxies we adopted
HR~296, HR~2429, and HR~2701 as kinematical templates for runs 1 and 2
and HR~296 and HR~510 for run 3. We considered the wavelength range
4817-6503 \AA\ in runs 1 and 2 and 5167-5378 \AA\ in run 3 around the
redshifted Mg lines of the galaxies. We derived for each galaxy
spectrum, the line-of-sight velocity distribution (LOSVD) along the
slit and measured the radial velocity $v$ and velocity dispersion
$\sigma$.  At each radius, they have been derived by fitting the LOSVD
with a Gaussian. The errors on the LOSVD moments were derived from photon
statistics and CCD read-out noise, calibrating them by Monte Carlo
simulations as done by BSG94. In general, errors are in the range
5--20 \kms, becoming larger in the outer regions of some galaxies
where $10 \leq S/N \leq 20$. These errors do not take into account the
possible systematic effects arising from template mismatch. The
measured stellar kinematics are plotted in Fig. \ref{fig:kinplot} and
given in Tab. \ref{tab:val_globtot}.

%%%%%%%%%%%%%%%%%%%%%%%%%%%%%%%%%%%%%%%%%%%%%%%%%%%%%%%%%%%%%%%%%%%%%%%%%%%%%%%%
\begin{table*}
\caption{The central values (averaged over 0.3 $r_{\rm e}$) of
  the velocity dispersion and line-strength of the measured Lick
  indices.}
\begin{tabular}{lrccccc}
\hline
\noalign{\smallskip}
\multicolumn{1}{c}{Galaxy} &
\multicolumn{1}{c}{$\sigma$} &
\multicolumn{1}{c}{\Fe} &
\multicolumn{1}{c}{\MgFe} &
\multicolumn{1}{c}{\Mgd} &
\multicolumn{1}{c}{\Mgb} &
\multicolumn{1}{c}{\Hb} \\
\multicolumn{1}{c}{ } &
\multicolumn{1}{c}{(\kms)} &
\multicolumn{1}{c}{(\AA)} &
\multicolumn{1}{c}{(\AA)} &
\multicolumn{1}{c}{(mag)} &
\multicolumn{1}{c}{(\AA)} &
\multicolumn{1}{c}{(\AA)} \\
\multicolumn{1}{c}{(1)} &
\multicolumn{1}{c}{(2)} &
\multicolumn{1}{c}{(3)} &
\multicolumn{1}{c}{(4)} &
\multicolumn{1}{c}{(5)} &
\multicolumn{1}{c}{(6)} &
\multicolumn{1}{c}{(7)} \\
\noalign{\smallskip}
\hline
\noalign{\smallskip}
ESO 358-50  & $  39.0 \pm 3.4$ & $2.45 \pm 0.24$  & $2.60 \pm 0.05$ &$ 0.004 \pm 2.719$  & $2.72 \pm 0.18$   & $2.35 \pm 0.15 $ \\
ESO 548-44  & $  63.8 \pm 7.0$ & $2.71 \pm 0.52$  & $2.91 \pm 0.30$ & $0.190 \pm 0.012$  & $3.04 \pm 0.40$   & $2.01 \pm 0.35 $ \\
IC 1993     & $ 182.8 \pm 15.0$ & $2.71 \pm 0.30$  & $2.94 \pm 0.10$ & $0.176 \pm 0.008$  & $3.07 \pm 0.23$   & $2.05 \pm 0.21 $ \\
IC 5267     & $ 203.1 \pm 15.1$ & $3.05 \pm 0.28$  & $3.80 \pm 0.14$ & $0.281 \pm 0.007$  & $4.65 \pm 0.26$   & $1.58 \pm 0.17 $ \\
IC 5309     & $ 108.0 \pm 15.9$ & $2.01 \pm 0.26$  & $2.15 \pm 0.05$ & $0.131 \pm 0.005$  & $2.27 \pm 0.20$   & $2.62 \pm 0.17 $ \\
NGC 1292    & $  31.7 \pm 3.4$  & $1.63 \pm 0.61$  & $1.67 \pm 0.23$ & $0.103 \pm 0.011$  & $1.61 \pm 0.46$   & $2.72 \pm 0.41 $ \\
NGC 1351    & $ 193.9 \pm 11.0$ & $2.85 \pm 0.23$  & $3.67 \pm 0.07$ & $0.279 \pm 0.005$  & $4.66 \pm 0.17$   & $1.60 \pm 0.14 $ \\
NGC 1366    & $ 175.8 \pm 9.2$  & $3.13 \pm 0.17$  & $3.67 \pm 0.04$ & $0.263 \pm 0.004$  & $4.23 \pm 0.13$   & $1.78 \pm 0.11 $ \\
NGC 1425    & $ 126.9 \pm 12.4$ & $2.59 \pm 0.21$  & $3.06 \pm 0.05$ & $0.214 \pm 0.004$  & $3.52 \pm 0.16$   & $1.72 \pm 0.14 $ \\
NGC 7515    & $ 157.7 \pm 15.6$ & $2.73 \pm 0.27$  & $3.15 \pm 0.10$ & $0.209 \pm 0.006$  & $3.56 \pm 0.22$   & $2.04 \pm 0.17 $ \\
NGC 7531    & $ 132.8 \pm 10.0$ & $2.75 \pm 0.17$  & $2.94 \pm 0.04$ & $0.199 \pm 0.004$  & $3.15 \pm 0.14$   & $2.00 \pm 0.11 $ \\
NGC 7557    & $ 104.6 \pm 15.0$ & $2.58 \pm 0.21$  & $2.78 \pm 0.04$ & $0.176 \pm 0.004$  & $2.91 \pm 0.16$   & $2.80 \pm 0.13 $ \\
NGC 7631    & $ 155.5 \pm 15.8$ & $2.45 \pm 0.37$  & $2.83 \pm 0.14$ & $0.170 \pm 0.008$  & $3.20 \pm 0.28$   & $1.93 \pm 0.24 $ \\
NGC 7643    & $ 116.9 \pm 13.0$ & $2.18 \pm 0.25$  & $2.46 \pm 0.06$ & $0.141 \pm 0.005$  & $2.71 \pm 0.19$   & $2.88 \pm 0.17 $ \\
\noalign{\smallskip}
\hline
\noalign{\bigskip}
\label{tab:centval_lickind}
\end{tabular}
\end{table*}
%%%%%%%%%%%%%%%%%%%%%%%%%%%%%%%%%%%%%%%%%%%%%%%%%%%%%%%%%%%%%%%%%%%%%%%%%%%%%%%%

Detailed measurements of kinematics is not only important to study the
dynamical properties of galaxies but also to derive the line strength
of the Lick indices. Following \citet{mehletal00}, we measured the Mg,
Fe, and \Hb\ line-strength indices (as defined by Faber et al. 1985
and Worthey et al. 1994) from the flux-calibrated spectra of run 1 and
2. Spectra were rebinned in the dispersion direction as well as in the
radial direction as before. We indicate the average Iron index with
$\rm{\left<Fe\right> = (Fe5270 + Fe5335)/2}$ \citep{gorgetal90}, and
the newly defined Magnesium-Iron index with $[{\rm
    MgFe}]^{\prime}=\sqrt{{\rm Mg}\,b\,(0.72\times {\rm Fe5270} +
  0.28\times{\rm Fe5335})}$ \citep{thmabe03}. We convolved all
  the spectra with a Gaussian with a proper $\sigma$ to degrade them
  to the fixed spectral resolution of the Lick system ($\simeq 9 $
  \AA). No focus correction was applied because the atmospheric
seeing was the dominant effect during observations \citep[see][for
  details]{mehletal98}. The errors on indices were derived from photon
statistics and CCD read-out noise, and calibrated by means of Monte
Carlo simulations. We calibrated our measurements to the Lick system
using the stars from \citet{wortetal94} we observed in run 1 and 2.
A well known problem when deriving age and metallicity of galaxy
stellar populations is the contamination of the \Hb\/ index by the
\Hb\/ emission line. To address this issue we adopted the code GANDALF
(Gas AND Absorbtion Line fitting) to fit the galaxy spectra with synthetic
population models as done by \citet{sarzetal06}.  The models were
built with different templates from the stellar libraries by
\citet{brucha03} and \citet{tremetal04}. We adopted the Salpeter
initial mass function \citep{salp55}, ages ranging between 1 Myr and
10 Gyr, and metallicities between 1 and 2.5 $(Z/\rm H)_\odot$. The
spectral resolution of the stellar templates (FWHM$\sim3$ \AA) was
degraded to match that our galaxy spectra. We simultaneously fitted
the observed spectra using emission lines in addition to the stellar
templates.
The \Hb\/ emission line was detected ($S/N > 3$) in NGC~1292,
NGC~7531, NGC~7631, NGC~7643, IC~5267, IC~5309. The equivalent with of
\Hb\/ emission line was ranging from 0 to 4 \AA\ depending on the
galaxy and radius (Tab. \ref{tab:val_globtot}). The emission line was
subtracted from the observed spectrum and we measured the \Hb\/
line-strength index from the resulting \Hb\/ absorption line. 

\begin{figure}
\centering
\includegraphics[angle=90.0,width=0.5\textwidth]{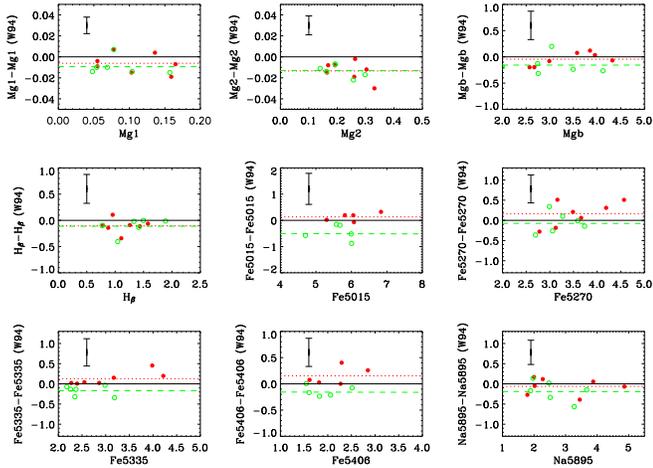}
\caption{Comparison of the values of the line-strength indices of the
  stars in common with \citet{wortetal94} (W94). The error bars in the
  upper left corner of each panel indicate the median errors of the
  measurements. Negligible offsets for all indices were obtained by
  assuming the the straight line averaging the data in run 1 (filled red
  dots) and 2 (empty green dots). }
\label{fig:indcomp}
\end{figure}

The measured values for the line-strength indices for the stars in
common with \citet{wortetal94} are shown in Fig. \ref{fig:indcomp}.
The agreement is good within the errors for all the indices and we did
not apply any zero-point correction.

%%%%%%%%%%%%%%%%%%%%%%%%%%%%%%%%%%%%%%%%%%%%%%%%%%%%%%%%%%%%%%%%%%%%%%%%%%%%%%%%
%% Figure 2
%%%%%%%%%%%%%%%%%%%%%%%%%%%%%%%%%%%%%%%%%%%%%%%%%%%%%%%%%%%%%%%%%%%%%%%%%%%%%%%%

\begin{figure}
\centering
\includegraphics[angle=0,width=0.49\textwidth]{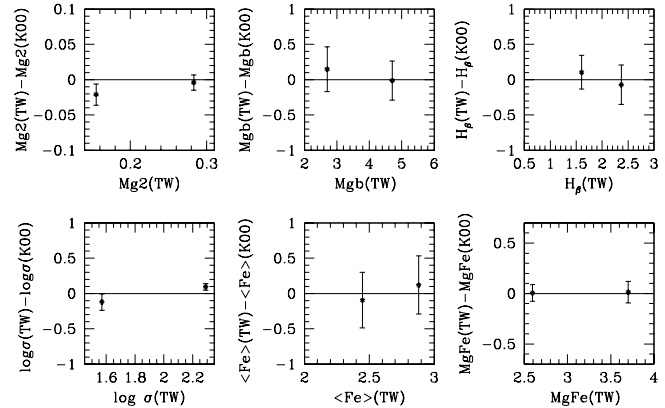}
\caption{Comparison of the values of the central velocity dispersion and 
  line-strength indices we and \citet{kuntetal00} (K00) measured for
  NGC~1351 and ESO~358-G50. Velocity dispersions are given in \kms ,
  \Mgd\ in mag, and all the other line-strength indices in \AA .}
\label{fig:conkun}
\end{figure}

\citet{kuntetal00} measured the central velocity dispersion and
line-strength indices for NGC~1351 and ESO~358-G50. The comparison is
shown in Fig. \ref{fig:conkun}. All the values are consistent with ours
within $3\sigma$.
%%%%%%%%%%%%%%%%%%%%%%%%%%%%%%%%%%%%%%%%%%%%%%%%%%%%%%%%%%%%%%%%%%%%%%%%%%%%%%%%
%% Figure 3
%%%%%%%%%%%%%%%%%%%%%%%%%%%%%%%%%%%%%%%%%%%%%%%%%%%%%%%%%%%%%%%%%%%%%%%%%%%%%%%%
The measured values of \Hb , \MgFe, \Fe, \Mgb, and \Mgd\ for all the
sample galaxies are plotted in Fig. \ref{fig:indices} and listed in
Tab. \ref{tab:val_globtot}.

%%%%%%%%%%%%%%%%%%%%%%%%%%%%%%%%%%%%%%%%%%%%%%%%%%%%%%%%%%%%%%%%%%%%%%%%%%%%%%%%
%% Figure 4
\begin{figure*}
\centering

\includegraphics[angle=0.0,width=0.431\textwidth]{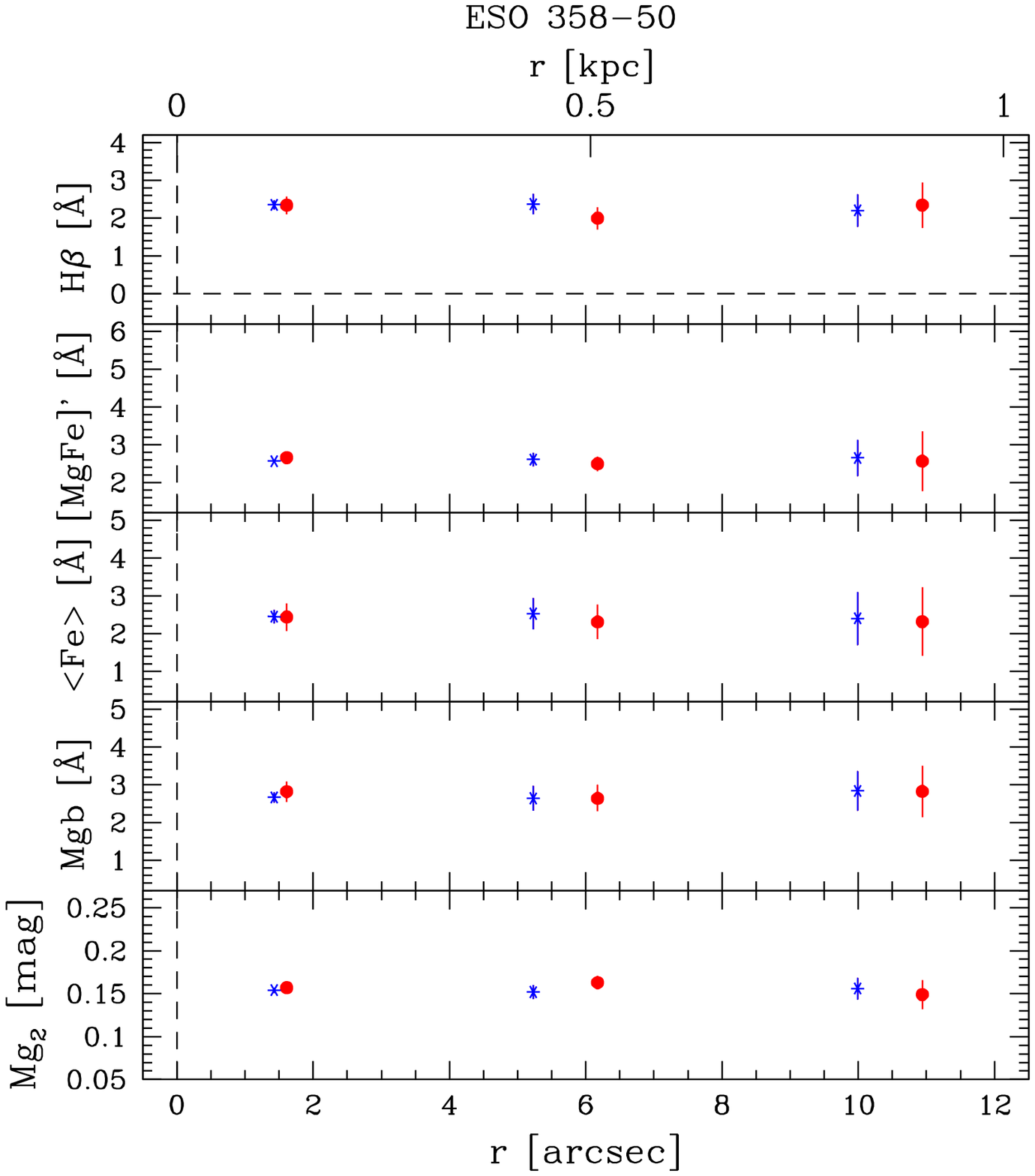}
\includegraphics[angle=0.0,width=0.431\textwidth]{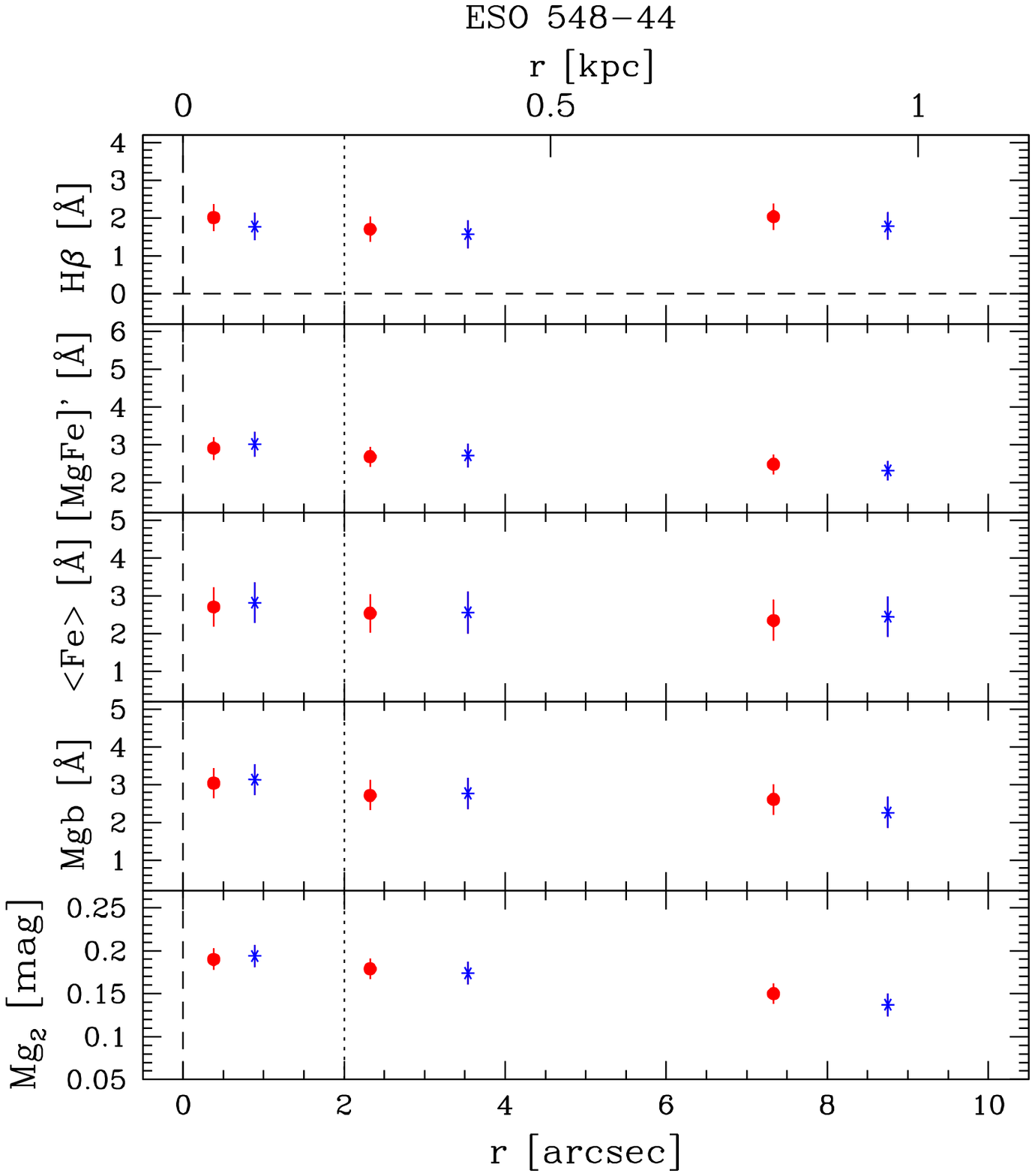}
\includegraphics[angle=0.0,width=0.431\textwidth]{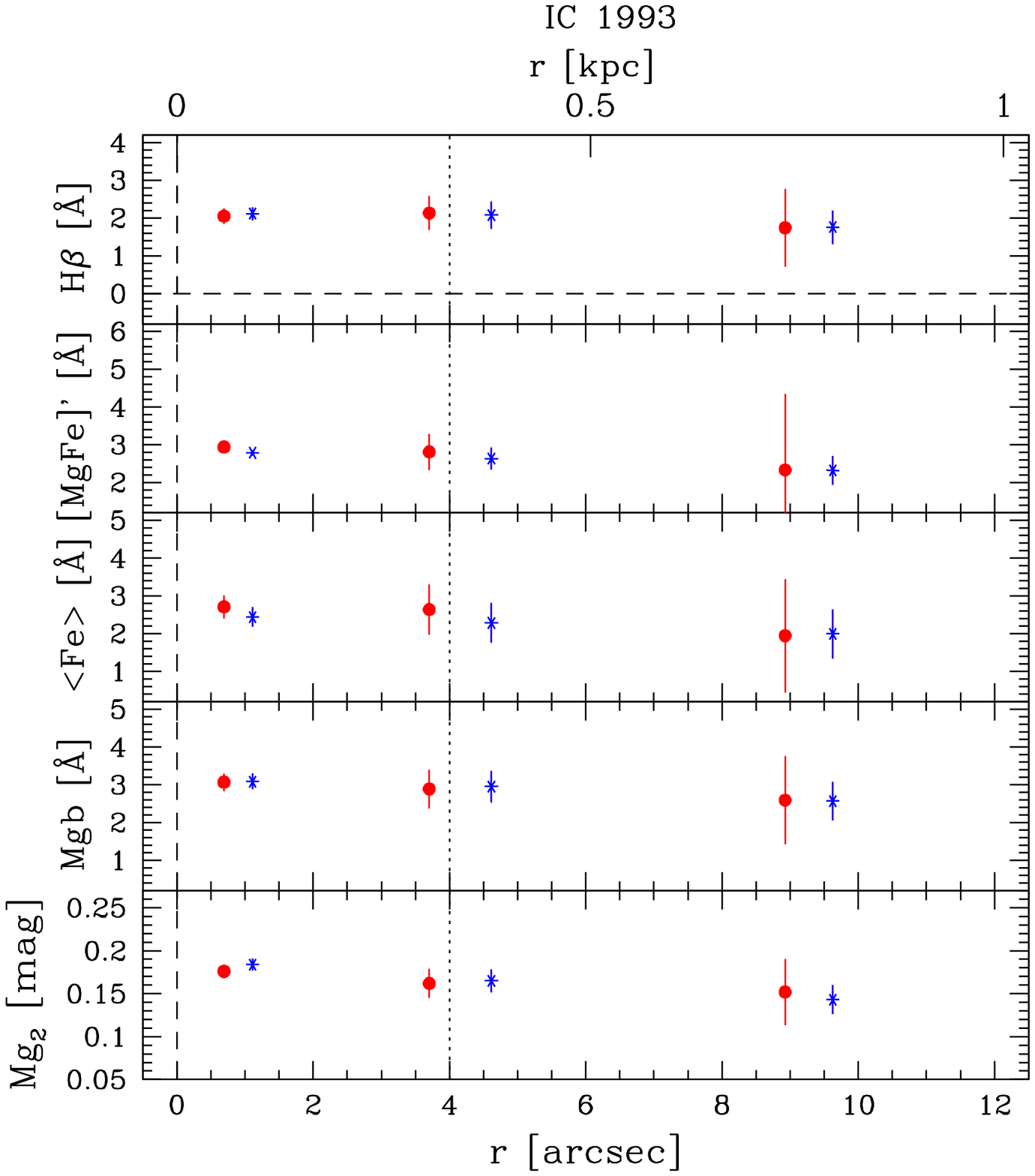}
\includegraphics[angle=0.0,width=0.431\textwidth]{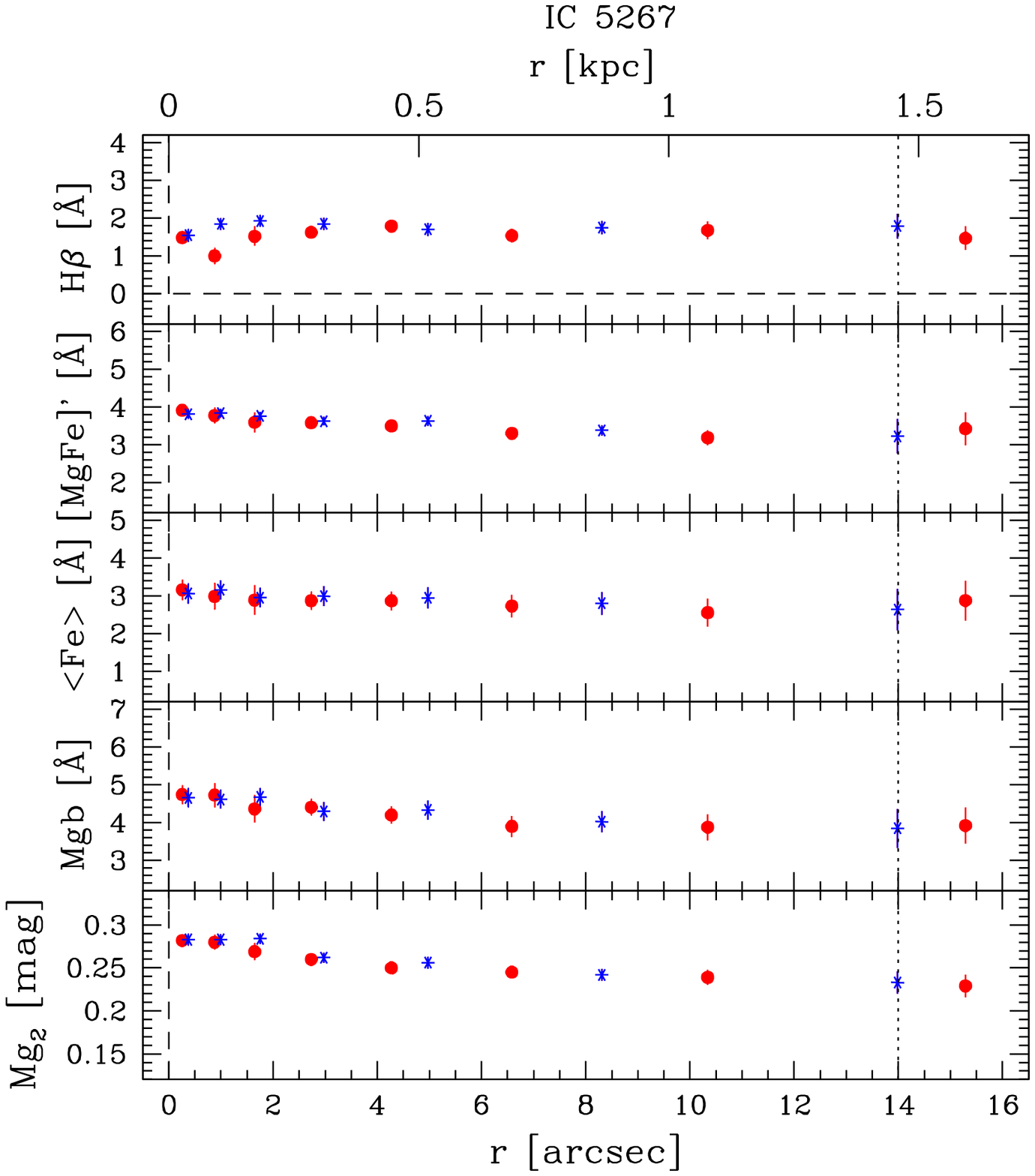}
\includegraphics[angle=0.0,width=0.431\textwidth]{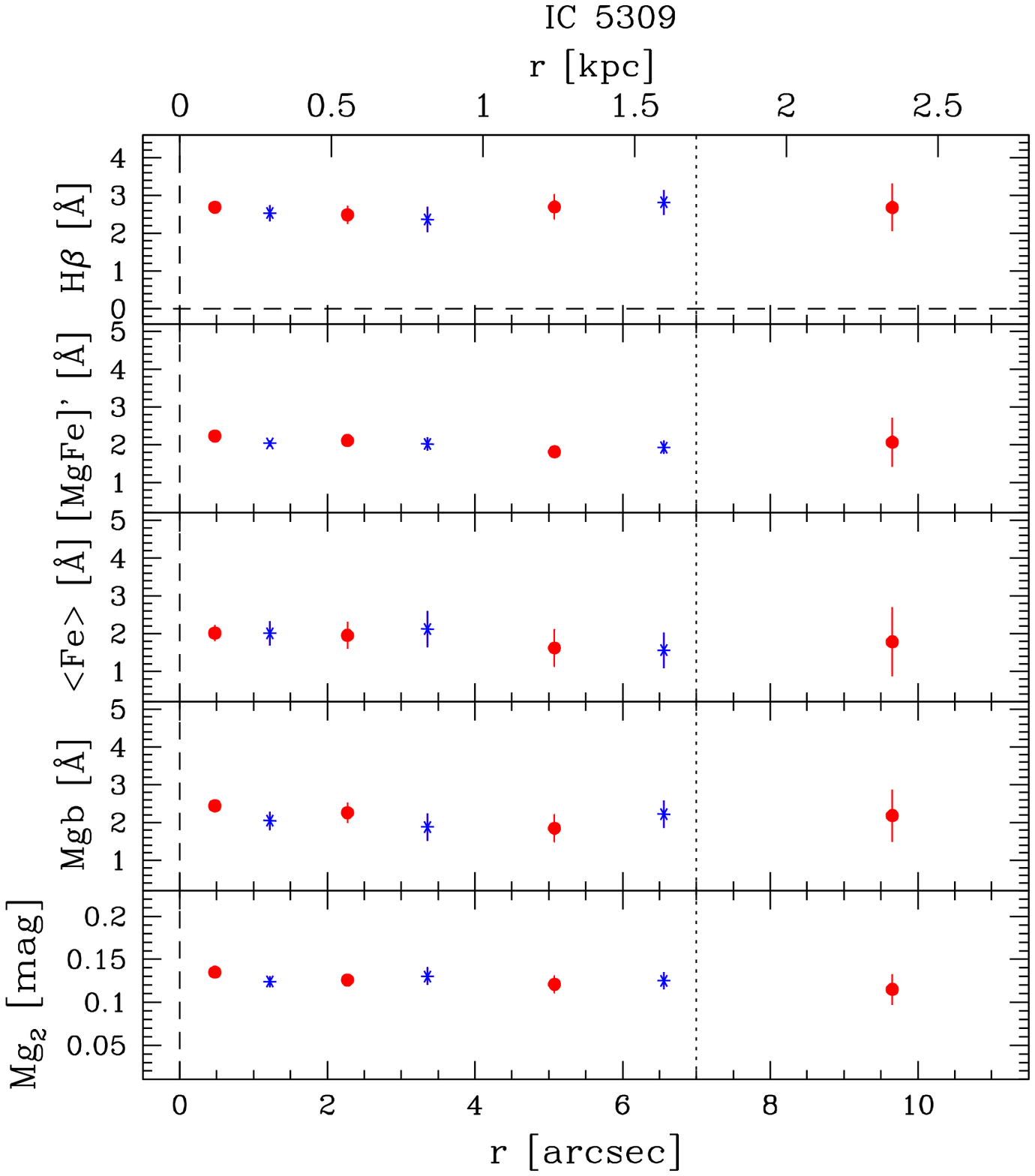}
\includegraphics[angle=0.0,width=0.431\textwidth]{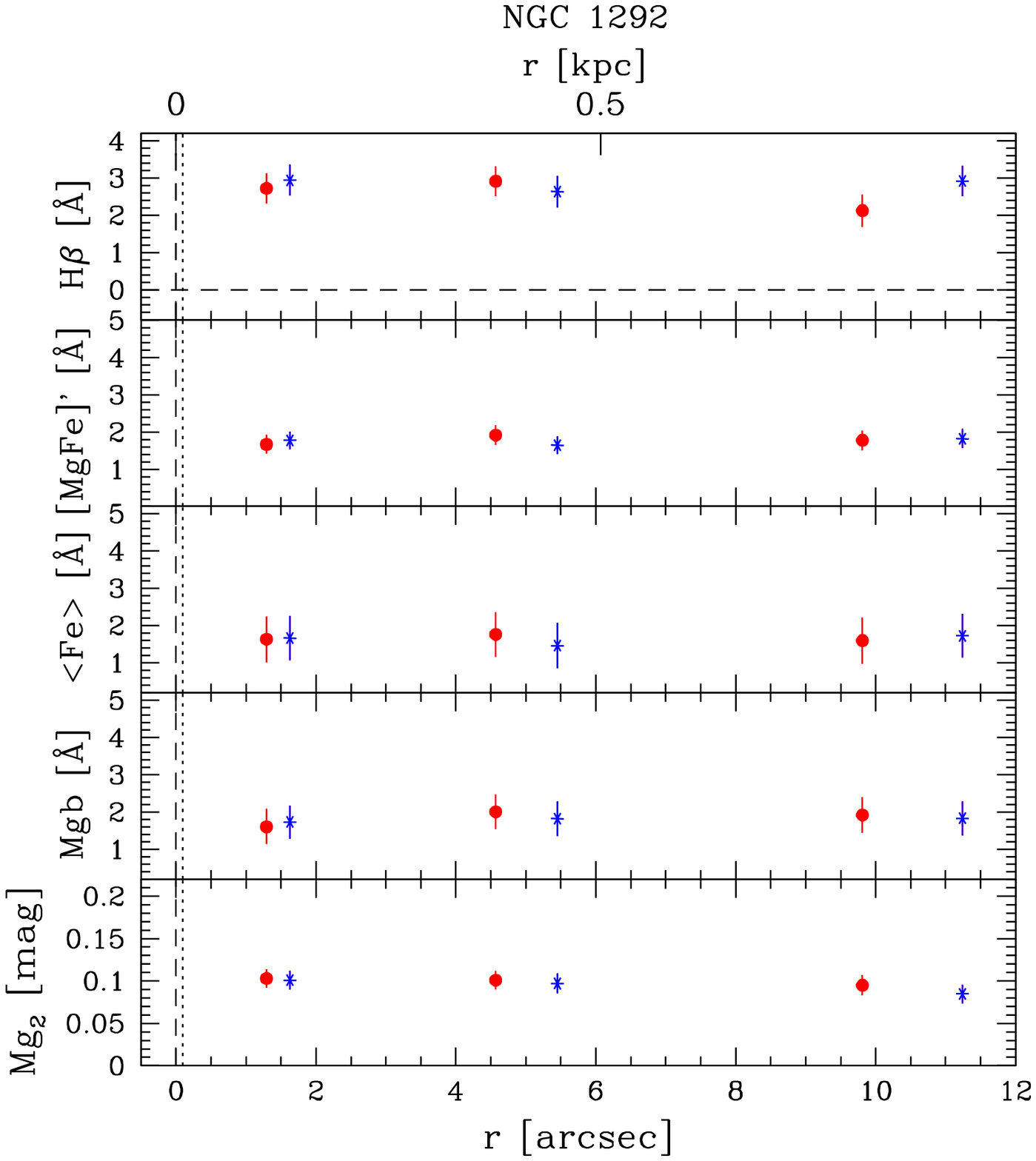}\\
\caption{The line-strength indices measured along the major axes of the
  sample galaxies. From top to bottom: folded radial profiles of \Hb,
  \MgFe, \Fe, \Mgb, and \Mgd. Asterisks and dots refer to the two
  sides (est/west) of the galaxy. The vertical dashed line indicates the
  radius(\Rdb) where the surface-brightness contributions of the bulge and disc
  are equal.}

\label{fig:indices}
\end{figure*}

\begin{figure*}
\centering
\includegraphics[angle=0.0,width=0.431\textwidth]{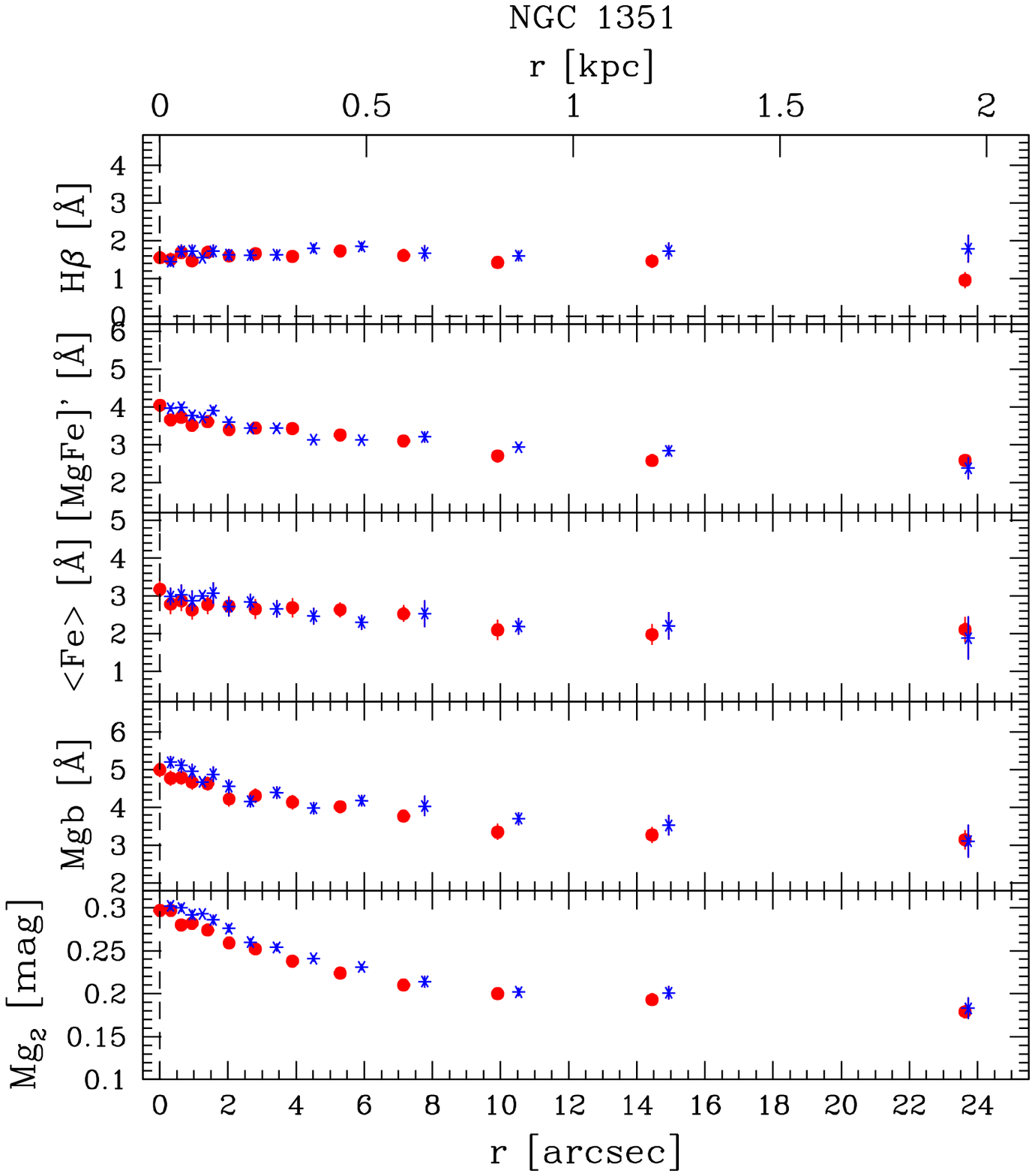}
\includegraphics[angle=0.0,width=0.431\textwidth]{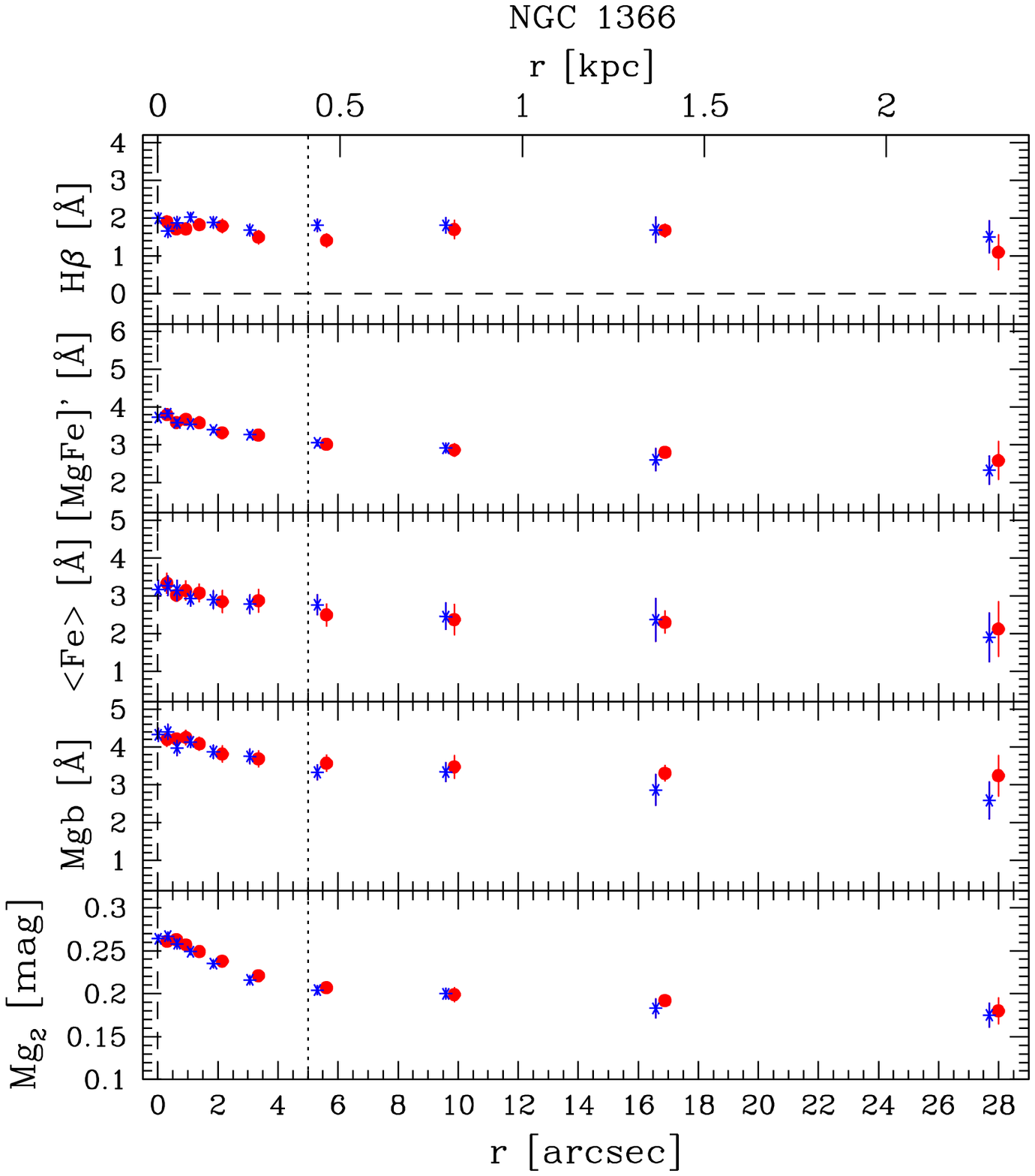}
\includegraphics[angle=0.0,width=0.431\textwidth]{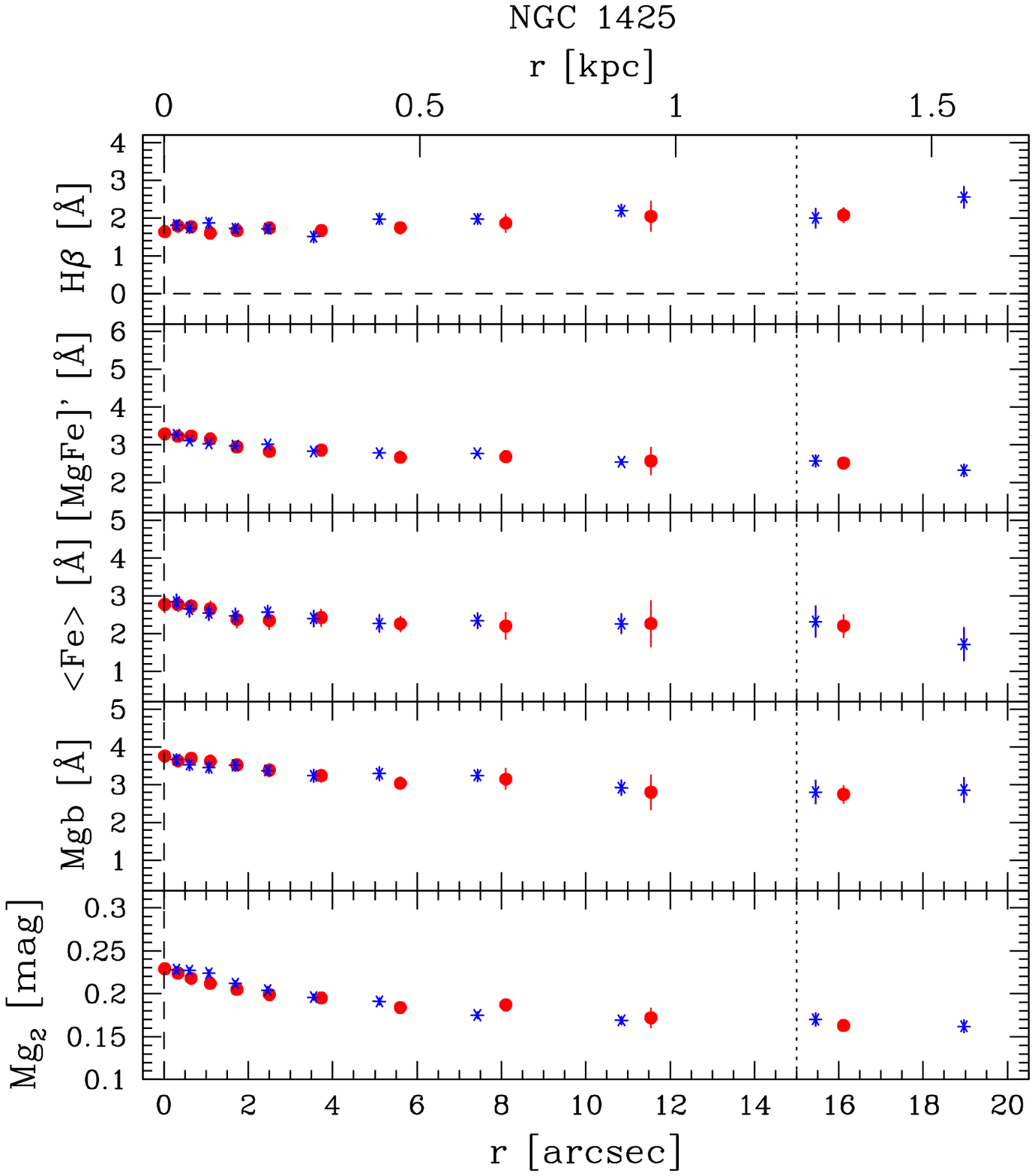}
\includegraphics[angle=0.0,width=0.431\textwidth]{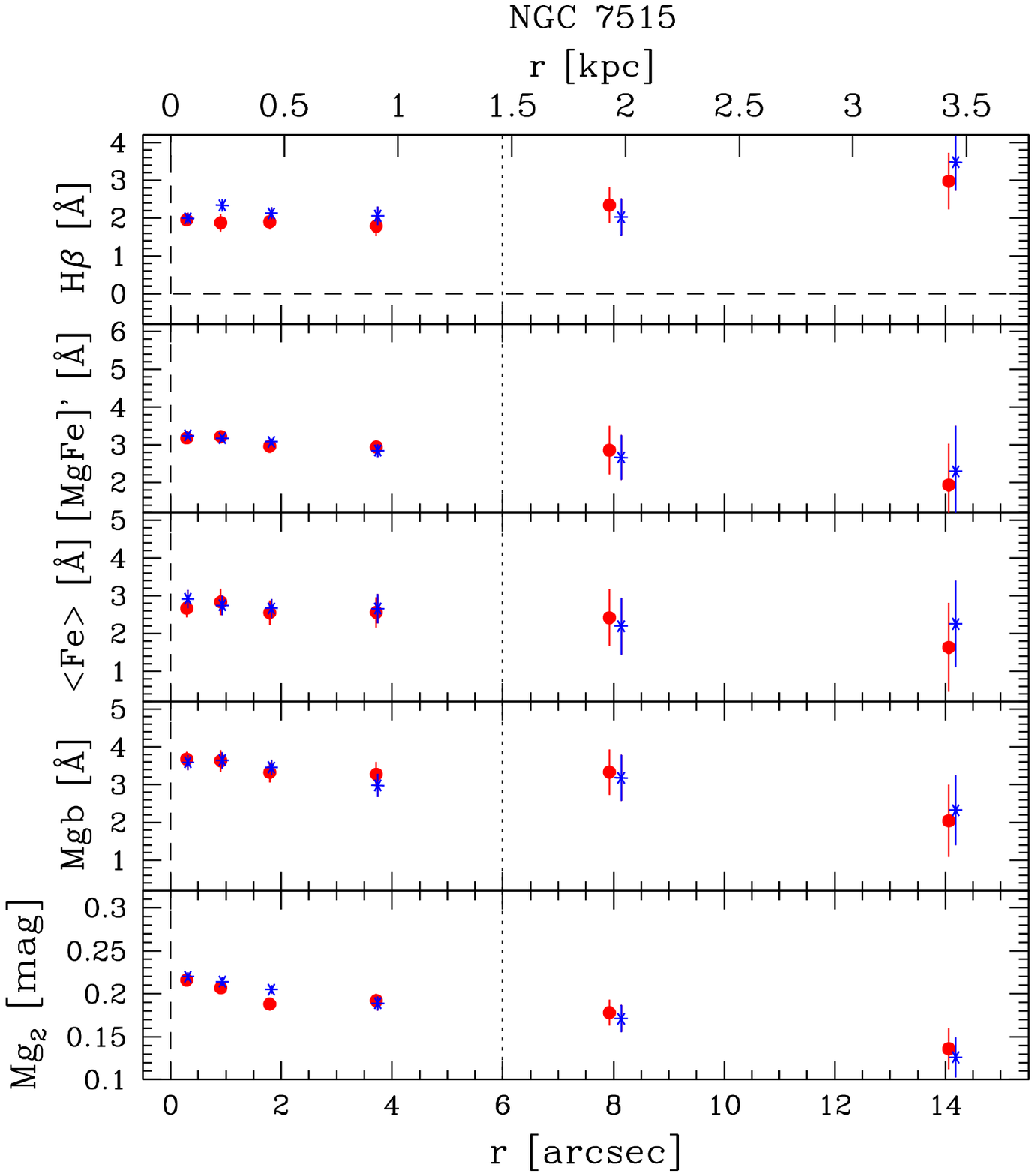}
\includegraphics[angle=0.0,width=0.431\textwidth]{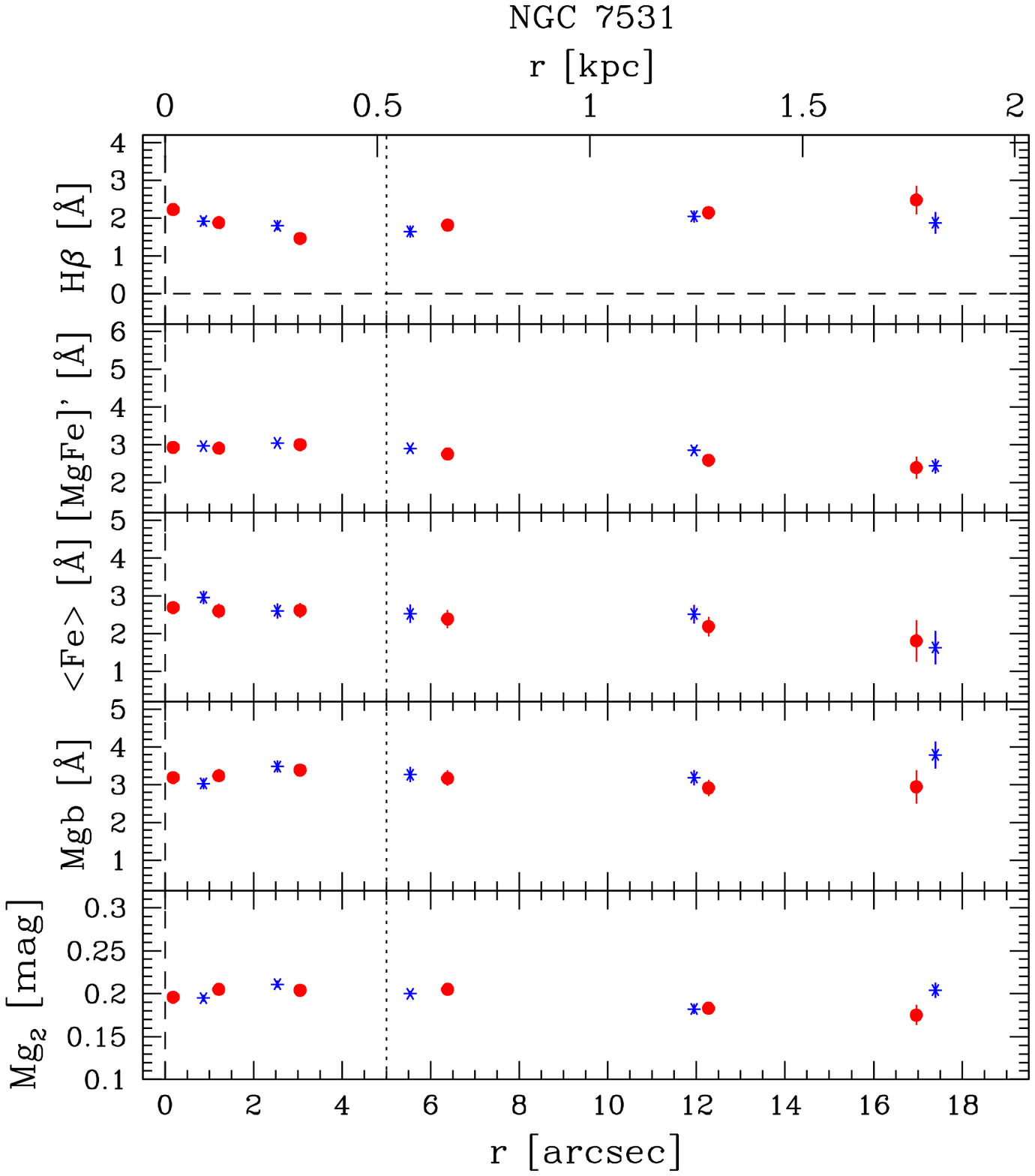}
\includegraphics[angle=0.0,width=0.431\textwidth]{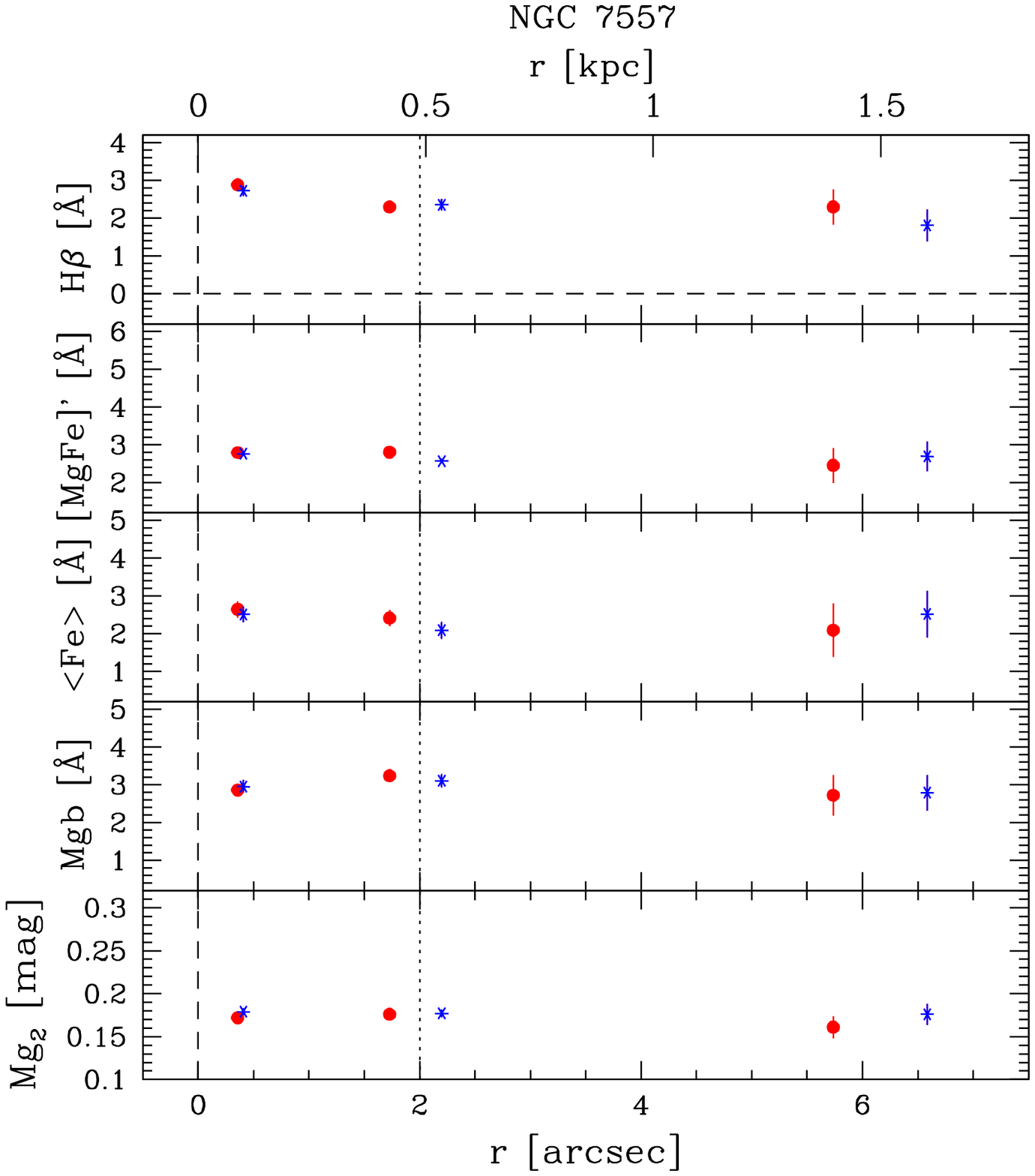}\\
\contcaption{}
\end{figure*}

\begin{figure*}
\centering
\includegraphics[angle=0.0,width=0.431\textwidth]{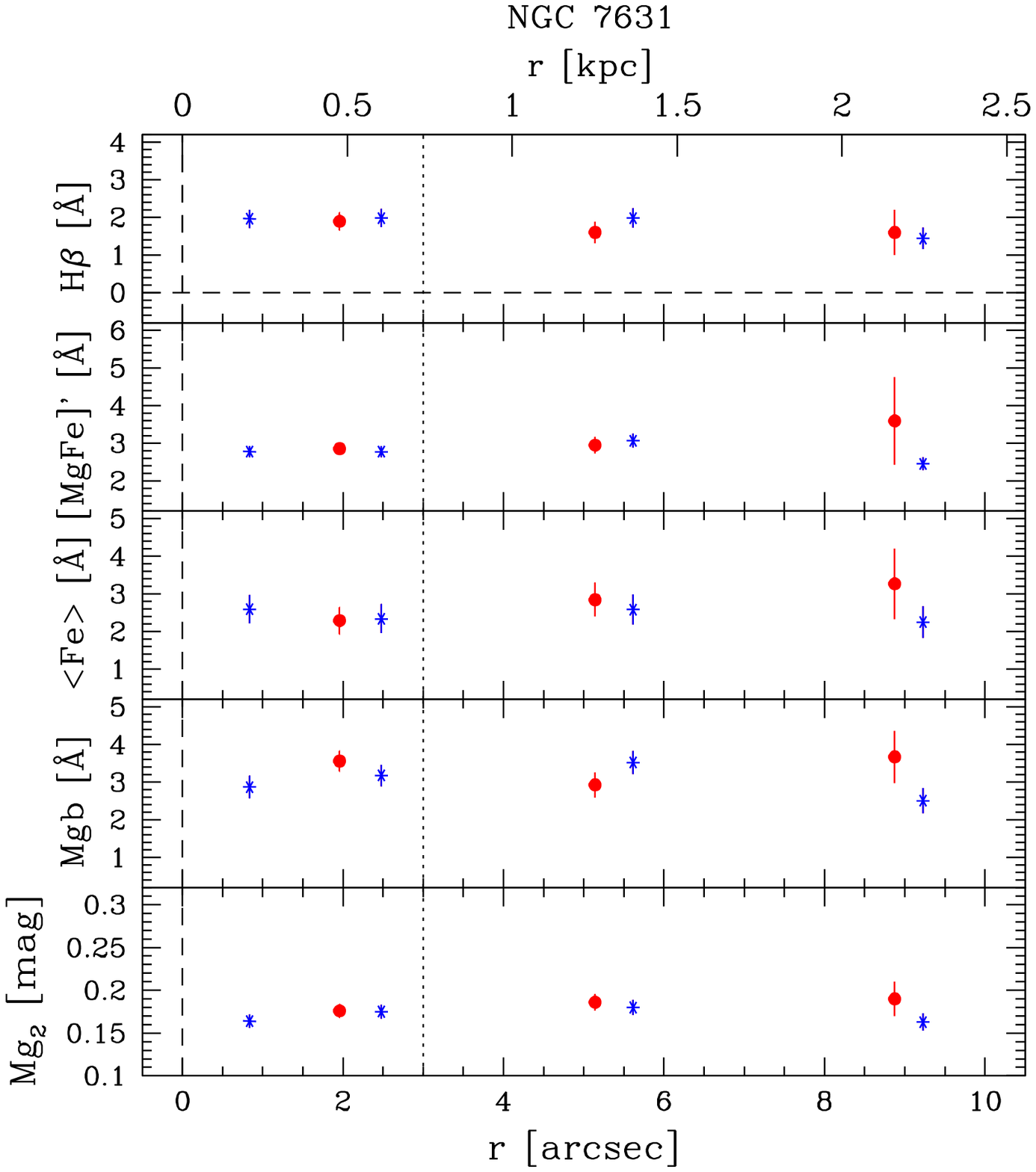}
\includegraphics[angle=0.0,width=0.431\textwidth]{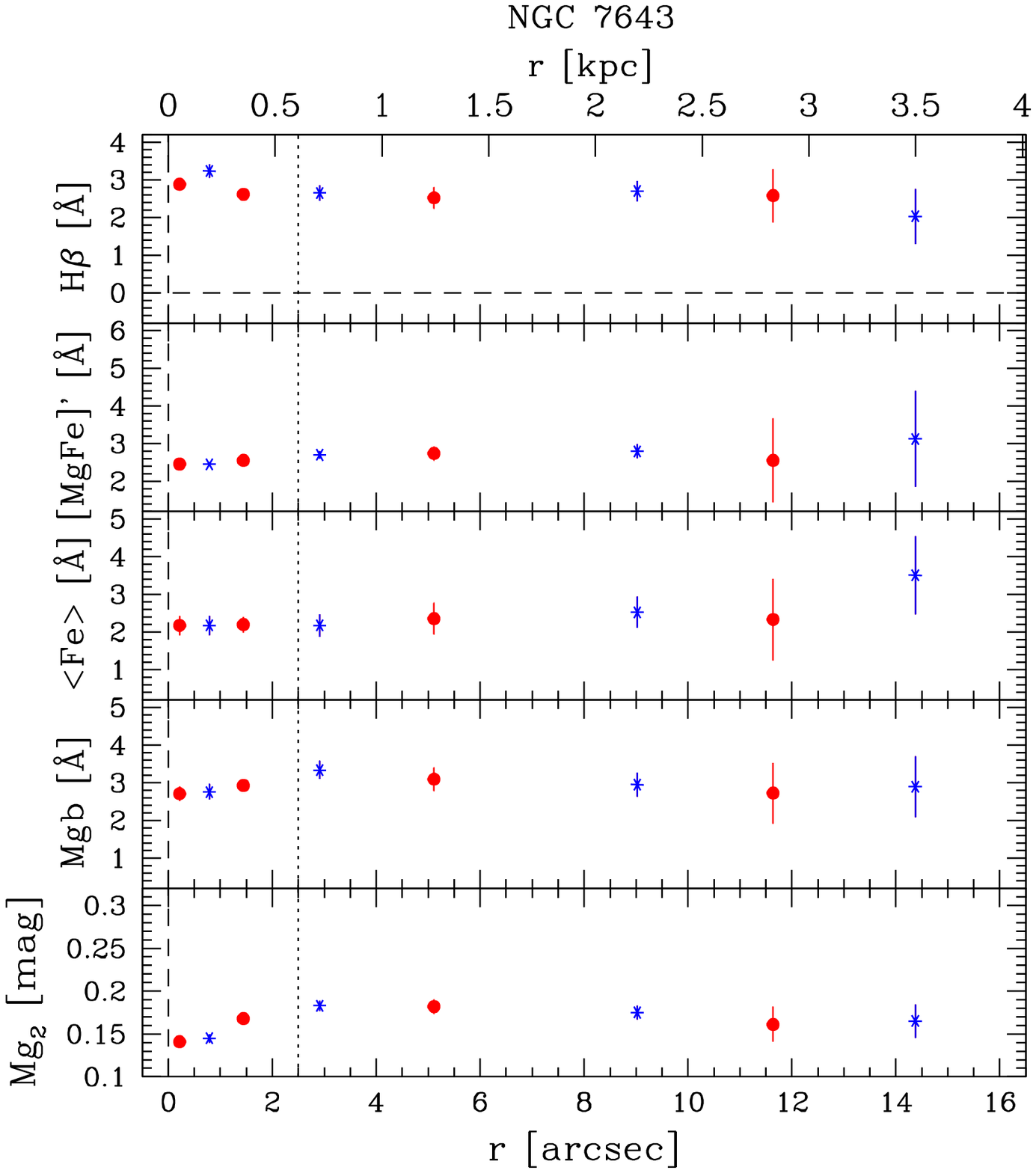}
\\
\contcaption{}
\end{figure*}
%%%%%%%%%%%%%%%%%%%%%%%%%%%%%%%%%%%%%%%%%%%%%%%%%%%%%%%%%%%%%%%%%%%%%%%%%%%%%%%%

\section{Line-strength indices: central values}
\label{sec:linestreng_cent}

Central values of velocity dispersion $\sigma$, \Mgb, \Mgd,
\Hb, \Fe, and \MgFe\/ line-strength indices were derived from the
major-axis profiles. The data points inside $0.3~r_{\rm e}$ were
averaged adopting a relative weight proportional to their $S/N$. The resulting
values are listed in Tab. \ref{tab:centval_lickind}.

%%%%%%%%%%%%%%%%%%%%%%%%%%%%%%%%%%%%%%%%%%%%%%%%%%%%%%%%%%%%%%%%%
\begin{figure}[h!]
\centering

\includegraphics[angle=90,width=0.47\textwidth]{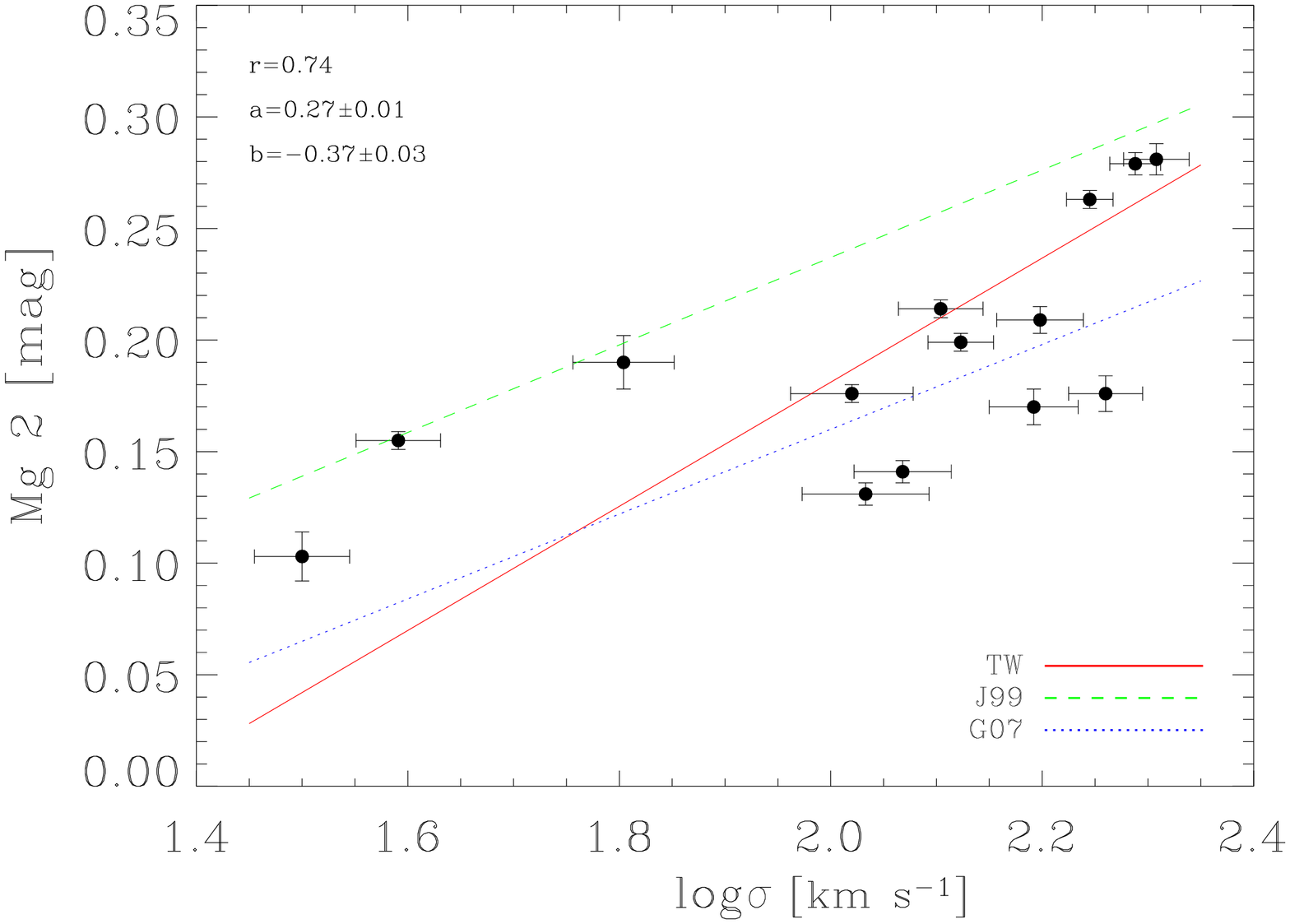}\\
\includegraphics[angle=90,width=0.47\textwidth]{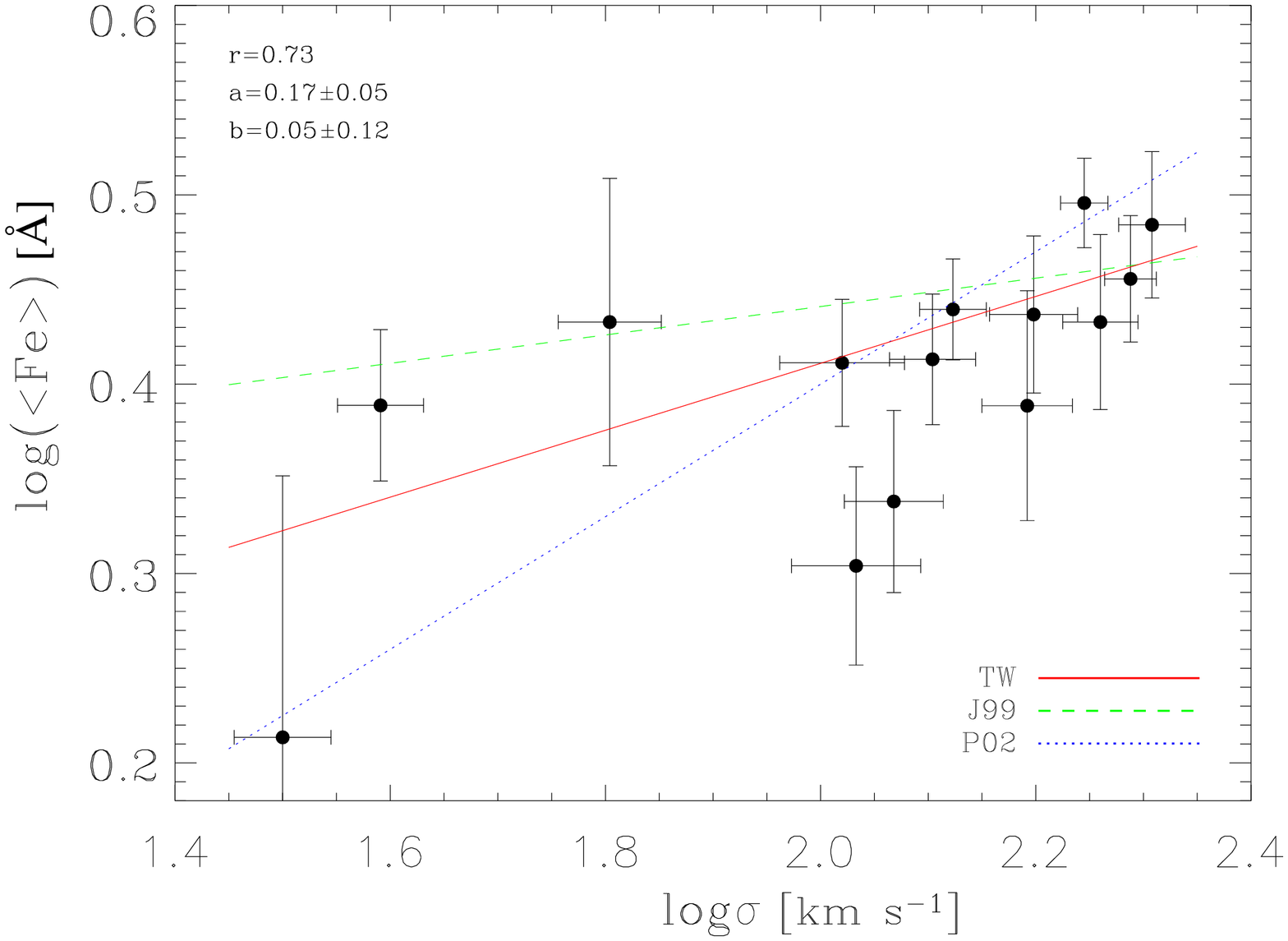}\\
\includegraphics[angle=90,width=0.47\textwidth]{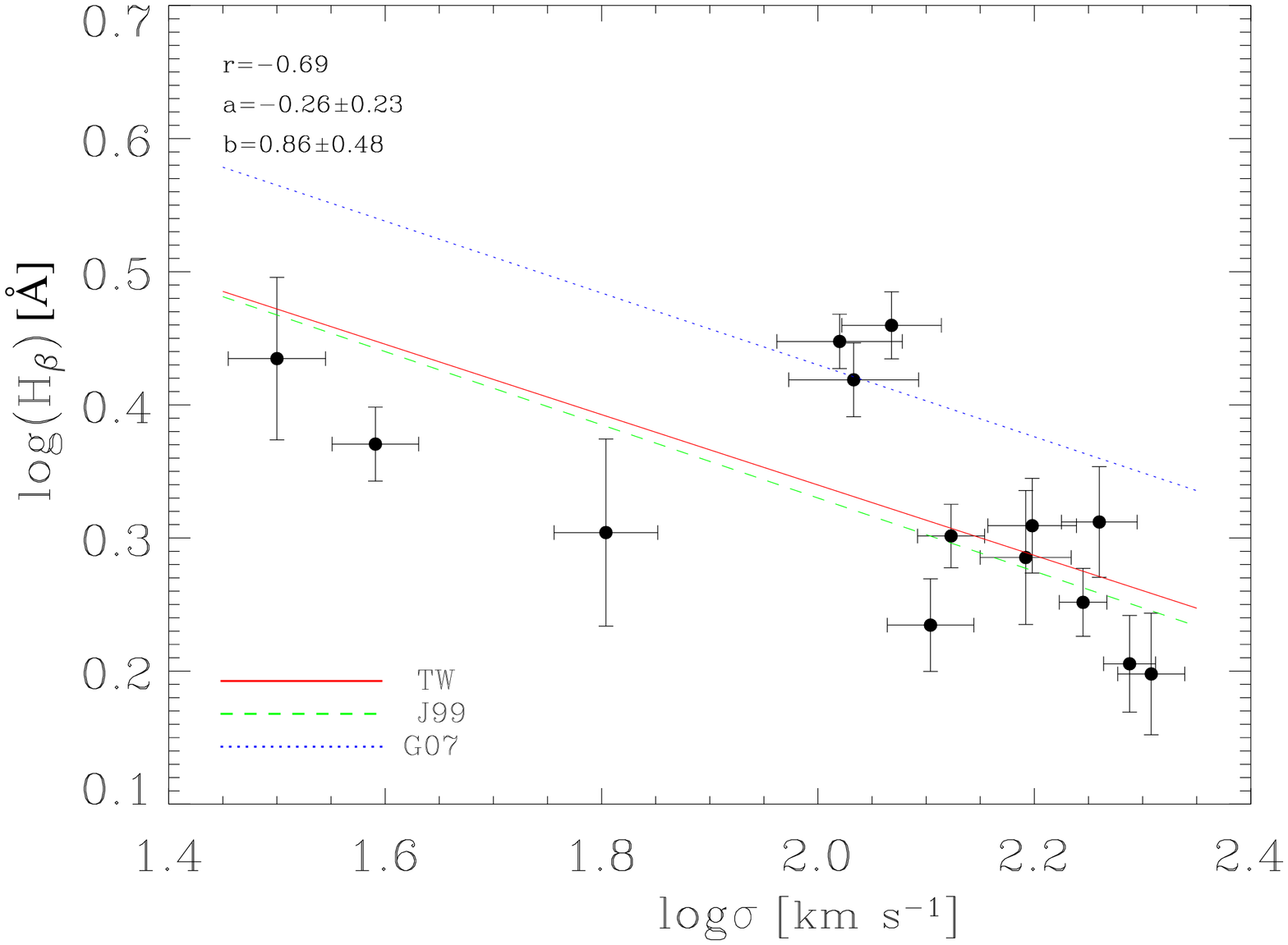}\\

\caption[The central value correlation \Mgd-$\sigma$ and
  \Fe-$\sigma$]{Central value of line strength indices \Mgd\ (upper
  panel), \Fe\ (central panel) and \Hb\ (lower panel) versus the
  velocity dispersion. In each panel the solid line represents the
  linear regression ($y=ax+b$) through all the data points. The
  Pearson correlation coefficient ($r$) and the results of the linear
  fit are given. Correlations found by differents authors are shown:
  TW = this work, J99 = \citep{jorgen99}, P02 = \citep{procetal02},
  G07 = \citep{gandetal07}.
\label{fig:censmg2fehb}}
\end{figure}
%%%%%%%%%%%%%%%%%%%%%%%%%%%%%%%%%%%%%%%%%%%%%%%%%%%%%%%%%%%%%%%%%%

The correlations between the velocity dispersion and \Mgd , \Hb, and
\Fe\/ line-strength indices for the sample galaxies are plotted in
Fig. \ref{fig:censmg2fehb}. They were extensively studied for
early-type galaxies
\citep[see][]{fishetal96,bernetal98,jorgen99,kuntetal00,mooretal02,mehletal03,rampetal05,sancetal06p,colletal06,annietal07}. 
  In Fig. \ref{fig:censmg2fehb} we also plotted for comparison
  the fitted correlations to different sample of elliptical/early-type
  galaxies and bulges.

The \Mgd\ line-strength index is generally adopted as the tracer of
the $\alpha$ elements giving an estimate of the enrichment, while
$\sigma$ is related with the gravitational potential. In elliptical
and S0 galaxies the \Mgd$-\sigma$ correlation shows that more massive
galaxies host a more metal rich stellar population
\citep[see][]{idiaetal96,bernetal98,jorgen99, mehletal03}. For
  bulges we measured a similar trend to that found by
  \citet{jabletal96}, \citet{prugetal01}, and \citet{gandetal07}. In
  agreement with \citet{gandetal07} we also found that the spirals lie
  below the \Mgd$-\sigma$ relation defined by the elliptical and
  early-type galaxies \citep{jorgen99}. 

Although the correlation between \Fe\ and $\sigma$ was predicted by
models of the dissipative collapse \citep[e.g.,][]{kodaetal98}, it was
not found to be very tight in early-type galaxies
\citep{fishetal96,jorgen99,tragetal98,mehletal03} except for those
studied by \citet{kuntetal00}. On the other hand, the \Fe$-\sigma$
relation is well defined for spiral bulges
\citep{idiaetal96,prugetal01,procetal02}. The fitted slope is slightly
steeper that the one traced for elliptical/early-type galaxies
\citep{jorgen99} and consistent with the relation by
\citet{procetal02}.

We also found a tight anti-correlation between \Hb\ and $\sigma$. For
large values of $\sigma$ our measurements are consistent with previous
results by \citet{jorgen99} and \citet{gandetal07}. At low $\sigma$
our fit follows closer the relation by \citet{jorgen99} due to the
presence of S0-Sa galaxies in our sample. Nevertheless, the spiral
galaxies are lying on the fitted line of the well selected Sb-Sc
sample given by \citet{gandetal07}.  The comparison with
\citet{jorgen99} could be affected by the possible contamination of
their \Hb\ line-strength indices by the \Hb\/ emission line due to the
presence of a young stellar component.

Young stellar population is often associated to dusts
\citep[e.g.,][]{peleetal07}. The dust structure is clearly visible in
the residual maps of the two-dimensional bulge-disc decompositions
(Fig.~\ref{fig:decomposition}). No correlation is found between dust
and central values of the line-strength indices. But, the dust is
located at larger radii than the regions mapped by the spectra
(Fig.~\ref{fig:indices}). Only for NGC~7515 and NGC~7643 the outer
radial bins of spectra cover a portion of the dust features observed
in the residual maps. The strong \Hb\/ emission line detected at these
radii is indicating the presence a local young stellar
population. But, this is not enough to drive a general conclusion
about the correlation of dust and stellar populations in the sample
galaxies.

\section{Ages, metallicities, and ${\bf \alpha}$/Fe enhancement:central values} 
 \label{agemet_cent}

From the central line-strength indices we derived the mean ages, total
metallicities, and total $\alpha/$Fe enhancements of the stellar populations
in the center of the sample bulges by using the stellar population
models by \citet{thmabe03}. These models predict the values of the
line-strength indices for a single stellar population as function of
the age, metallicity and \aFe\ ratios.

\begin{figure*}
\centering
\includegraphics[angle=90,width=0.49\textwidth]{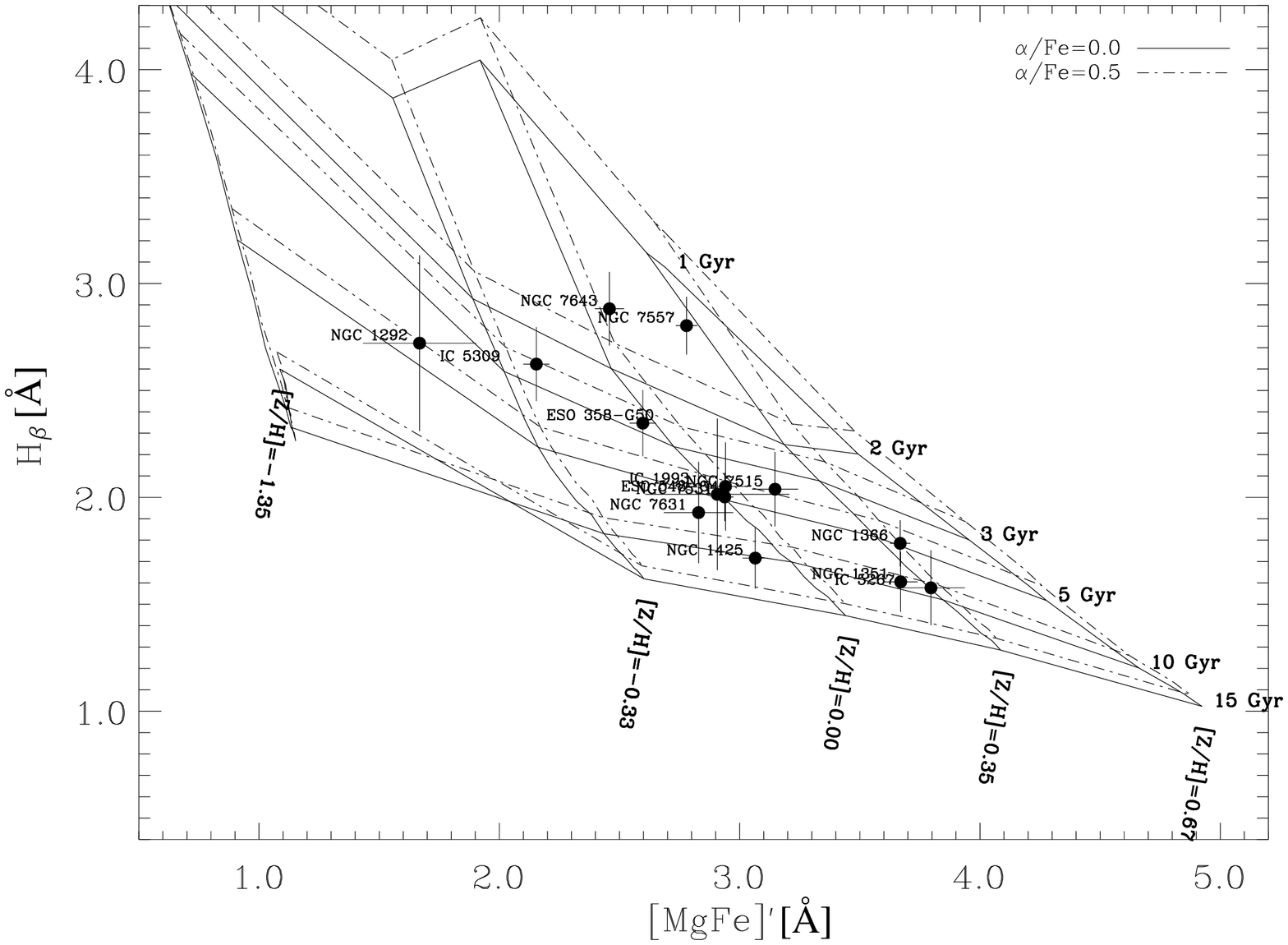}
\includegraphics[angle=90,width=0.49\textwidth]{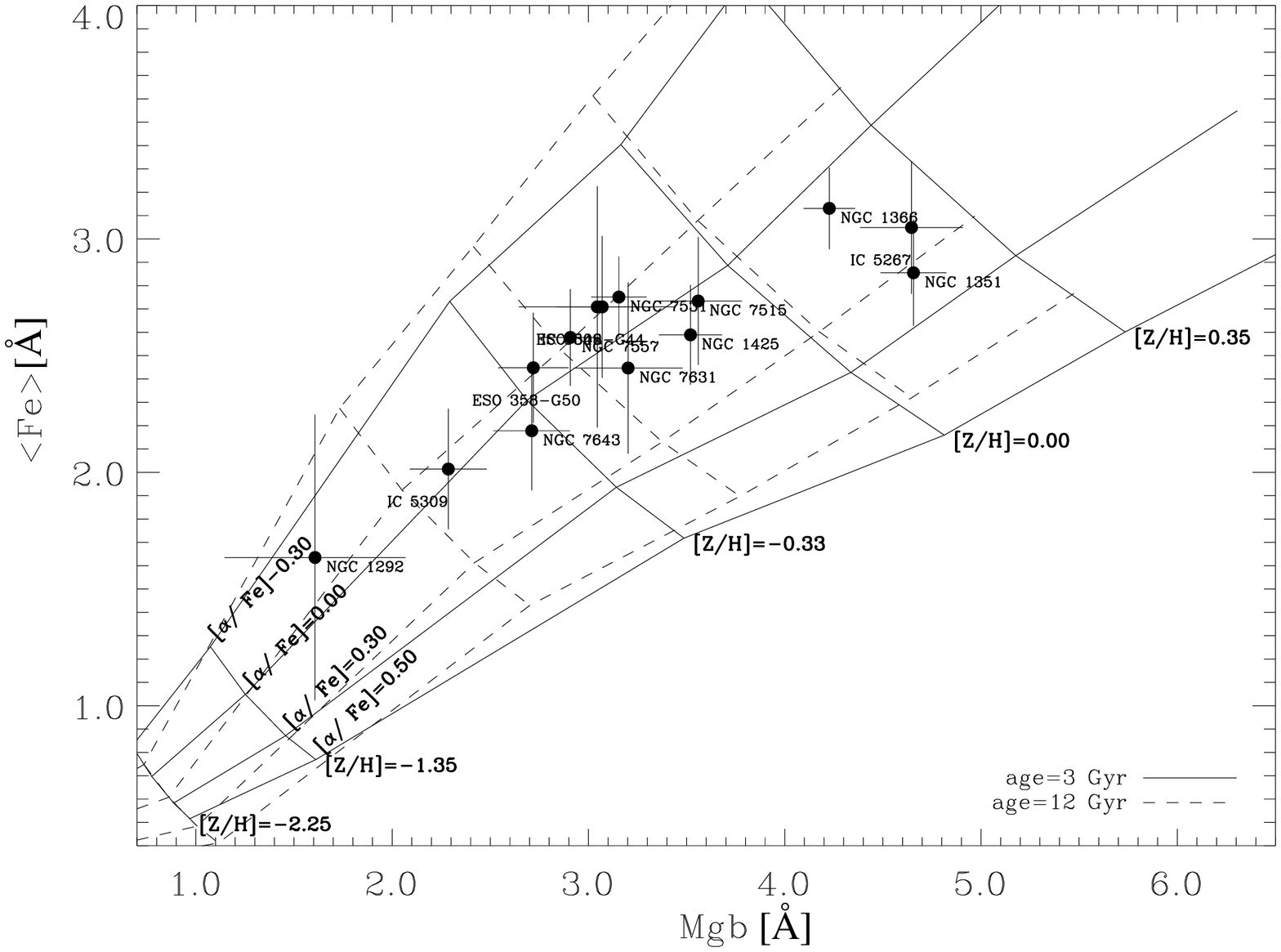}
\caption[Distribution \Hb-\MgFe\/ and \Mgb-\Fe]{The distribution of 
  the central values of \Hb\/ and \MgFe\/ indices (left panel) and \Fe
  and \Mgb\/ indices (right panel) averaged over $0.3~r_{\rm e}$ for
  the 15 sample galaxies . The lines indicate the models by
  \cite{thmabe03}. In the left panel the age-metallicity grids are
  plotted with two different $\alpha$/Fe enhancements: \aFe$\;=0.0$
  dex (continuous lines) and \aFe$\;=0.5$ dex (dashed lines).  In the
  right panel the \aFe-metallicity grids are plotted with two different
  ages: 3 Gyr (continuous lines) and 12 Gyr (dashed lines).
\label{fig:hbemgfemgbcent}}
\end{figure*}

\begin{figure}%[hp!]
\centering
\includegraphics[angle=90,width=0.5\textwidth]{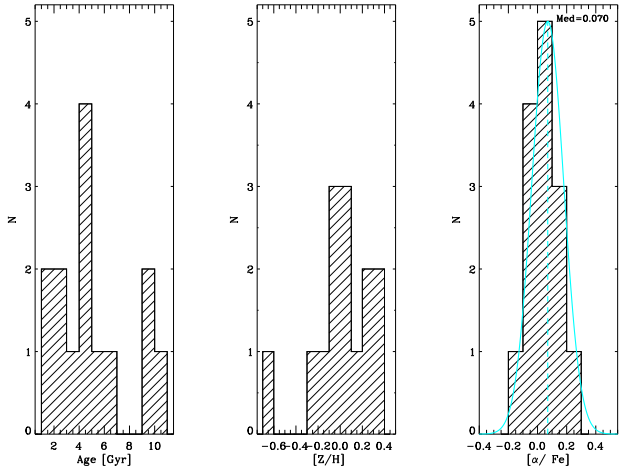}\\
\caption[]{Distribution of age (left panel), metallicity
  (central panel) and \aFe\/ enhancement (right panel) for the central
  regions of the sample galaxies. The solid
  line in the right panel represents a Gaussian centered in the median value\aFe$\;=0.07$ of
  distribution. Its $\sigma=~0.11$ is approximated by the value
  containing the 68\% of the objects of the distribution.
\label{fig:hist_ama}}
\end{figure}

The distribution of the central \Hb\ and \MgFe\ with the stellar
population models by \citet{thmabe03} is shown in
Fig. \ref{fig:hbemgfemgbcent} (left panel). The models are plotted for two fixed
\aFe\ ratios (0 and 0.5 dex) corresponding to stellar populations with 
solar and supersolar $\alpha/$Fe enhancements, respectively. In this
parameters space the mean age and total metallicity appear to be
almost insensitive to the variations of the $\alpha/$Fe enhancement.
The distribution of the central \Mgb\ and \Fe\ with the stellar
population models by \citet{thmabe03} is shown in
Fig. \ref{fig:hbemgfemgbcent} (right panel). The models are plotted
for two fixed ages (3 and 12 Gyr) corresponding to intermediate-age
and old stellar populations, respectively. In this parameters space
the $\alpha/$Fe enhancement and total metallicity appear to be almost
insensitive to the variations of age.
Central age, metallicity and total $\alpha/$Fe enhancement of each
bulge were derived by a linear interpolation between the model points
using the iterative procedure described in \citet{mehletal03}. The
derived values and their corresponding errors are listed in
Tab. \ref{tab:agemetalfa}. The histograms of their number distribution
are plotted in Fig. \ref{fig:hist_ama}. Even though the number of
galaxies does not allow us to trace a firm statistical conclusion,
three classes of objects were identified.  according to their age and
metallicity.  A similar result was found by \citet{mooretal06}.
The young bulges are scattered about an
average age of 2 Gyr with hints of star formation as shown by the
presence of the \Hb\ emission line in their spectra. The
intermediate-age bulges spans the age range between 4 and 8 Gyr. They
are characterized by solar metallicity (\ZH$\;=0.0$ dex).  Finally,
the old bulges have a narrow distribution in age around 10 Gyr and
high metallicity (\ZH$\;=0.30$ dex).
\citet{kuntetal00} and \citet{mehletal03} found that elliptical
galaxies in Fornax and Coma clusters are on average older and more
metal rich than S0 galaxies.
%
%We did not observe any correlation between the age and metallicity of
%bulge stellar population and the galaxy morphology. The same is true for
%the lenticular and spiral galaxies studied by \citet{thda06} and
%\citet{gandetal07}. This was interpreted as an indication of
%independent evolution of the stellar populations of bulges and discs
%\citep{thda06}.

\begin{figure}
\centering
\includegraphics[angle=90,width=0.5\textwidth]{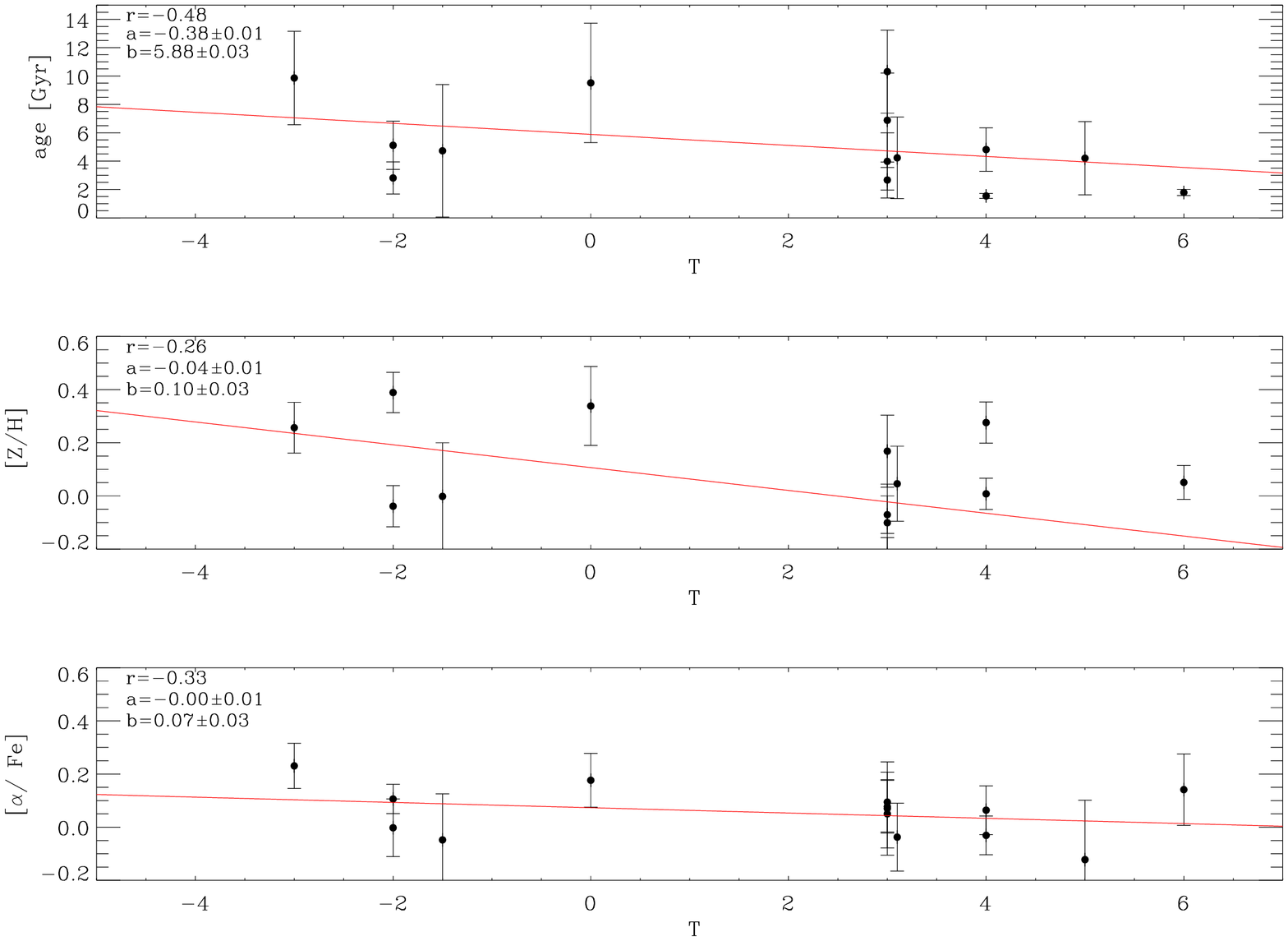}\\
\caption[]{Correlation between the central values of age, metallicity, and
  $\alpha$-enhancement with morphological type of the sample galaxies. In
  each panel the solid line represents the linear regression
  ($y=ax+b$) through all the data points. The Pearson correlation
  coefficient ($r$) and the results of the linear fit are given.
\label{fig:T_agemetalfa}}
\end{figure}

\citet{thda06} did not observe any correlation between the age and
metallicity of the bulge stellar population and galaxy
morphology. This was interpreted as an indication of independent
evolution of the stellar populations of the bulges and discs.  In the
sample studied by \citet{peleetal07} the early-type spirals often show
young stellar populations in their central regions, but there are also
objects that are as old as the oldest ellipticals. Young central
populations are seen in all the late-type spirals of
\citet{gandetal07}. The correlation between the morphological type
(col. 3, Tab. \ref{tab:sample}) and the age, metallicity and
$\alpha/$Fe enhancement of our sample galaxies are shown in
Fig. \ref{fig:T_agemetalfa}. Although the correlations are not
statistically very strong (in particular for the $\alpha/$Fe
enhancement), the elliptical and S0 galaxies ($T<0$) are older and
more metal rich than the spirals ($T>0$).

Most of the sample bulges display Solar $\alpha/$Fe enhancements with
the median of the distribution  at (\aFe$\;=0.07$). A few have a central
super-solar enhancement (\aFe$\;=0.3$). These values are similar to
those found for the elliptical galaxies in cluster
\citep{tesipeletier,jorg99,tragetal00,kuntetal00,kuntetal02} and imply
a star-formation timescale ranging from 1 to 5 Gyr.

\begin{figure}
\centering
\includegraphics[angle=90,width=0.5\textwidth]{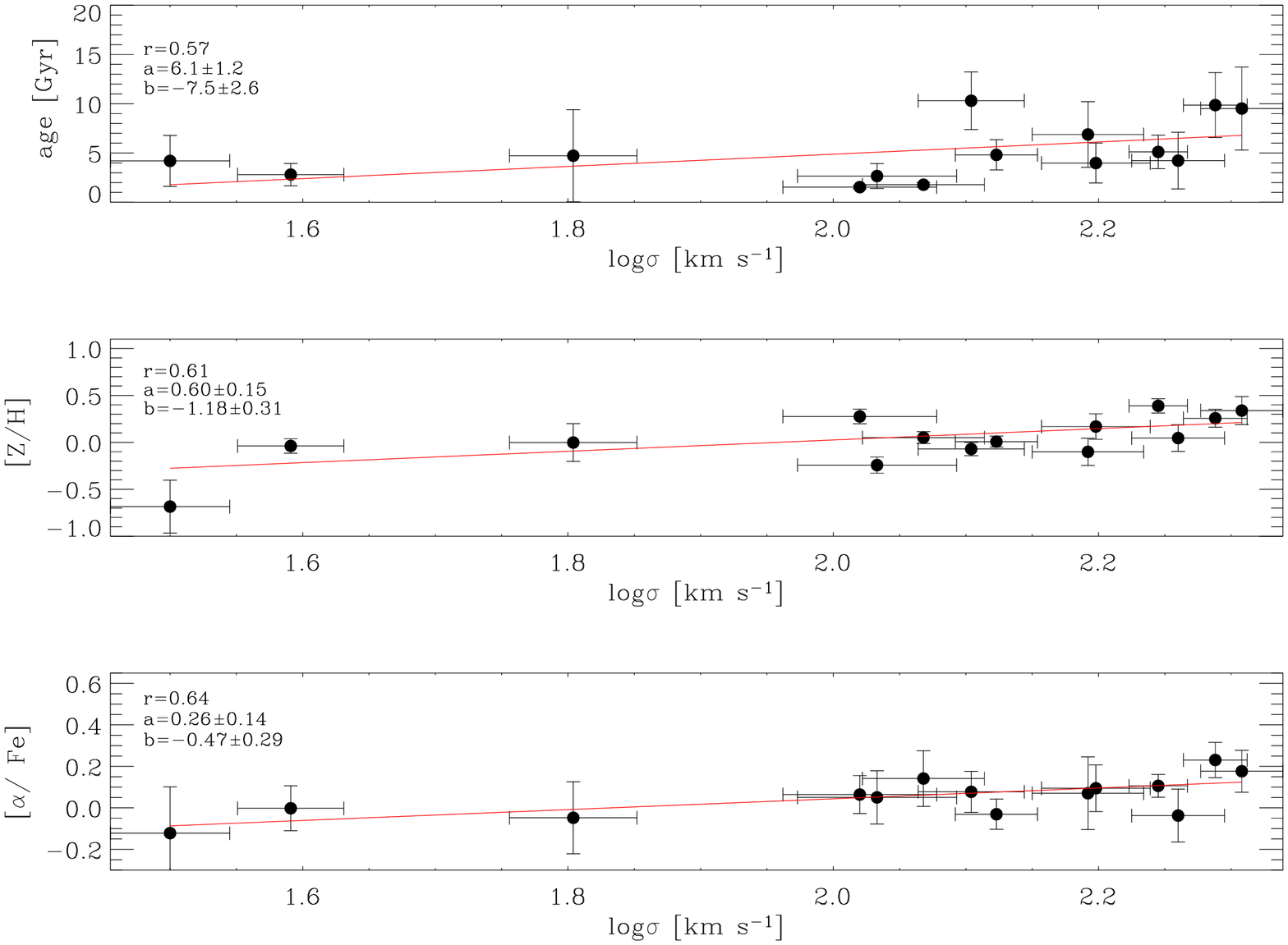}\\
\caption[]{Correlation between the central values of age, metallicity and
  $\alpha$-enhancement with central velocity dispersion of the sample
  galaxies. In each panel the solid line represents the linear
  regression ($y=ax+b$) through all the data points. The Pearson
  correlation coefficient ($r$) and the results of the linear fit are
  given.

\label{fig:sig_agemetalfa}}
\end{figure}

\begin{figure*}%[hp!]
\centering
\includegraphics[angle=90,width=1\textwidth]{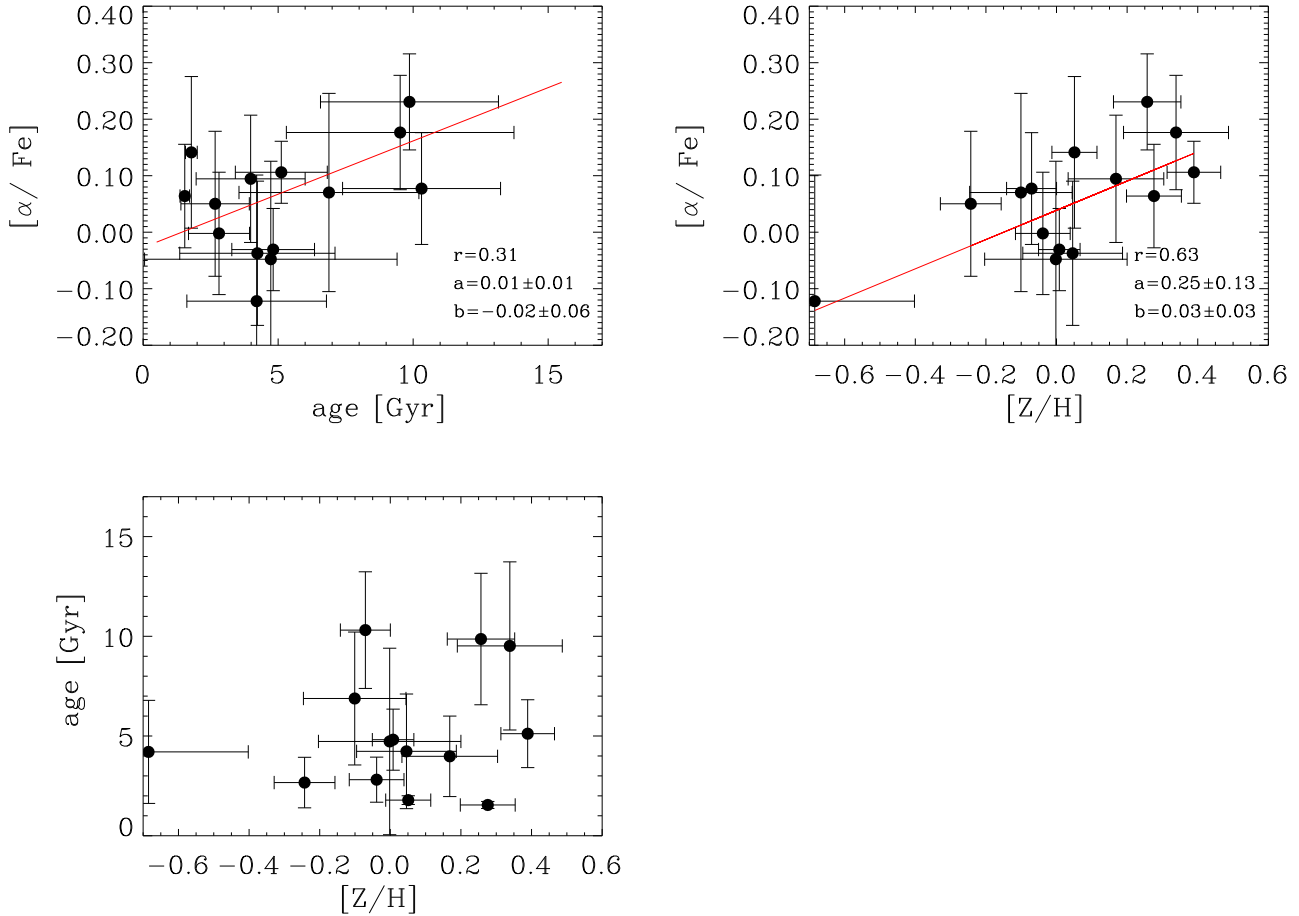}\\
\caption[]{The central ages, metallicities and $\alpha/$Fe enhancements
  of the sample bulges listed in Tab.\ref{tab:agemetalfa}.
  In each panel the solid line represents the linear
  regression ($y=ax+b$) through all the data points. The 
  Pearson correlation coefficient ($r$) and the results of the linear
  fit are given.
\label{fig:mgbfe}}
\end{figure*}

Age, metallicity, and $\alpha/$Fe enhancement correlate with velocity
dispersion (Fig. \ref{fig:sig_agemetalfa}). In early-type galaxies the
metallicity and $\alpha$/Fe enhancement are well correlated with the
central velocity dispersion, while the correlation is less evident
with age \citep{mehletal03,sancetal06p,denietal05}. In our bulges both the
metallicity and $\alpha$/Fe enhancement correlate with the velocity
dispersion and this is probably driving the correlation observed
between the metallicity and $\alpha$/Fe enhancement
(Fig. \ref{fig:mgbfe}). Age is mildly correlated with velocity
dispersion (Fig. \ref{fig:sig_agemetalfa}), and $\alpha$/Fe
enhancement (Fig.\ref{fig:mgbfe}). Recently, \citet{thda06} proved
that these correlations are tighter when the age estimation is based on
bluer Balmer line-strenght indices instead of \Hb .
We conclude that the more massive bulges of our sample galaxies are
older, more metal rich and characterized by a fast star
formation. Since we did not found any correlation with galaxy
morphology we exclude a strong interplay between the bulge and the
disc components.

%%%%%%%%%%%%%%%%%%%%%%%%%%%%%%%%%%%%%%%%%%%%%%%%%%%%%%%%%%%%%%%%%%%%%%%%%
%%% TABLE Indeces central values
\begin{table}
\caption{The central ages, metallicities and $\alpha/$Fe enhancements
of the sample bulges derived from the line-strength indices listed in
Tab. \ref{tab:centval_lickind} using the stellar population models by
\citet{thmabe03}.}
\begin{center}
\begin{small}
\begin{tabular}{lrrr}
\hline
\noalign{\smallskip}
\multicolumn{1}{c}{Galaxy} &
\multicolumn{1}{c}{\ZH} &
\multicolumn{1}{c}{Age} &
\multicolumn{1}{c}{\aFe} \\
\noalign{\smallskip}
\multicolumn{1}{c}{} &
\multicolumn{1}{c}{} &
\multicolumn{1}{c}{[Gyr]} &
\multicolumn{1}{c}{} \\
\noalign{\smallskip}
\multicolumn{1}{c}{(1)} &
\multicolumn{1}{c}{(2)} &
\multicolumn{1}{c}{(3)} &
\multicolumn{1}{c}{(4)} \\
\noalign{\smallskip}
\hline
\hline
\noalign{\smallskip}  

ESO 358-50 & $-0.04 \pm 0.08$  & $ 2.8 \pm 1.1$&  $ -0.00 \pm 0.11$ \\
ESO 548-44 & $-0.00 \pm 0.20$  & $ 4.7 \pm 4.6$&  $ -0.05 \pm 0.17$ \\
IC  1993   & $ 0.04 \pm 0.14$  & $ 4.2 \pm 2.8$&  $ -0.04 \pm 0.13$ \\
IC  5267   & $ 0.34 \pm 0.15$  & $ 9.5 \pm 4.2$&  $  0.18 \pm 0.10$ \\
IC  5309   & $-0.24 \pm 0.08$  & $ 2.7 \pm 1.3$&  $  0.05 \pm 0.13$ \\
NGC 1292   & $-0.68 \pm 0.28$  & $ 4.2 \pm 2.6$&  $ -0.12 \pm 0.22$ \\
NGC 1351   & $ 0.25 \pm 0.09$  & $ 9.9 \pm 3.3$&  $  0.23 \pm 0.08$ \\
NGC 1366   & $ 0.39 \pm 0.08$  & $ 5.1 \pm 1.7$&  $  0.11 \pm 0.05$ \\
NGC 1425   & $-0.07 \pm 0.07$  & $10.3 \pm 2.9$&  $  0.08 \pm 0.10$ \\
NGC 7515   & $0.17  \pm 0.13$  & $ 4.0 \pm 2.0$&  $  0.09 \pm 0.11$ \\
NGC 7531   & $0.01  \pm 0.06$  & $ 4.8 \pm 1.5$&  $ -0.03 \pm 0.07$ \\
NGC 7557   & $0.27  \pm 0.08$  & $ 1.5 \pm 0.2$&  $  0.06 \pm 0.09$ \\
NGC 7631   & $-0.10 \pm 0.14$  & $ 6.9 \pm 3.3$&  $  0.07 \pm 0.17$ \\
NGC 7643   & $ 0.05 \pm 0.06$  & $ 1.8 \pm 0.2$&  $  0.14 \pm 0.13$ \\

\noalign{\smallskip}
\hline
\noalign{\medskip}
\end{tabular}
\end{small}
\label{tab:agemetalfa}
\end{center}
\end{table}
%%%%%%%%%%%%%%%%%%%%%%%%%%%%%%%%%%%%%%%%%%%%%%%%%%%%%%%%%%%%
%

%
%%%%%%%%%%%%%%%%%%%%%%%%%%%%%%%%%%%%%%%%%%%%%%%%%%%%%%%%%%%%
%griglie  age met alfafe
%%%%%%%%%%%%%%%%%%%%%%%%%%%%%%%%%%%%%%%%%%%%%%%%%%%%%%%%%%%%

%%%%%%%%%%%%%%%%%%%%%%%%%%%%%%%%%%%%%%%%%%%%%%%%%%%%%%%%%%%%
% Figure histogramma valori eta metallicita alfa enhancement
%%%%%%%%%%%%%%%%%%%%%%%%%%%%%%%%%%%%%%%%%%%%%%%%%%%%%%%%%%%%

%%%%%%%%%%%%%%%%%%%%%%%%%%%%%%%%%%%%%%%%%%%%%%%%%%%%%%%%%%%%
%figure di sigma contro eta metallicita alfa
%%%%%%%%%%%%%%%%%%%%%%%%%%%%%%%%%%%%%%%%%%%%%%%%%%%%%%%%%%%%%%%%%%%%%%%%%
%%%%%%%%%%%%%%%%%%%%%%%%%%%%%%%%%%%%%%%%%%%%%%%%%%%%%%%%%%%%
%figure di T contro eta metallicita alfa
%%%%%%%%%%%%%%%%%%%%%%%%%%%%%%%%%%%%%%%%%%%%%%%%%%%%%%%%%%%%%%%%%%%%%%%%%

%%%%%%%%%%%%%%%%%%%%%%%%%%%%%%%%%%%%%%%%%%%%%%%%%%%%%%%%%%%%%%%%%%%%%%%%%
% --- Figure eta metallicita alfe all VS all
%%%%%%%%%%%%%%%%%%%%%%%%%%%%%%%%%%%%%%%%%%%%%%%%%%%%%%%%%%%%%%%%%%%%%%%%%

\section{Ages, metallicities, and ${\bf \alpha}$/Fe enhancement: gradients} 
\label{sec:agemetalpha_grad}

Different formation scenarios predict different radial trends of age,
metallicity, and $\alpha$/Fe enhancement. Therefore the radial
gradients of the properties of the stellar populations of bulges are a
key information to understand the processes of their formation and
evolution.

The total metallicity of a stellar population depends only on the
efficiency of the star formation and on the amount of gas transformed
in star \citep{tins80}. In the monolithic collapse scenario gas
dissipation toward the galaxy centre with subsequent occurrence of star
formation and blowing of galactic winds produce a steep metallicity
gradient \citep{eglbsa62,lars74,aryo87}. A strong gradient in
$\alpha$/Fe enhancement is expected too \citep{fesi02}.
The predictions for bulges forming through a long time-scale processes
as dissipation-less secular evolution are more contradictory. In this
scenario the bulge is formed by redistribution of disc stars.  The
gradients eventually present in the progenitor disk, could be either
amplified since the resulting bulge has a smaller scale length than
the progenitor or erased as a consequence of disc
heating \citep{mooretal06}.

\begin{figure}%[hp!]
\centering
\includegraphics[angle=90,width=0.5\textwidth]{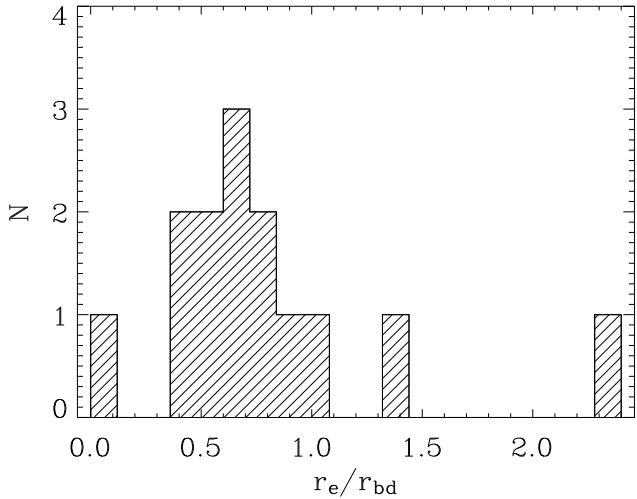}\\
\caption[]{Distribution of the ratio between $r_{\rm e}$ and $r_{\rm
    bd}$; For most of the galaxies ($r_{\rm e}$/$r_{\rm bd}$=0.7) the
  gradient evaluated up to $r_{\rm bd}$ is mapping a region slightly
  larger than $r_{\rm e}$.
\label{fig:hist_rdb_re}}
\end{figure}

An issue in measuring the gradients of age, metallicity and
$\alpha/$Fe enhancement in bulge, could be the contamination of their
stellar population by the light coming from the underlying disc
stellar component.  This effect is negligible in the galaxy center but
it could increase going to the outer regions of the bulge, where the
light starts to be dominated by the disc component.  In order to
reduce the impact of disc contamination and extend as much as possible
the region in which deriving gradients, we map them inside $r_{\rm
bd}$, the radius where the bulge and disc give the same contribution to
the total surface-brightness. This is a region slightly larger than
$r_{\rm e}$ of the galaxy (Fig. \ref{fig:hist_rdb_re}). Deriving
gradients in the bulge dominated region with this approach, will not
remove completely the contamination by the disc stellar population but
it will assure always a similar degree of contamination in comparing the
gradients of different galaxies.

%%%%%%%%%%%%%%%%%%%%%%%%%%%%%%%%%%%%%%%%%%%%%%%%%%%%%%%%%%%%
% Figure histogramma rapporto re/rdb
%%%%%%%%%%%%%%%%%%%%%%%%%%%%%%%%%%%%%%%%%%%%%%%%%%%%%%%%%%%%

For each galaxy, we derived the \Mgd , \Hb, and \Fe\/ line-strength
indices at the radius $r_{\rm bd}$ (see
Tab. \ref{tab:parameters}). The ages, metallicities and $\alpha/$Fe
enhancements were derived by using the stellar population models by
\citet{thmabe03} as done for the central values.

The gradients were set as the difference between the values at centre
and \Rdb and their corresponding errors were calculated through Monte
Carlo simulations taking into account the errors in the gradients and
the logarithmic fit to the adopted indices. For NGC~1292 no gradients
were obtained since it was not possible to estimate $r_{\rm bd}$. The
galaxy bulge is fainter than the disc at all radii
(Fig. \ref{fig:decomposition}). Indeed, the surface-brightness radial
profile of the galaxy could be fitted by adopting only an exponential
disc (Tab. \ref{tab:parameters}).

\begin{figure}%[hp!]
\centering
\includegraphics[angle=90,width=0.49\textwidth]{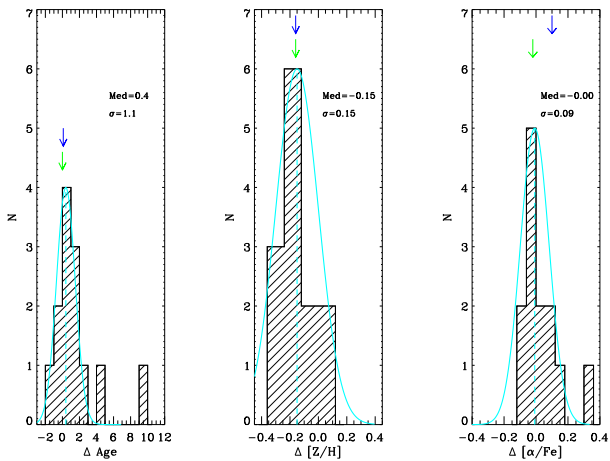}\\
\caption[]{Distribution of the gradients of age (left panel),
  metallicity (central panel) and \aFe\/ enhancement (right panel) for
  the sample galaxies. Dashed line represents the median of the
  distribution and its values is reported.Solid line represents a
  Gaussian centered in the median value of distribution. Their
  $\sigma$ approximated by the value containing the 68\% of the
  objects of the distribution are reported.The green and blue arrows show
  the average gradient found for early-type galaxies and bulges
  by \citet{mehletal03} and \citet{jabletal07}, respectively.
\label{fig:histgrad_ama}}
\end{figure}

The final gradients of age, metallicity and $\alpha/$Fe enhancement
and their errors are listed in Tab. \ref{tab:agemetalfa_grad}.  In
Tab. \ref{tab:agemetalfa_grad} we also report the age, metallicity and
$\alpha/$Fe enhancement gradients rescaled to a fix radius value of 1
kpc. For two galaxies (ESO~358-G50 and NGC~7531) the small values of
$r_{\rm bd}$ combined with the big values of gradients give
meaningless values of age extrapolated at 1 kpc and we omit them in the
table.  The histograms of their number distribution are plotted in
Fig. \ref{fig:histgrad_ama}.

%%%%%%%%%%%%%%%%%%%%%%%%%%%%%%%%%%%%%%%%%%%%%%%%%%%%%%%%%%%%%%%%%%%%%%%%%
%%% TABLE age metallicity alfa/fe gradients  values
\begin{table*}
\caption{The columns show the following: (2),(3),(4) the gradients of the ages, metallicities and $\alpha/$Fe enhancements
  of the sample bulges derived from the central values and values at
  $r_{\rm bd}$ listed Tab. \ref{tab:parameters}. (5),(6),(7) the gradients of the ages, metallicities and $\alpha/$Fe enhancements
  of the sample bulges rescaled at 1 kpc. Stellar population
  models by \citet{thmabe03} were used.}
\begin{center}
\begin{small}
\begin{tabular}{lrrrrrr}
\hline
\noalign{\smallskip}
\multicolumn{1}{c}{Galaxy} &
\multicolumn{1}{c}{$\Delta$(\ZH)} &
\multicolumn{1}{c}{$\Delta$(Age)} &
\multicolumn{1}{c}{$\Delta$(\aFe)} &
\multicolumn{1}{c}{$\Delta$(\ZH)$_{kpc}$} &
\multicolumn{1}{c}{$\Delta$(Age)$_{kpc}$} &
\multicolumn{1}{c}{$\Delta$(\aFe)$_{kpc}$}\\
\noalign{\smallskip}
\multicolumn{1}{c}{} &
\multicolumn{1}{c}{} &
\multicolumn{1}{c}{[Gyr]} &
\multicolumn{1}{c}{}&
\multicolumn{1}{c}{} &
\multicolumn{1}{c}{[Gyr]} &
\multicolumn{1}{c}{} \\
\noalign{\smallskip}
\multicolumn{1}{c}{(1)} &
\multicolumn{1}{c}{(2)} &
\multicolumn{1}{c}{(3)} &
\multicolumn{1}{c}{(4)} &
\multicolumn{1}{c}{(5)} &
\multicolumn{1}{c}{(6)} &
\multicolumn{1}{c}{(7)} \\

\noalign{\smallskip}
\hline
\hline
\noalign{\smallskip}  
ESO 358-50 & $-0.15 \pm 0.23$  & $ 2.95  \pm 3.47 $&  $ -0.008  \pm 0.25$& $-0.06 \pm  0.10$& $ 1.34\pm  1.63$ & $-0.003\pm 0.11$  \\
ESO 548-44 & $-0.01 \pm 0.35$  & $ 1.04  \pm 10.80 $&  $ -0.078 \pm0.26$& $-0.05 \pm  1.44$& $ ...        $ & $-0.325\pm 1.02$  \\
IC  1993   & $-0.14 \pm 0.26$  & $ 0.20  \pm 4.30 $&  $  0.098  \pm 0.23$& $-0.35 \pm  0.62$& $ 0.51\pm 11.09$ & $ 0.251\pm 0.63$  \\
IC  5267   & $-0.22 \pm 0.21$  & $-0.76  \pm 7.77 $&  $ -0.024  \pm 0.20$& $-0.15 \pm  0.13$& $-0.52\pm  5.29$ & $-0.016\pm 0.14$  \\
IC  5309   & $-0.31 \pm 0.22$  & $ 1.48  \pm 4.67 $&  $ -0.087  \pm 0.22$& $-0.18 \pm  0.10$& $ 0.87\pm  2.85$ & $-0.051\pm 0.12$  \\
NGC 1351   & $-0.17 \pm 0.26$  & $-0.73  \pm 6.43 $&  $ -0.043  \pm 0.15$& $-0.06 \pm  0.09$& $-0.27\pm  2.38$ & $-0.016\pm 0.05$  \\
NGC 1366   & $-0.29 \pm 0.24$  & $ 4.58  \pm 4.95 $&  $ -0.038  \pm 0.13$& $-0.71 \pm  0.52$& $11.16\pm 12.97$ & $-0.092\pm 0.31$  \\
NGC 1425   & $-0.07 \pm 0.17$  & $-1.73  \pm 6.24 $&  $  0.053  \pm 0.18$& $-0.05 \pm  0.13$& $-1.39\pm  4.91$ & $ 0.042\pm 0.15$  \\
NGC 7515   & $-0.15 \pm 0.18$  & $ 0.40  \pm 4.89 $&  $ -0.040  \pm 0.21$& $-0.10 \pm  0.11$& $ 0.27\pm  3.38$ & $-0.027\pm 0.14$  \\
NGC 7531   & $-0.21 \pm 0.14$  & $ 9.54  \pm 3.93 $&  $  0.029  \pm 0.12$& $-0.40 \pm  0.23$& $   ...        $ & $ 0.055\pm 0.24$  \\
NGC 7557   & $-0.31 \pm 0.08$  & $ 1.43  \pm 3.90 $&  $  0.164  \pm 0.22$& $-0.60 \pm  0.10$& $ 2.81\pm  8.85$ & $ 0.321\pm 0.57$  \\ 
NGC 7631   & $ 0.02 \pm 0.19$  & $ 0.89  \pm 5.82 $&  $  0.324  \pm 0.23$& $ 0.02 \pm  0.26$& $ 1.22\pm  8.29$ & $ 0.443\pm 0.43$  \\
NGC 7643   & $ 0.04 \pm 0.11$  & $ 0.19  \pm 0.44 $&  $  0.080  \pm 0.17$& $ 0.06 \pm  0.19$& $ 0.31\pm  0.79$ & $ 0.133\pm 0.31$  \\

\noalign{\smallskip}
\hline
\noalign{\medskip}
\end{tabular}
\end{small}
%\begin{minipage}{8cm}
%NOTE -- 
%Col.(1):Name; Col.(2): metallicity
%Col.(4):
%\end{minipage}
\label{tab:agemetalfa_grad}
\end{center}
\end{table*}

%%%%%%%%%%%%%%%%%%%%%%%%%%%%%%%%%%%%%%%%%%%%%%%%%%%%%%%%%%%%
% --- Figure histogramma valori gradienti eta ------------------------------------------------------

%%%%%%%%%%%%%%%%%%%%%%%%%%%%%%%%%%%%%%%%%%%%%%%%%%%%%%%%%%%%%%%%%%%%%%%%%

Most of the sample galaxies show no gradient in age (median=0.4), in
agreement with the earlier findings by \citet{mehletal03} and
\citet{sancetal06s} for the early-type galaxies, and by
\citet{jabletal07} for bulges. Only NGC~1366 and NGC~7531 display a
steep age gradient (see Fig. \ref{fig:histgrad_ama}).

Negative gradients of metallicity were observed in the sample
bulges. The number distribution show a clear peak at
$\Delta$(\ZH)$=-0.15$. This was already known for the stellar
populations in early-type galaxies
\citep{procetal02,mehletal03,sancetal06s}, and it has been recently
found for spiral bulges too by \citet{jabletal07}.
The presence of negative gradient in the metallicity radial profile
favors a scenario with bulge formation via dissipative collapse
\citep{lars74}.

Dissipative collapse implies strong inside-out formation that should
give rise to a negative gradient in the $\alpha$/Fe enhancement too
\citep{fesi02}. But no gradient was measured in the \aFe\/ radial
profiles for almost all the galaxies. Only 1 object of 14 are out of 3
$\sigma$ of distribution (Fig. \ref{fig:histgrad_ama}). All the
deviations from the median values of the other objects can be
explained by their errors only (Tab. \ref{tab:agemetalfa_grad}). 
  The same conclusion has been found by \citet{jabletal07} where they
  state that the changes in $\alpha$/Fe enhancement 
  are small and it is rather constant among their bulges.

The absence of gradients in $\alpha$/Fe enhancement is not in contrast
with the presence of a metallicity gradient and could be due to the
different enrichment of the material fuelling the star formation.

\begin{figure*}%[hp!]
\centering
\includegraphics[angle=90,width=0.9\textwidth]{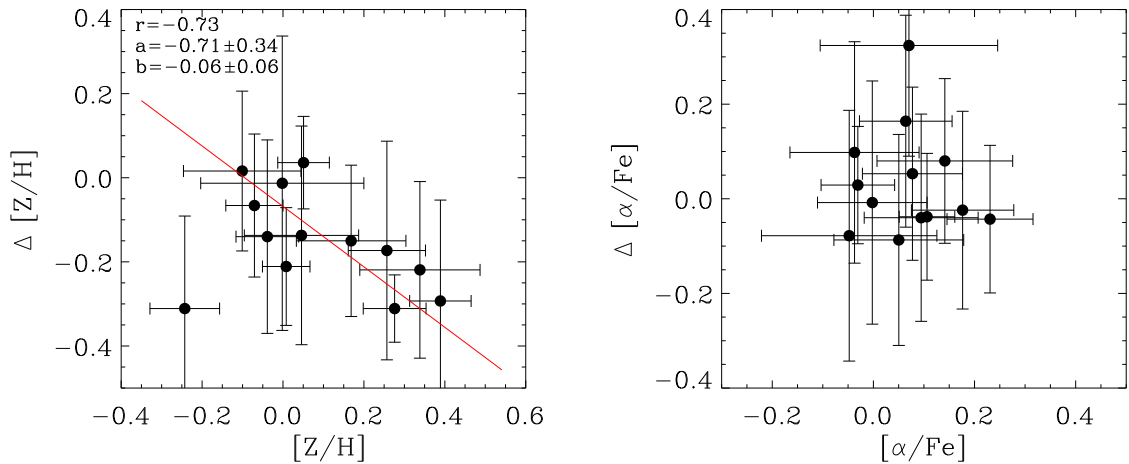}\\
\caption[]{The gradient and central values of metallicity  (left panel)
  and $\alpha$/Fe enhancement (right panel). In the left panel the solid line
  represents the linear regression ($y=ax+b$) through all the data
  points. The Pearson correlation coefficient ($r$) and the results of
  the linear fit are given.
\label{fig:grad_central_met_alfa}}
\end{figure*}

No correlation was found between the central value and gradient of
$\alpha$/Fe enhancement as in \citet{redaetal07}, while the central
value and gradient of metallicity are related in spite of the large
error-bars (Fig. \ref{fig:grad_central_met_alfa}).
All these hints suggest that a pure dissipative collapse is not able
to explain formation of bulges and that other phenomena like mergers
or acquisition events need to be invoked \citep{besh99,cobari99}.

%%%%%%%%%%%%%%%%%%%%%%%%%%%%%%%%%%%%%%%%%%%%%%%%%%%%%%%%%%%%
% --- Figure gradienti met-grad(met) alfafe-grad(alfafe) ------------------------------------------------------

%%%%%%%%%%%%%%%%%%%%%%%%%%%%%%%%%%%%%%%%%%%%%%%%%%%%%%%%%%%%%%%%%%%%%%%%%

The peculiar gradients observed for the stellar population of the
bulges of NGC~1366 and NGC~7531 suggest they have different
characteristic with respect to the rest of the sample. NGC~1366 shows
a steep positive age gradient (from about $5$ to $11$ Gyr) a strong
negative metallicity gradient (from about 0.5 to 0 dex) within 5" from
the centre. In this region a sharp rotation is measured while
immediately further out the galaxy is almost not rotating. Moreover,
the rotation in the innermost regions is opposite to that observed at
large radii (Fig. \ref{fig:kinplot}). Even if a disc-bulge decoupling
\citep{bertetal99,sarzetal04,matdeg04} could give a similar
kinematical signature the stellar population analysis is more
suggestive of the presence counter-rotating nuclear stellar disc
similar to those already observed in both elliptical
\citep[e.g.,][]{moreetal04} and spiral galaxies
\citep[e.g.,][]{corsetal99,pizzetal02,krajaf03,emsetal04,mcdetal06}.
The nuclear disc is younger than the rest of host bulge and formed by
enriched material probably acquired via interaction or minor merging.
The age of the stellar population of NGC~7531 rises from a central
value of about 2 Gyr to 12 Gyr at 4 arcsec. Further out it decreases
to 2 Gyr at 6 arcsec from the centre. Despite this change, no gradient
in both metallicity and $\alpha/$Fe enhancement was found.  We suggest
that this is due to the presence of a component which is corotating
but structurally decoupled with respect to the rest of the galaxy
\citep[e.g.,][]{mcdetal06}.

\section{Pseudobulges}
\label{slowfastrotator}

The current picture of bulge demography reveals that disc galaxies can
host bulges with different photometric and kinematic
properties \citep[see][for a review]{korken04}. Classical bulges are
similar to low-luminosity ellipticals and are thought to be formed by
mergers and rapid collapse. Pseudobulges are disc or bar
components which were slowly assembled by acquired material,
efficiently transferred to galaxy centre where it formed
stars. 
Pseudobulges can be identified according to their morphologic,
photometric, and kinematic properties, following the list of
characteristics compiled by \citet{korken04}. The more apply, the
safer the classification becomes.

The apparent flattening of the bulge is similar to that of the disc in
NGC~1292, NGC~1351 (Tab. \ref{tab:parameters}). Moreover, most of the
sample bulges have a S\'ersic index $n\leq2$. Only IC~5267, NGC~1351,
NGC~1425, NGC~7515, and NGC 7631 have $n>2$
(Tab. \ref{tab:parameters}).

\begin{figure}%[hp!]
\centering
\includegraphics[angle=90.0,width=0.49\textwidth,height=0.24\textheight]{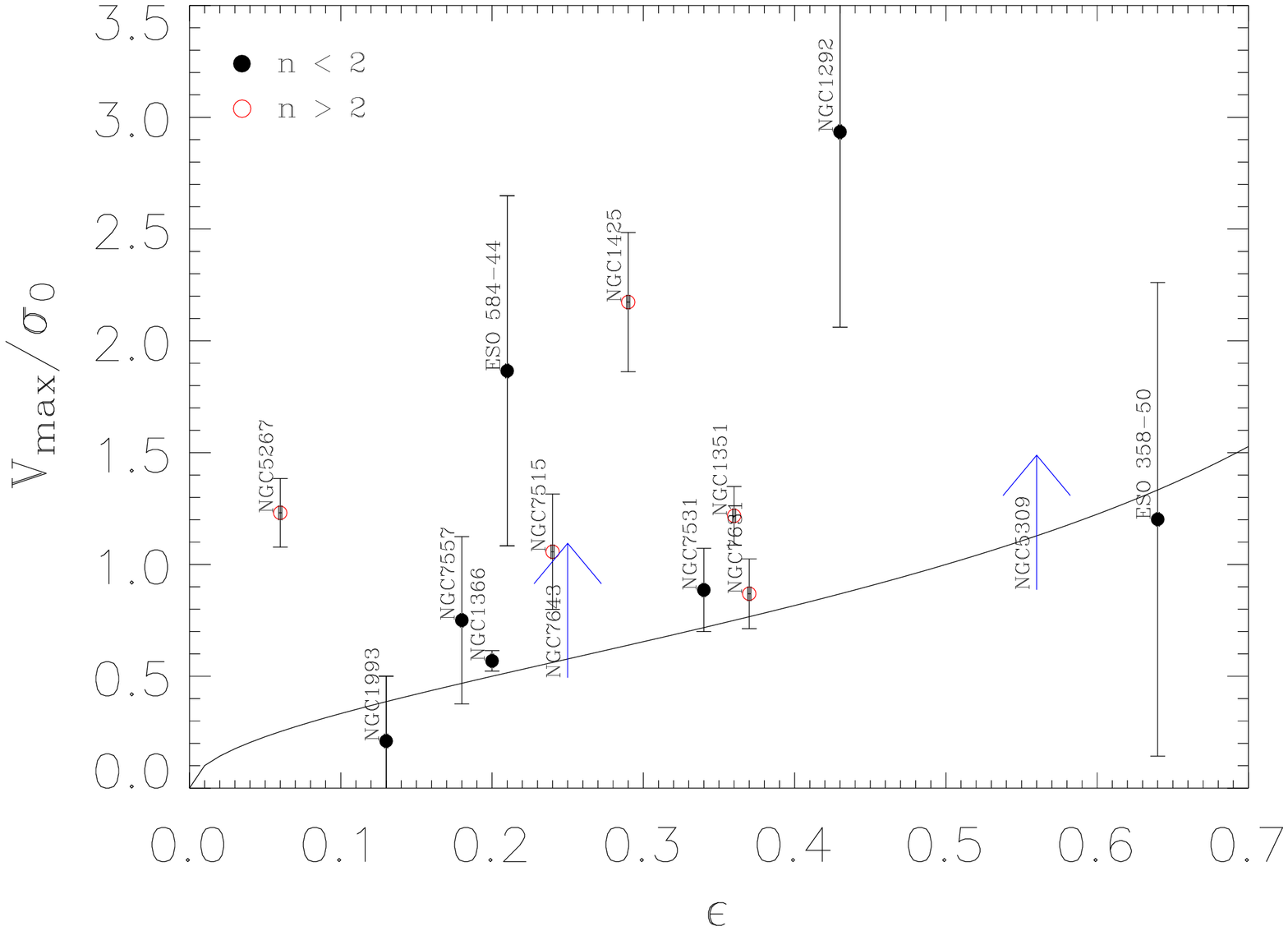}
\caption[]{The location of the sample bulges in the $(V_{\rm max}/\sigma_0,\epsilon)$
  plane. Filled and open circles correspond to bulges with S\'ersic
  index $n\leq2$ and $n>2$, respectively. The continuous line
  corresponds to oblate-spheroidal systems that have isotropic
  velocity dispersions and that are flattened only by rotation.
\label{fig:VSigma}} 
\end{figure}

Pseudobulges are expected to be more rotation-dominated than classical
bulges which are more rotation-dominated than giant elliptical
galaxies \citep{korken04}. We measured the maximum rotation velocity
$V_{\rm max}$ within $r_{\rm bd}$ from the stellar velocity curve and
central velocity dispersion $\sigma_0$ from the velocity dispersion
profile. The ellipticity $\epsilon$ of the bulge was measured by the
photometric decomposition (Tab. \ref{tab:parameters}).
For each galaxy we derived the ratio $V_{\rm max}/\sigma_0$. In
  Fig. \ref{fig:VSigma} we compared it to the value predicted for an
  oblate spheroid with isotropic velocity dispersion and the same
  observed ellipticity \citep{binney78,binney80,bitr87}. The large
  error-bars on the $V_{\rm max}/\sigma_0$ are driven by uncertainties
  on the central velocity dispersion.
 Most of the sample bulges rotate as fast as both bulges of
unbarred
\citep{kormendy82,korm93,kormetal82} and barred galaxies
\citep{kormendy82,aguetal05b}. However, the values of $V_{\rm
    max}/\sigma_0$ measured for NGC~1292, NGC~1425, ESO~548-44, and
  NGC~5267 are significantly larger than that
  predicted for the oblate spheroids.

\begin{figure}%[hp!]
\centering
\includegraphics[angle=90.0,width=0.49\textwidth,height=0.24\textheight]{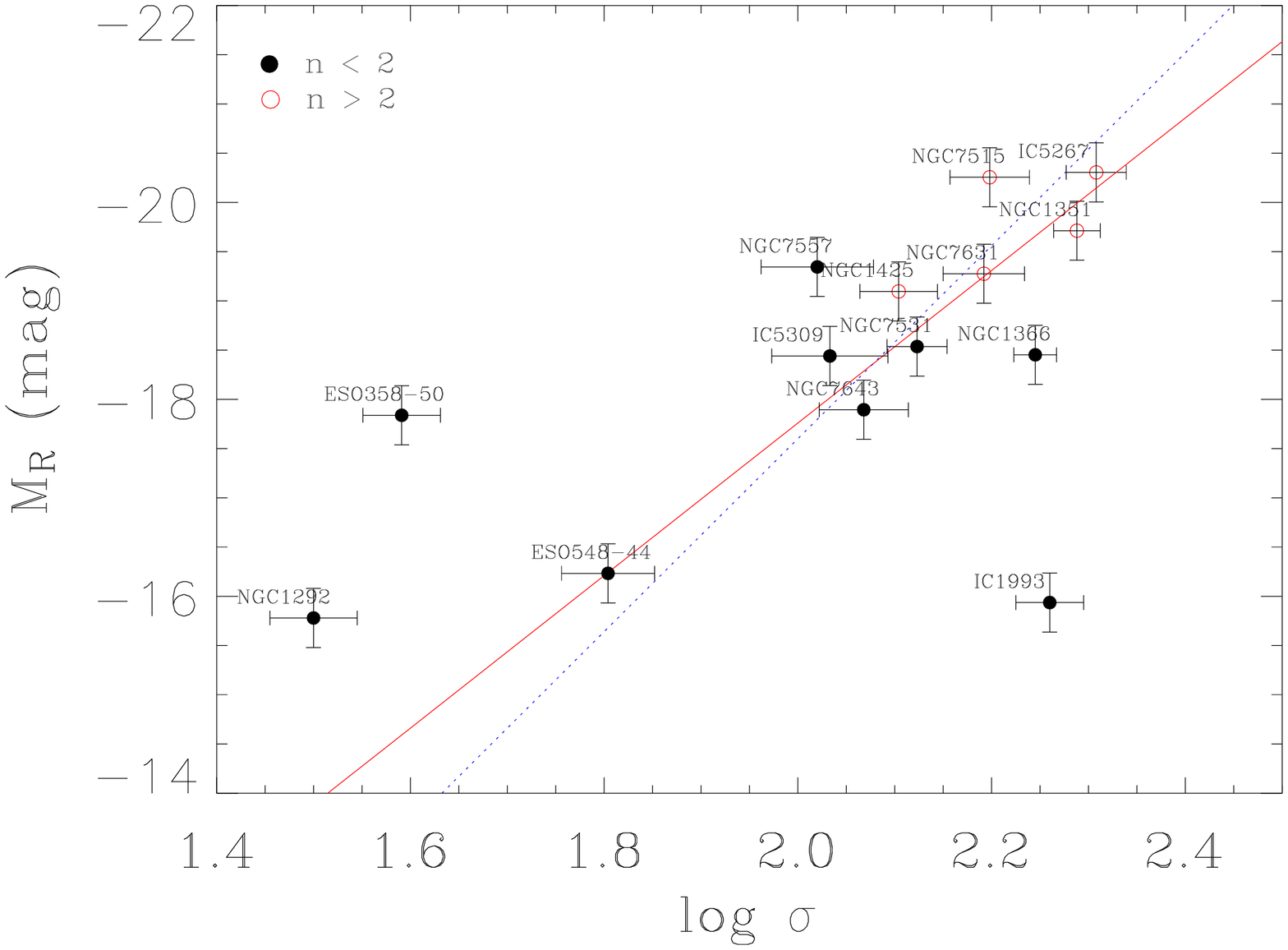}
\caption[]{The location of the sample bulges with respect to the
    FJ relation by Forbes \& Ponman (1999, blue dashed line). Filled
    and open circles correspond to bulges with S\'ersic index $n\leq2$
    and $n>2$, respectively and the linear fit is shown (red continuous
    line).
\label{fig:FJ}} 
\end{figure}

An other characteristic listed by \citet{korken04} is the position of
pseudobulges with respect to the Faber-Jackson relation.  The
Faber-Jackson relation \citep[FJ,][]{faje76} relates the luminosity of
the elliptical galaxies and early-type bulges to their central
velocity dispersion. The pseudobulges fall above the correlation
\citep{korken04}. Sample bulges, except for ESO~358-G50 and NGC~1292,
are consistent with the $R-$band FJ relation we built from
\citet[][$L\propto\sigma^{3.92}$]{forpon99}. They are characterized by
a lower velocity dispersion or equivalently a higher luminosity with
respect to their early-type counterparts (Fig. \ref{fig:FJ}).

According to the prescriptions by \citet{korken04}, the bulge of
NGC~1292 is the most reliable pseudobulges in our sample. Information
about its stellar population give more constraints on its nature and
formation process. In fact, the bulge population has a intermediate
age (3 Gyr) and low metal content (\ZH$=-0.7$ dex). The $\alpha/$Fe
enhancement is the lowest in our sample (\aFe$=-0.12$ dex) suggesting
a prolonged star formation history. The presence of emission lines in
the spectrum shows that star formation is still
ongoing. These properties are consistent with a slow buildup of the
bulge of NGC~1292 within a scenario of secular evolution.

%%%%%%%%%%%%%%%%%%%%%%%%%%%%%%%%%%%%%%%%%%%%%%%%%%%%%%%%%%%%
% --- Figure V/S-Ell con n ----------------------------------

%%%%%%%%%%%%%%%%%%%%%%%%%%%%%%%%%%%%%%%%%%%%%%%%%%%%%%%%%%%%
% --- Figure FJ --------- ----------------------------------

\section{Conclusions}
\label{conclusion}

The structural parameters and properties of the stellar population of
the bulges of sample of 14 S0 and spiral galaxies of the Fornax,
Eridanus and Pegasus cluster, and NGC~7582 group were investigated to
constrain the dominant mechanism at the epoch of their assembly.

\begin{itemize} 
 
\item The bulge and disc parameters of the sample galaxies were
  derived performing a two-dimensional photometric decomposition of
  their $R-$band images. The surface-brightness distribution of the
  galaxy was assumed to be the sum of the contribution of a S\'ersic
  bulge and an exponential disk. The two components were characterized
  by elliptical and concentric isophotes with constant (but possibly
  different) ellipticity and position angles. Most of the bulges have
  a S\'ersic index $n\leq2$ and for few of them the apparent
  flattening of the bulge is similar to that of the disc. According to
  \citet{korken04} the disc-like flattening and radial profile are the
  photometric signature of the pseudobulge.

\item The central values of velocity dispersion $\sigma$ and \Mgb,
  \Mgd, \Hb, \Fe, and \MgFe\/ line-strength indices were derived from
  the major-axis spectra. Correlations between \Mgd , \Fe , \Hb , and
  $\sigma$ were found. The \Mgd$-\sigma$ and \Hb$-\sigma$,
  correlations are steeper than those found for early-type galaxies
  \citep[e.g.,][]{bernetal98,jorgen99,kuntetal00,mehletal03}.  The
  \Fe$-\sigma$ correlation is consistent with previous findings for
  spiral bulges \citep{idiaetal96,prugetal01,procetal02}.

\item The mean ages, total metallicities, and total $\alpha/$Fe
      enhancements in the center of the sample bulges were derived by
      using the stellar population models by \citet{thmabe03}.
      The youngest bulges
      have an average age of 2 Gyr. They are characterized by ongoing
      star formation.  The stellar population of intermediate-age
      bulges is 4 to 8 Gyr old. It has solar metallicity (\ZH$\;=0.0$
      dex).  The older bulges have a narrow distribution in age around
      10 Gyr and high metallicity (\ZH$\;=0.30$ dex). Most of the
      sample bulges display solar $\alpha/$Fe enhancements. A few have
      a central super-solar enhancement (\aFe$\;=0.3$).

\item There is no correlation between age, metallicity, and
      $\alpha/$Fe enhancement of bulges with the membership of the host
      galaxy to different cluster. There is a correlation with the
      velocity dispersion.
      The more massive bulges of our sample galaxies are older, more
      metal rich and characterized by a fast star formation. Since we
      did not find any correlation with galaxy morphology we exclude
      a strong interplay between the bulge and disc components.

\item Most of the sample galaxies show no gradient in age
      and a negative gradient of metallicity. This is in
      agreement with the earlier findings by \citet{mehletal03} and
      \citet{sancetal06s} for the early-type galaxies, and by
      \citet{jabletal07} for bulges. 
      The presence of negative gradient in the metallicity radial
      profile favors a scenario with bulge formation via dissipative
      collapse. This implies strong inside-out formation that should
      give rise to a negative gradient in the $\alpha$/Fe enhancement
      too \citep{fesi02}. But, no gradient was measured in the \aFe\/
      radial profiles for all the galaxies, except for NGC~1366 and
      NGC~7531.
      Moreover, the correlation between the central value and gradient
      of metallicity can not be built by pure dissipative collapse
      \citep{besh99,cobari99} and suggests that mergers or acquisition
      events need to be invoked during the bulge assembly.

\item The peculiar gradients observed for the stellar population of
  the bulges of NGC~1366 and NGC~7531 suggest they host a
  substructure.  Very interestingly, in NGC~1366 we found the presence
  of a kinematically-decoupled component. It is younger than the host
  bulge and formed by enriched material probably acquired via
  interaction or minor merging.
      
\item According to the prescriptions by \citet{korken04} the bulge of
  NGC~1292 is a pseudobulge. The properties of its stellar population
  are consistent with a slow buildup within a scenario of secular
  evolution. Indeed, the bulge of NGC~1292 has a intermediate age (3
  Gyr) and low metal content (\ZH$=-0.7$ dex). The $\alpha/$Fe
  enhancement is the lowest in our sample (\aFe$=-0.12$ dex)
  suggesting a prolonged star formation history. The presence of
  emission lines in the spectrum is a signature of ongoing star
  formation.

\end{itemize} 
 
\section*{Acknowledgments}

We thank David Burstein, Stephane Courteau, Reynier Peletier, and
Alexandre Vazdekis for useful discussion and suggestions. LM is
supported by grant (CPDR061795/06) by Padova University.
University. FB, EMC, and AP receive support from grant PRIN2005/32 by
Istituto Nazionale di Astrofisica (INAF) and from grant CPDA068415/06
by Padua University. LM and EMC acknowledge the Max-Planck-Institut
fuer extraterrestrische Physik for hospitality while this paper was in
progress.

\bibliographystyle{mn2e}
\bibliography{bib}

\appendix

\section{Kinematics results}
%%%%%%%%%%%%%%%%%%%%%%%%%%%%%%%%%%%%%%%%%%%%%%%%%%%%%%%%%%%%%%%%%%%%%%%%%
%%% Kinematics results

\begin{scriptsize}
\begin{figure*} 
\caption[Kinematical plot]{Kinematic parameters measured along the
  major axis of the galaxies. For each galaxy the curve is folded
  around the nucleus. Asterisks and dots refer to the two sides (est/west) of the
  galaxy. The radial profiles of the line-of-sight velocity ($v$) after
  the subtraction of the systemic velocity and velocity dispersion
  ($\sigma$) are shown from top to bottom. For NGC~1292, IC~1993,
  ESO~358-50 and ESO~548-44, only the central value of $\sigma$ was measured and it
  is tabulated in Tab. \ref{tab:centval_lickind}. The vertical dashed
line indicates the radius (\Rdb) where the surface-brightness
contributions of bulge and disc are the same.  \centering}
\includegraphics[angle=0.0,width=0.431\textwidth,height=0.24\textheight]{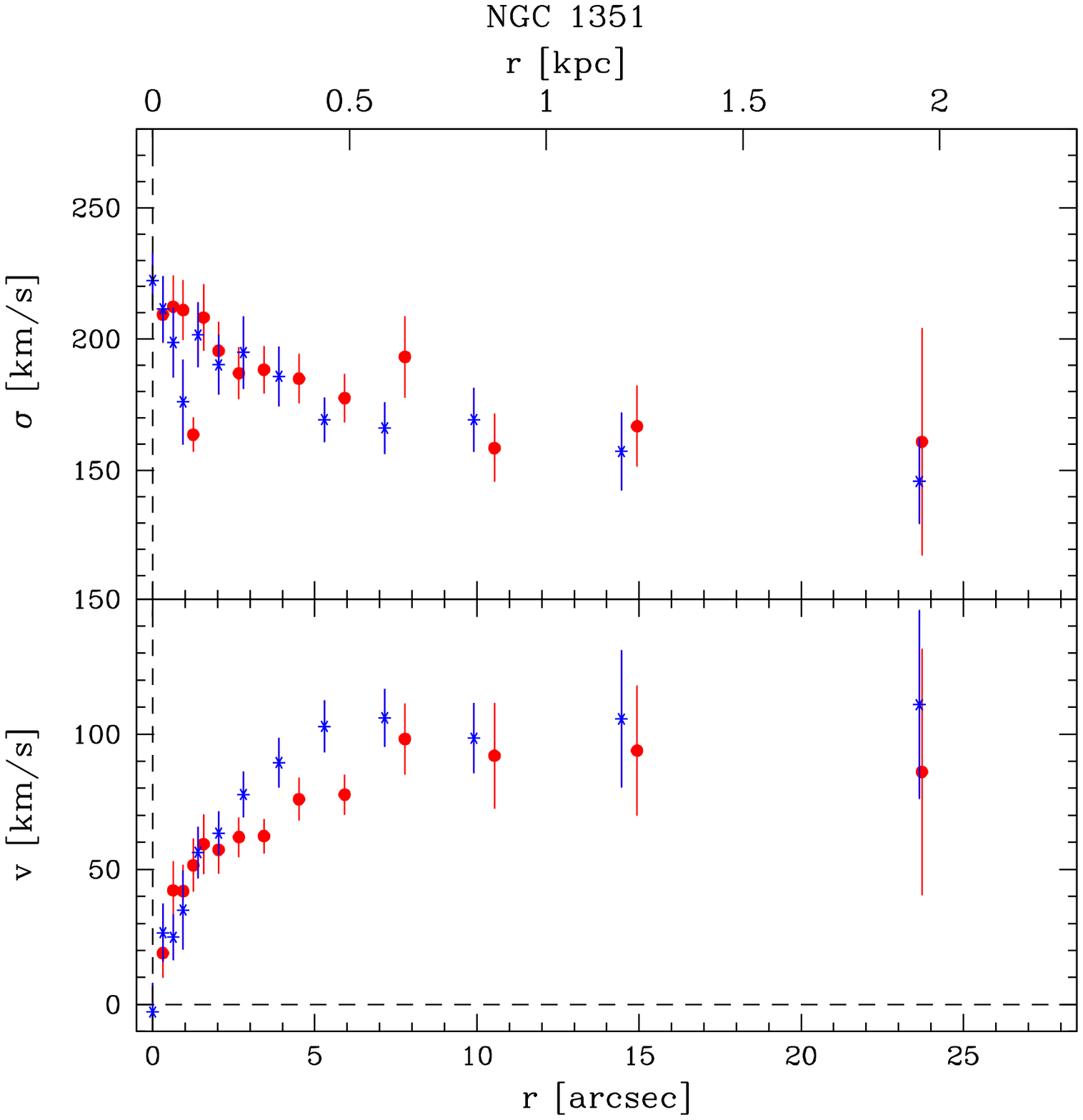}
\includegraphics[angle=0.0,width=0.431\textwidth,height=0.24\textheight]{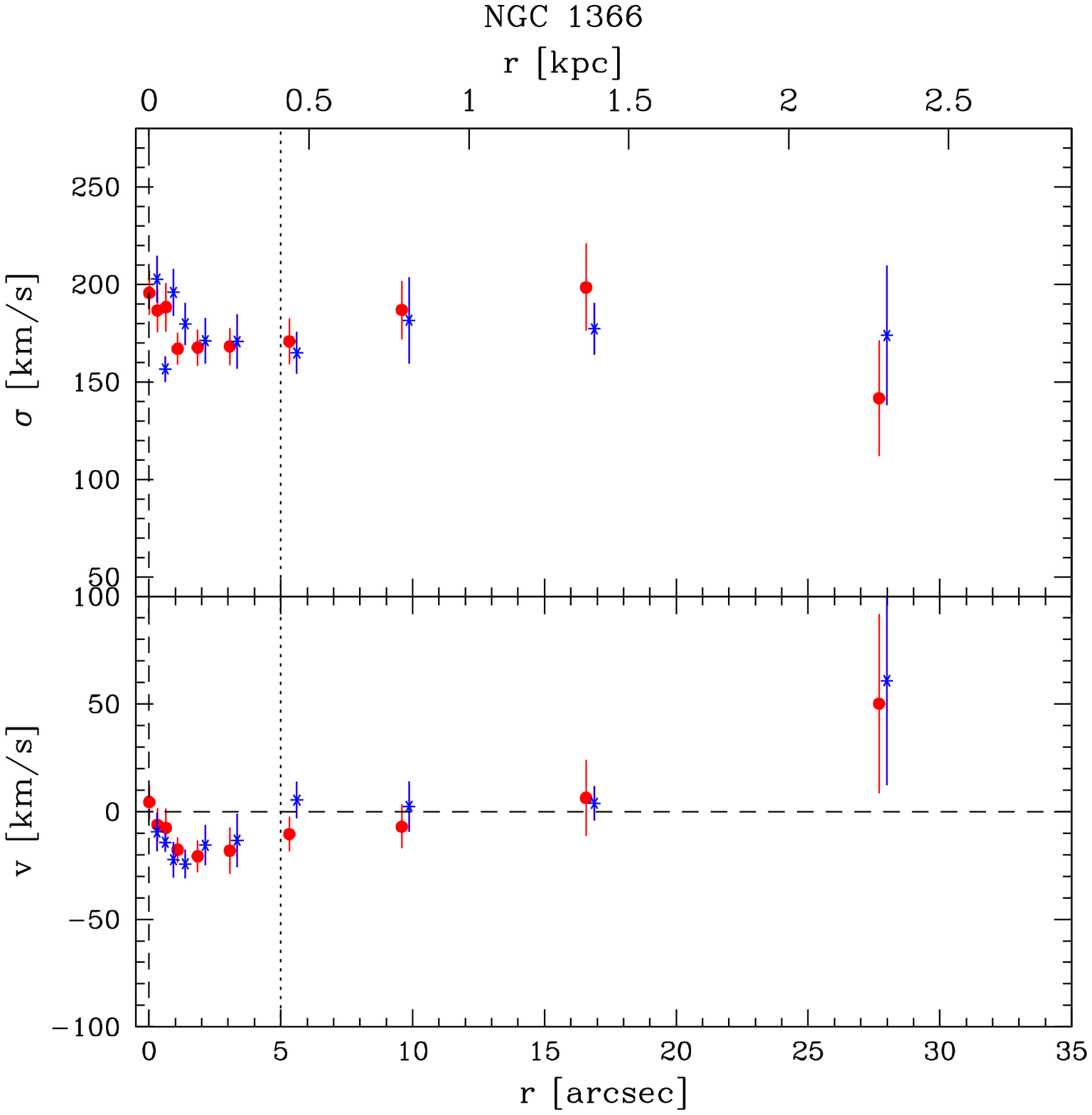}
\includegraphics[angle=0.0,width=0.431\textwidth,height=0.24\textheight]{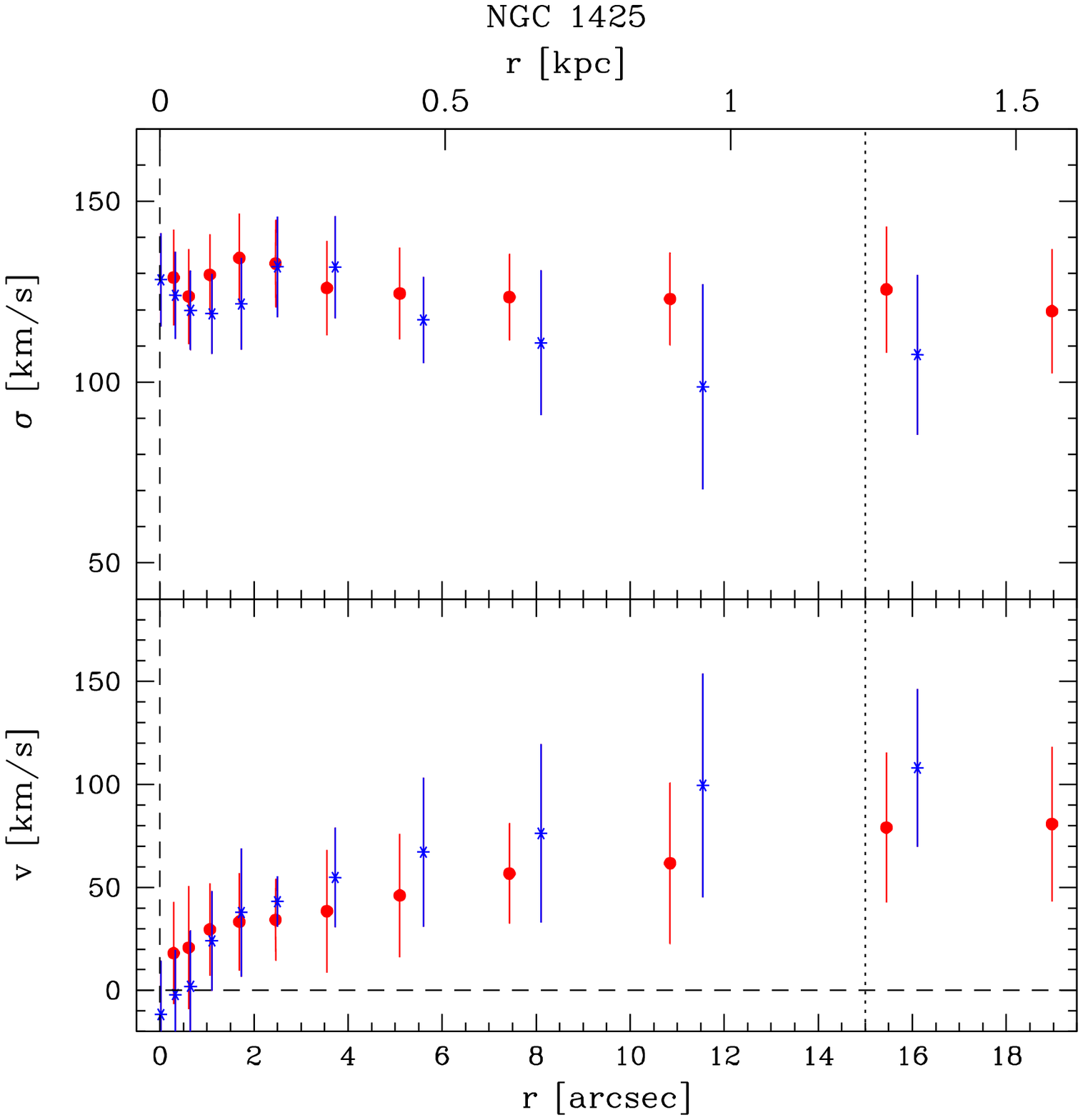}
\includegraphics[angle=0.0,width=0.431\textwidth,height=0.24\textheight]{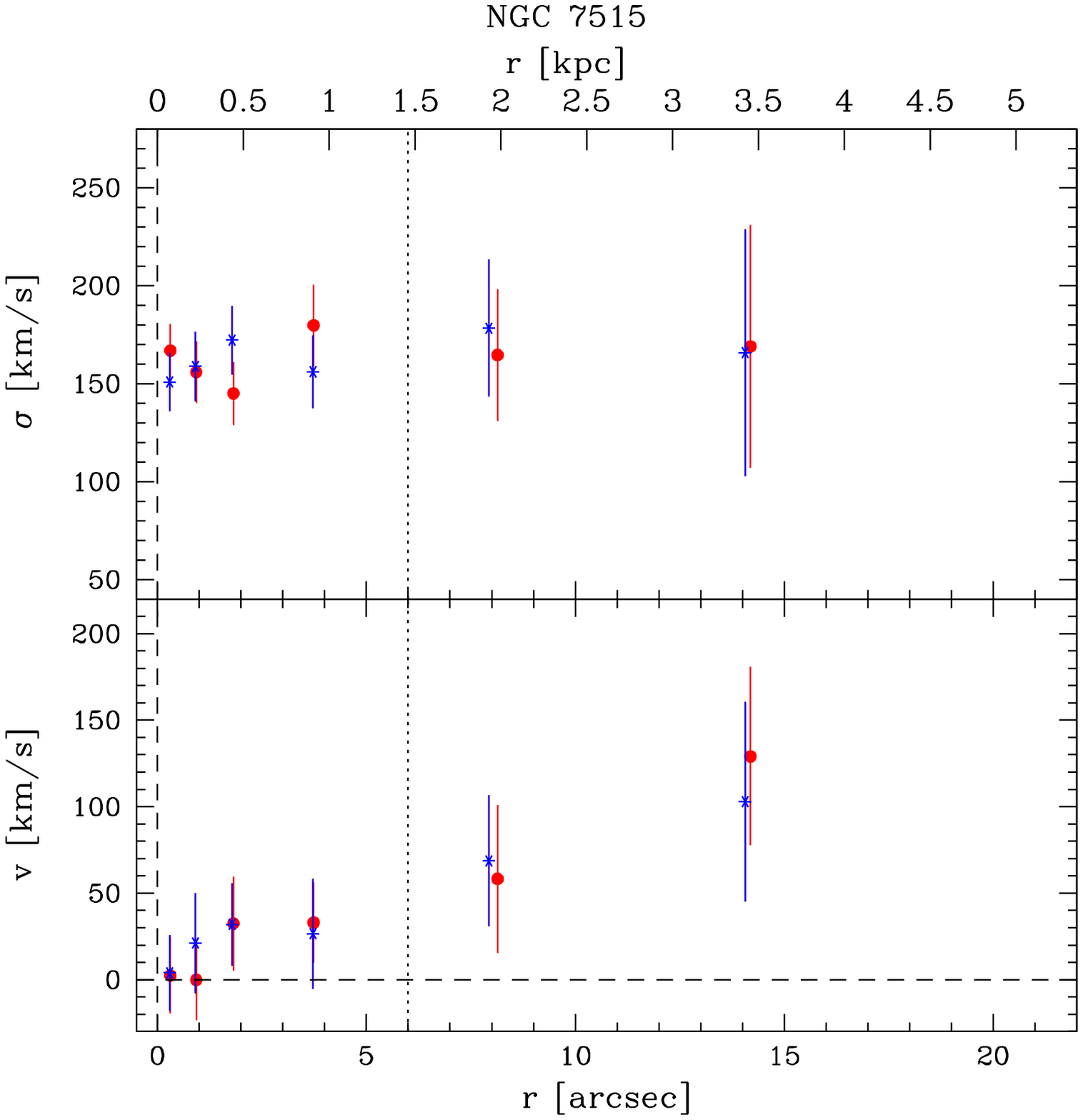}
\includegraphics[angle=0.0,width=0.431\textwidth,height=0.24\textheight]{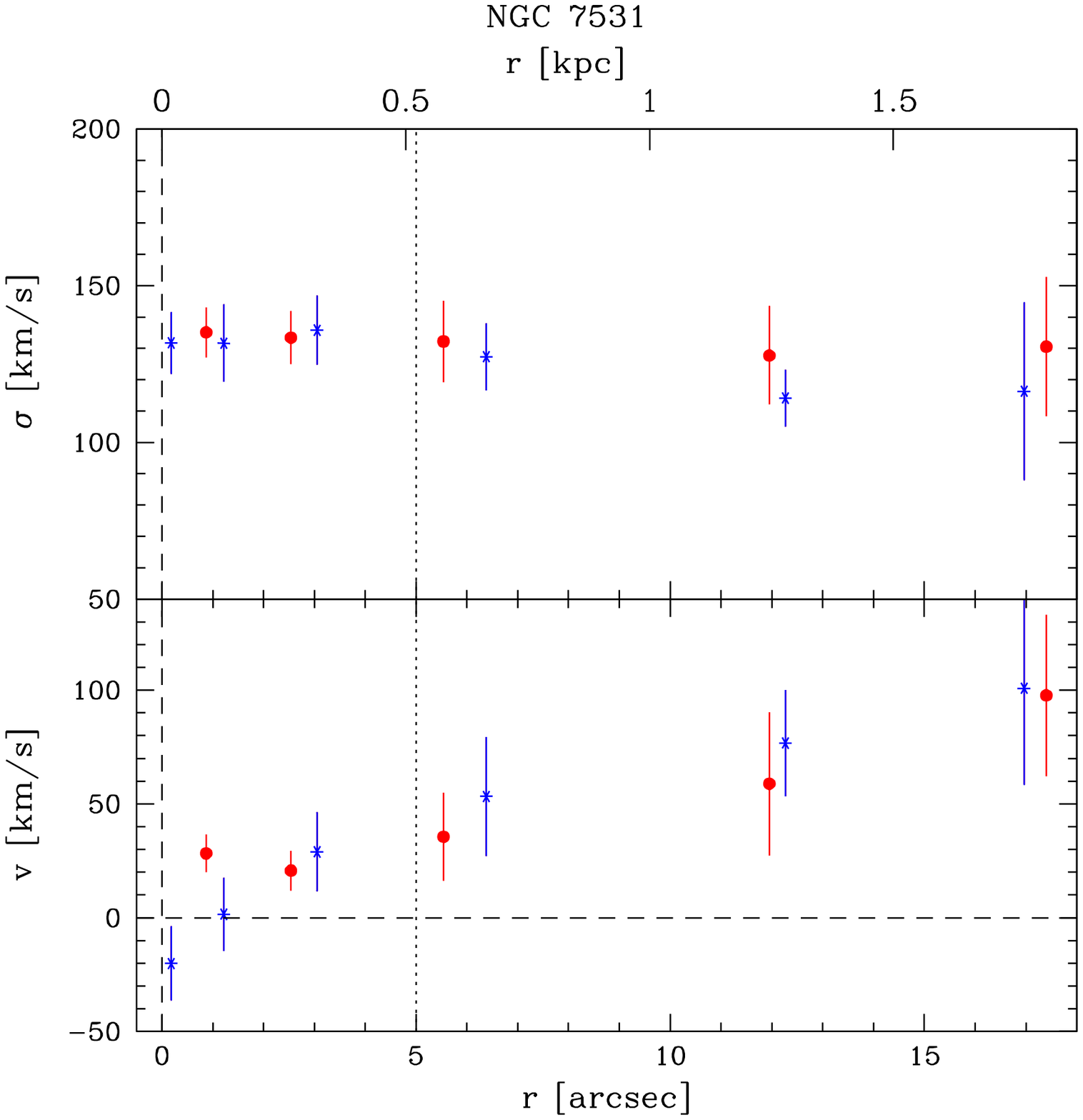}
\includegraphics[angle=0.0,width=0.431\textwidth,height=0.24\textheight]{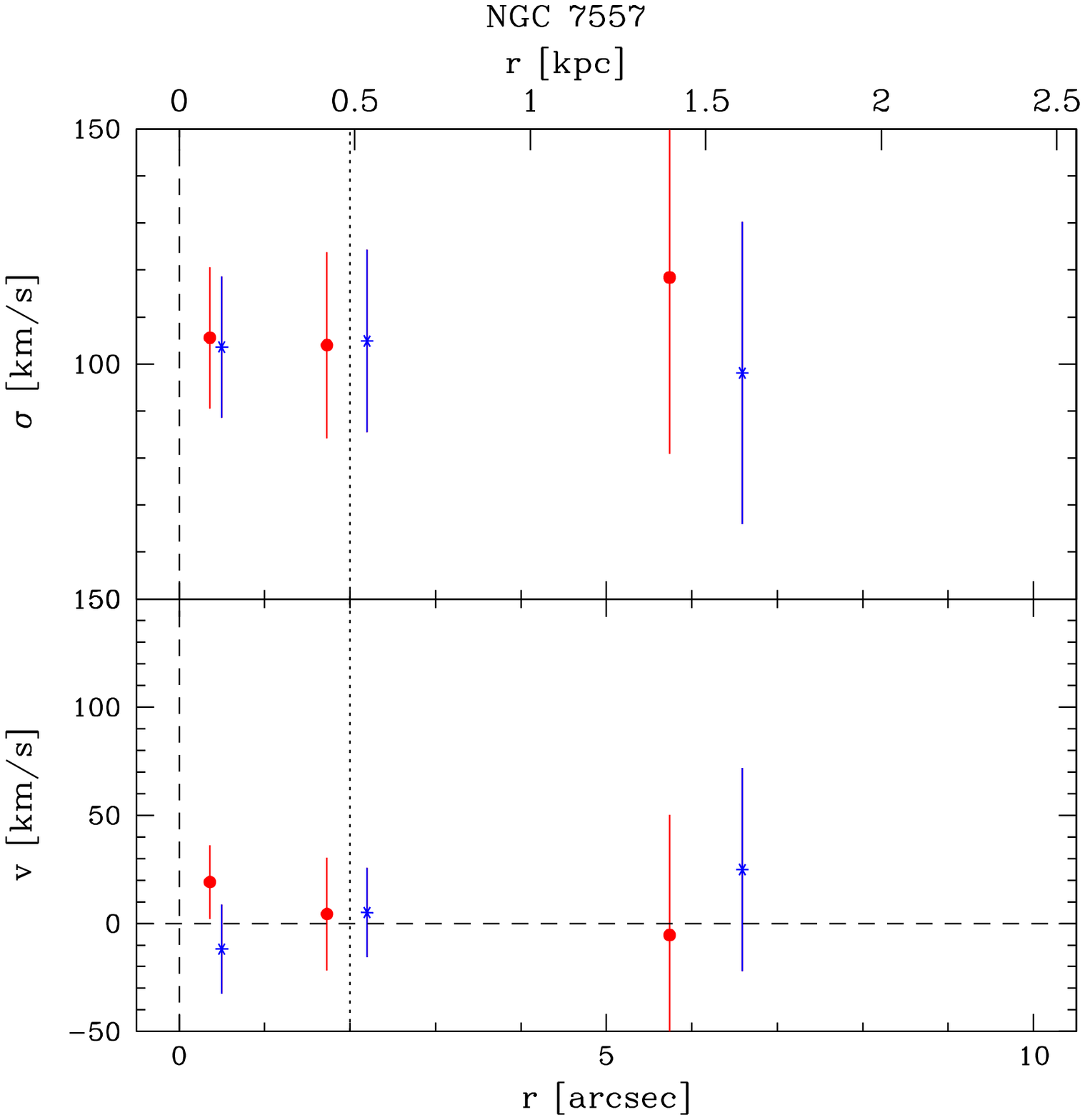}
\includegraphics[angle=0.0,width=0.431\textwidth,height=0.24\textheight]{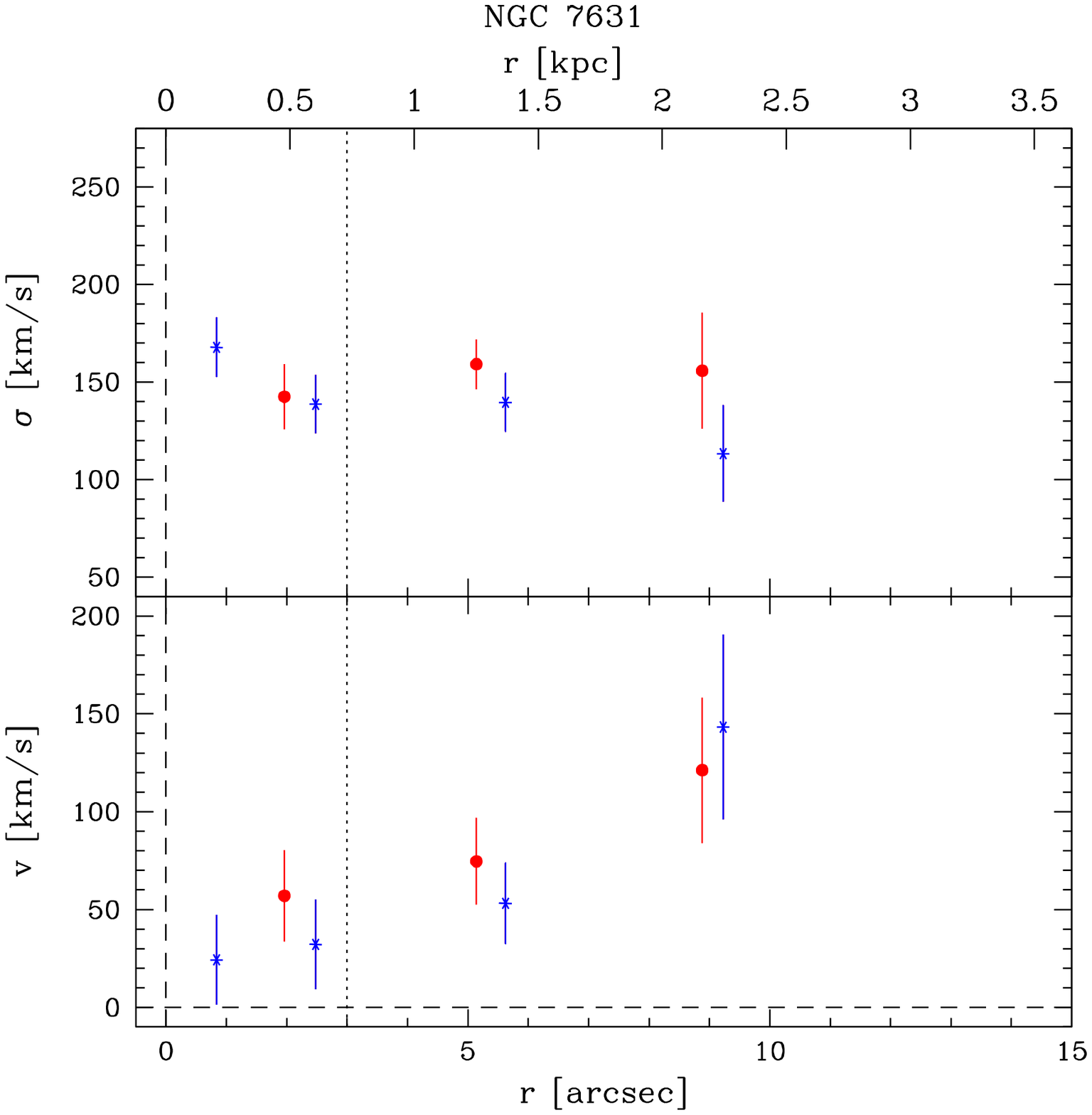}
\includegraphics[angle=0.0,width=0.431\textwidth,height=0.24\textheight]{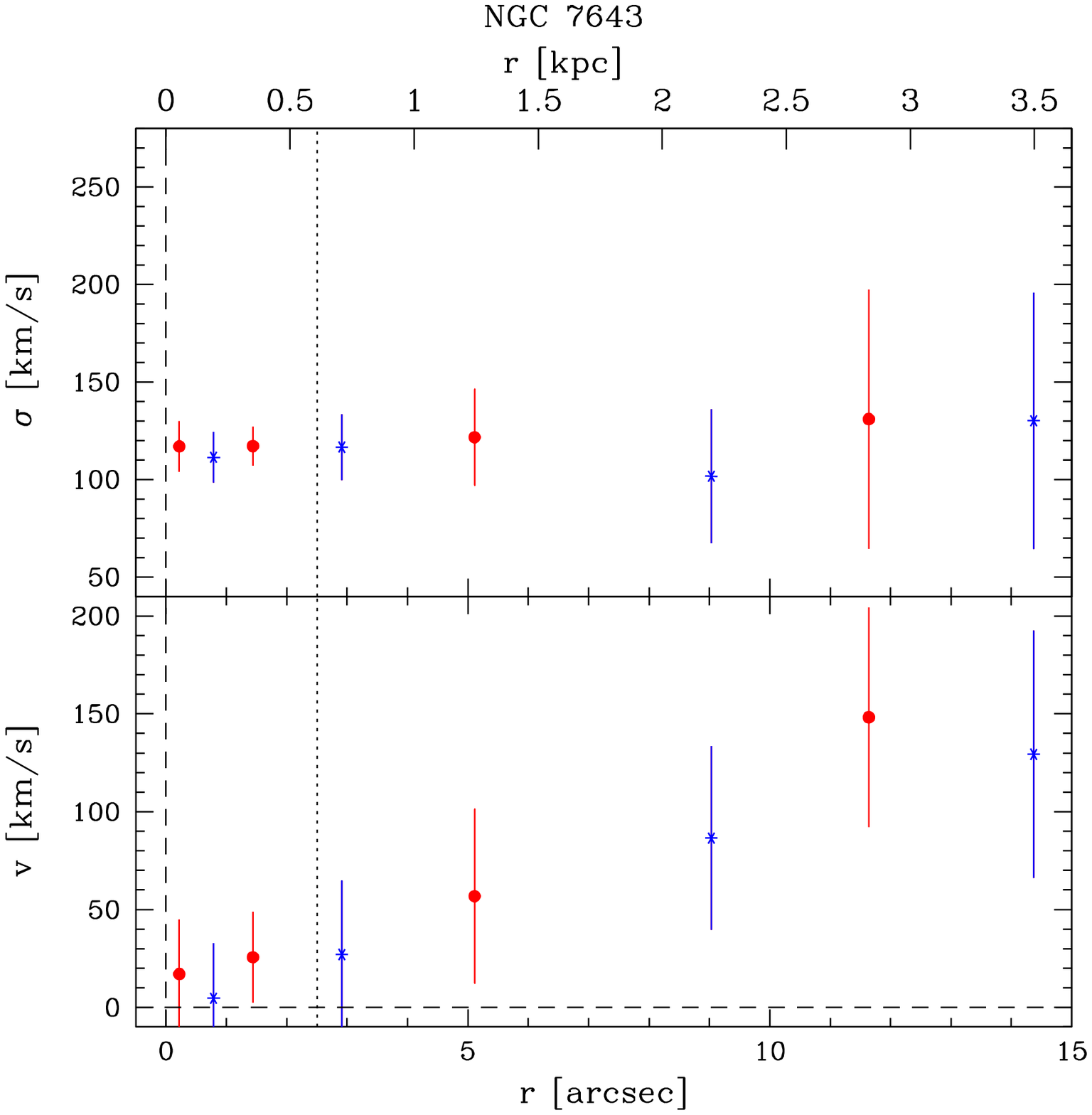}
\label{fig:kinplot}
\end{figure*}
\end{scriptsize}
\begin{figure*}
\centering
\includegraphics[angle=0.0,width=0.431\textwidth,height=0.24\textheight]{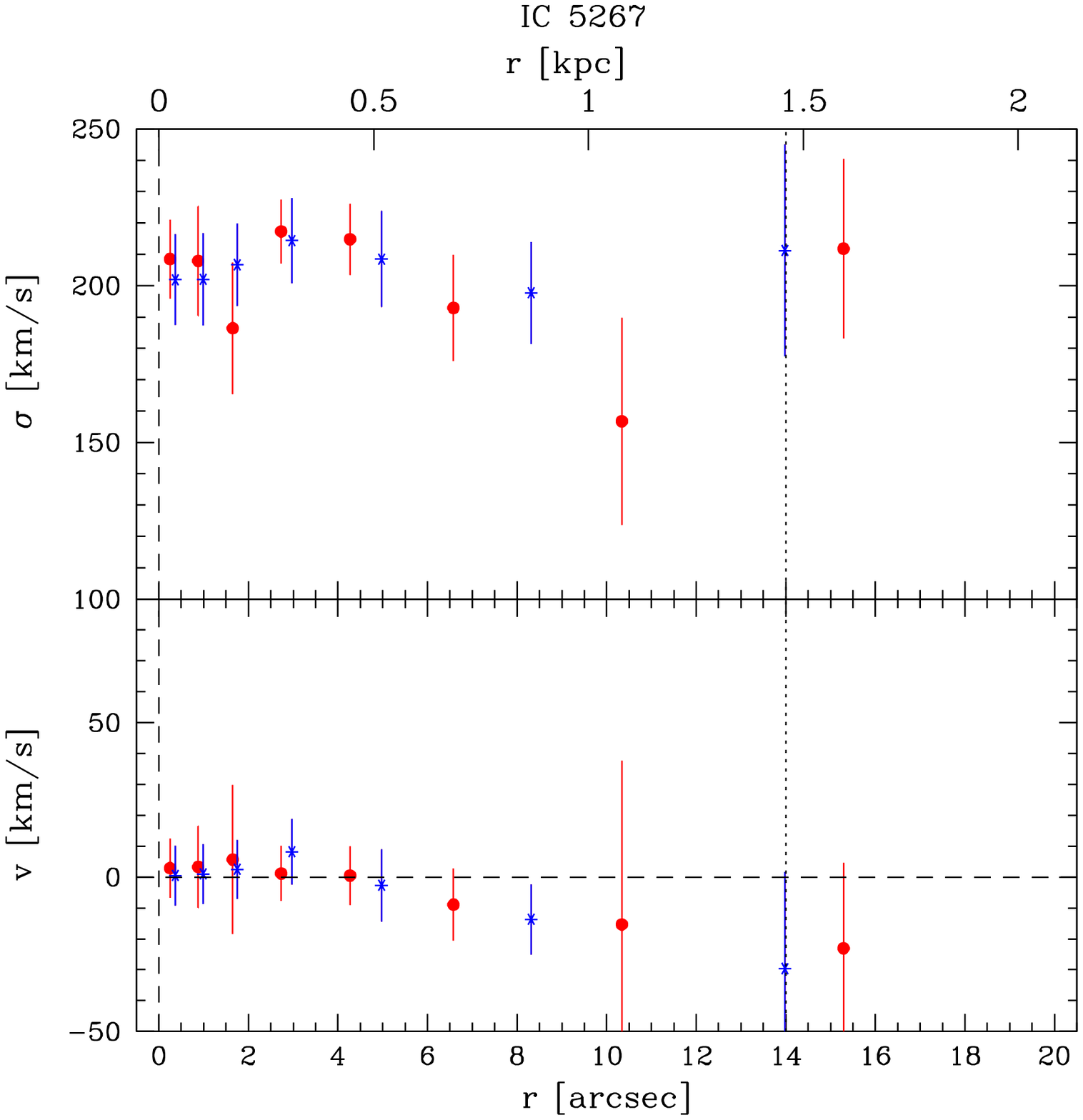}
\includegraphics[angle=0.0,width=0.431\textwidth,height=0.24\textheight]{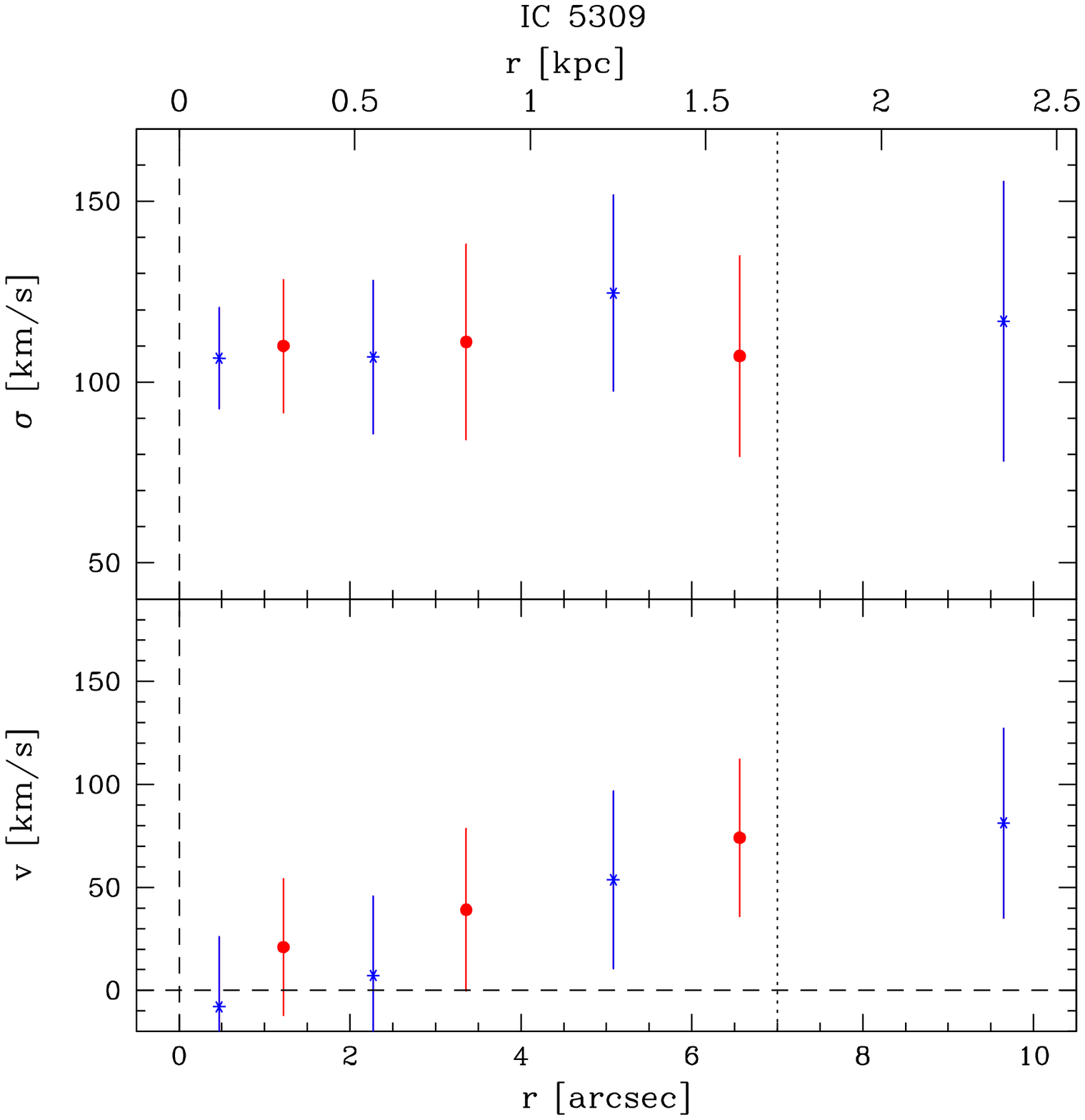}
\contcaption{}
\end{figure*}
%%%%%%%%%%%%%%%%%%%%%%%%%%%%%%%%%%%%%%%%%%%%%%%%%%%%%%%%%%%%%%%%%%%%%%%%%
%%% KINEMATICAL and indices table

\clearpage
\section{Stellar kinematic and line strength indices.}
\begin{table*}
\caption{Stellar kinematic and line strength indices of the sample
  galaxies.The columns show the following: (1) radius along the major
  axis, (2) rotation velocity, subtracted of the galaxy systemic
  velocity, along the major axis, (3) rotation velocity dispersion,
  (4), (5), (6), (7), (8) measured values for \Mgd, \Mgb, {\rm Fe5270}, 
{\rm Fe5335}, \Hb\/ respectively. Positive and negative
  values are respectively for absorption and emission line
  strength.Column (9) shows the applied correction to \Hb\/ absorption index.}
\begin{tabular}{rrrrrrrrr}
\hline
\noalign{\smallskip}
\multicolumn{1}{c}{r} &
\multicolumn{1}{c}{$V$} &
\multicolumn{1}{c}{$\sigma$} &
\multicolumn{1}{c}{\Mgd} &
\multicolumn{1}{c}{\Mgb} &
\multicolumn{1}{c}{{\rm Fe5270}} &
\multicolumn{1}{c}{{\rm Fe5335}} &
\multicolumn{1}{c}{\Hb} &
\multicolumn{1}{c}{$\Delta$ \Hb } \\
\multicolumn{1}{c}{(arcsec) } &
\multicolumn{1}{c}{(\kms)} &
\multicolumn{1}{c}{(\kms)} &
\multicolumn{1}{c}{(mag)} &
\multicolumn{1}{c}{(\AA)} &
\multicolumn{1}{c}{(\AA)} &
\multicolumn{1}{c}{(\AA)} &
\multicolumn{1}{c}{(\AA)} &
\multicolumn{1}{c}{(\AA)} \\
\multicolumn{1}{c}{(1)} &
\multicolumn{1}{c}{(2)} &
\multicolumn{1}{c}{(3)} &
\multicolumn{1}{c}{(4)} &
\multicolumn{1}{c}{(5)} &
\multicolumn{1}{c}{(6)} &
\multicolumn{1}{c}{(7)} &
\multicolumn{1}{c}{(8)} &
\multicolumn{1}{c}{(9)} \\
\noalign{\smallskip}
\hline
\noalign{\smallskip}
{\bf ESO~358-50} \\
\noalign{\smallskip}
\hline
\noalign{\smallskip}
$-10.9 $ &  $-25.1 \pm 43.1  $ & $ ...~~~~~~~~~ $  &  $0.149  \pm  0.017  $ &  $2.826  \pm 0.683 $ &  $2.337 \pm 0.816 $ &  $2.304 \pm 0.996 $  &  $ 2.35 \pm 0.59 $ & 0.00 \\
$ -6.1 $ &  $-21.9 \pm 36.3  $ & $ ...~~~~~~~~~ $  &  $0.163  \pm  0.008  $ &  $2.641  \pm 0.341 $ &  $2.422 \pm 0.408 $ &  $2.199 \pm 0.506 $  &  $ 1.99 \pm 0.29 $ & 0.00 \\
$ -1.6 $ &  $ -5.5 \pm 31.4  $ & $ ...~~~~~~~~~ $  &  $0.157  \pm  0.007  $ &  $2.822  \pm 0.270 $ &  $2.587 \pm 0.327 $ &  $2.297 \pm 0.400 $  &  $ 2.34 \pm 0.23 $ & 0.00 \\
$  1.4 $ &  $ 10.0 \pm 18.1  $ & $ ...~~~~~~~~~ $  &  $0.154  \pm  0.003  $ &  $2.669  \pm 0.135 $ &  $2.491 \pm 0.157 $ &  $2.412 \pm 0.189 $  &  $ 2.35 \pm 0.11 $ & 0.00 \\	
$  5.2 $ &  $ 10.1 \pm 38.5  $ & $ ...~~~~~~~~~ $  &  $0.152  \pm  0.008  $ &  $2.645  \pm 0.324 $ &  $2.643 \pm 0.378 $ &  $2.419 \pm 0.462 $  &  $ 2.37 \pm 0.27 $ & 0.00 \\
$  9.9 $ &  $ 32.4 \pm 41.8  $ & $ ...~~~~~~~~~ $  &  $0.156  \pm  0.012  $ &  $2.841  \pm 0.506 $ &  $2.572 \pm 0.606 $ &  $2.235 \pm 0.768 $  &  $ 2.20 \pm 0.41 $ & 0.00 \\
\noalign{\smallskip}
\hline
\noalign{\smallskip}
{\bf ESO~548-44} \\
\noalign{\smallskip}
\hline
\noalign{\smallskip}
$-8.7 $ &  $35.0  \pm  36.8 $ & $ ...~~~~~~~~~ $ &  $0.137  \pm 0.013  $ &  $  2.262 \pm 0.406 $ &  $2.269 \pm 0.484 $ &  $2.635 \pm  0.591 $  &  $ 1.78 \pm 0.35  $ & 0.00 \\
$-3.5 $ &  $20.9  \pm  36.1 $ & $ ...~~~~~~~~~ $ &  $0.174  \pm 0.013  $ &  $  2.774 \pm 0.415 $ &  $2.788 \pm 0.502 $ &  $2.333 \pm  0.614 $  &  $ 1.58 \pm 0.37  $ & 0.00 \\
$-0.8 $ &  $ 7.9  \pm  43.7 $ & $ ...~~~~~~~~~ $ &  $0.194  \pm 0.013  $ &  $  3.139 \pm 0.409 $ &  $2.994 \pm 0.490 $ &  $2.644 \pm  0.577 $  &  $ 1.77 \pm 0.36  $ & 0.00 \\	
$ 0.3 $ &  $ 0.1  \pm  48.1 $ & $ ...~~~~~~~~~ $ &  $0.190  \pm 0.012  $ &  $  3.045 \pm 0.399 $ &  $2.861 \pm 0.476 $ &  $2.557 \pm  0.558 $  &  $ 2.01 \pm 0.35  $ & 0.00 \\
$ 2.3 $ &  $33.7  \pm  51.0 $ & $ ...~~~~~~~~~ $ &  $0.179  \pm 0.012  $ &  $  2.719 \pm 0.392 $ &  $2.789 \pm 0.455 $ &  $2.295 \pm  0.552 $  &  $ 1.71 \pm 0.33  $ & 0.00 \\
$ 7.3 $ &  $30.5  \pm  38.4 $ & $ ...~~~~~~~~~ $ &  $0.150  \pm 0.012  $ &  $  2.615 \pm 0.405 $ &  $2.363 \pm 0.480 $ &  $2.335 \pm  0.584 $  &  $ 2.04 \pm 0.34  $ & 0.00 \\
\noalign{\smallskip}
\hline
\noalign{\smallskip}
{\bf IC~1993} \\
\noalign{\smallskip}
\hline
\noalign{\smallskip}
$-9.6 $ &  $-15.0  \pm  45.2 $ &$ ...~~~~~~~~~ $  &  $0.143 \pm  0.017  $ &  $2.569  \pm 0.503$ &  $2.222 \pm  0.597 $ &  $ 1.776 \pm 0.696 $  &  $1.76  \pm 0.44  $ & 0.00 \\
$-4.6 $ &  $-10.2  \pm  49.5 $ &$ ...~~~~~~~~~ $  &  $0.165 \pm  0.013  $ &  $2.955  \pm 0.407$ &  $2.419 \pm  0.478 $ &  $ 2.155 \pm 0.557 $  &  $2.08  \pm 0.35  $ & 0.00 \\
$-1.1 $ &  $  5.8  \pm  38.2 $ &$ ...~~~~~~~~~ $  &  $0.184 \pm  0.007  $ &  $3.084  \pm 0.198$ &  $2.617 \pm  0.235 $ &  $ 2.268 \pm 0.275 $  &  $2.12  \pm 0.17  $ & 0.00 \\	
$ 0.6 $ &  $ 19.0  \pm  40.3 $ &$ ...~~~~~~~~~ $  &  $0.176 \pm  0.008  $ &  $3.070  \pm 0.234$ &  $2.955 \pm  0.276 $ &  $ 2.464 \pm 0.332 $  &  $2.05  \pm 0.20  $ & 0.00 \\
$ 3.7 $ &  $ 12.1  \pm  44.7 $ &$ ...~~~~~~~~~ $  &  $0.162 \pm  0.017  $ &  $2.892  \pm 0.515$ &  $2.853 \pm  0.610 $ &  $ 2.423 \pm 0.714 $  &  $2.13  \pm 0.45  $ & 0.00 \\
$ 8.9 $ &  $-11.5  \pm  63.1 $ &$ ...~~~~~~~~~ $  &  $0.152 \pm  0.038  $ &  $2.595  \pm 1.150$ &  $2.291 \pm  1.365 $ &  $ 1.606 \pm 1.605 $  &  $1.74  \pm 1.01  $ & 0.00 \\
\noalign{\smallskip}
\hline
\noalign{\smallskip}
{\bf IC~5267} \\
\noalign{\smallskip}
\hline
\noalign{\smallskip}
$-13.9 $ &  $  29.5 \pm  31.3 $ & $211.2 \pm 33.8 $  &  $0.233 \pm 0.013  $ &  $ 3.849 \pm 0.515 $ &  $2.779 \pm  0.481 $ &  $2.501 \pm 0.619 $  &  $1.73  \pm 0.33  $ & $0.04$ \\
$ -8.3 $ &  $  13.6 \pm  11.3 $ & $197.7 \pm 16.3 $  &  $0.242 \pm 0.007  $ &  $ 4.025 \pm 0.273 $ &  $2.891 \pm  0.258 $ &  $2.714 \pm 0.334 $  &  $1.73  \pm 0.17  $ & $0.01$ \\
$ -4.9 $ &  $   2.6 \pm  11.5 $ & $208.5 \pm 15.1 $  &  $0.256 \pm 0.007  $ &  $ 4.333 \pm 0.255 $ &  $3.178 \pm  0.240 $ &  $2.715 \pm 0.315 $  &  $1.14  \pm 0.16  $ & $0.55$ \\	
$ -2.9 $ &  $  -8.2 \pm  10.6 $ & $214.4 \pm 13.6 $  &  $0.262 \pm 0.006  $ &  $ 4.299 \pm 0.245 $ &  $3.148 \pm  0.230 $ &  $2.840 \pm 0.303 $  &  $0.99  \pm 0.16  $ & $0.85$ \\
$ -1.7 $ &  $  -2.5 \pm   9.5 $ & $206.7 \pm 13.1 $  &  $0.284 \pm 0.006  $ &  $ 4.678 \pm 0.236 $ &  $3.112 \pm  0.222 $ &  $2.797 \pm 0.291 $  &  $1.07  \pm 0.15  $ & $0.85$ \\
$ -0.9 $ &  $  -1.0 \pm   9.7 $ & $202.0 \pm 14.7 $  &  $0.283 \pm 0.006  $ &  $ 4.617 \pm 0.237 $ &  $3.231 \pm  0.223 $ &  $3.097 \pm 0.290 $  &  $1.13  \pm 0.15  $ & $0.70$ \\
$ -0.3 $ &  $  -0.5 \pm   9.7 $ & $201.9 \pm 14.5 $  &  $0.283 \pm 0.006  $ &  $ 4.660 \pm 0.237 $ &  $3.182 \pm  0.223 $ &  $2.949 \pm 0.290 $  &  $1.07  \pm 0.15  $ & $0.47$ \\
$  0.2 $ &  $   2.8 \pm   9.6 $ & $208.5 \pm 12.6 $  &  $0.282 \pm 0.006  $ &  $ 4.743 \pm 0.236 $ &  $3.310 \pm  0.221 $ &  $3.014 \pm 0.290 $  &  $1.11  \pm 0.15  $ & $0.36$ \\
$  0.8 $ &  $   3.2 \pm  13.2 $ & $207.9 \pm 17.4 $  &  $0.280 \pm 0.008  $ &  $ 4.728 \pm 0.309 $ &  $3.049 \pm  0.291 $ &  $2.926 \pm 0.380 $  &  $1.06  \pm 0.20  $ & $0.06$ \\	
$  1.6 $ &  $   5.6 \pm  24.1 $ & $186.4 \pm 20.9 $  &  $0.269 \pm 0.010  $ &  $ 4.363 \pm 0.366 $ &  $3.055 \pm  0.344 $ &  $2.726 \pm 0.443 $  &  $1.08  \pm 0.24  $ & $0.43$ \\
$  2.7 $ &  $   1.1 \pm   8.9 $ & $217.3 \pm 10.2 $  &  $0.260 \pm 0.005  $ &  $ 4.408 \pm 0.210 $ &  $2.965 \pm  0.198 $ &  $2.782 \pm 0.262 $  &  $1.18  \pm 0.13  $ & $0.43$ \\
$  4.2 $ &  $   0.4 \pm   9.5 $ & $214.8 \pm 11.3 $  &  $0.250 \pm 0.006  $ &  $ 4.200 \pm 0.230 $ &  $2.974 \pm  0.217 $ &  $2.764 \pm 0.286 $  &  $1.31  \pm 0.15  $ & $0.46$ \\
$  6.5 $ &  $  -8.9 \pm  11.7 $ & $192.9 \pm 16.9 $  &  $0.245 \pm 0.007  $ &  $ 3.900 \pm 0.276 $ &  $2.875 \pm  0.260 $ &  $2.592 \pm 0.336 $  &  $1.52  \pm 0.18  $ & $0.01$ \\
$ 10.3 $ &  $ -15.3 \pm  53.0 $ & $156.7 \pm 33.1 $  &  $0.239 \pm 0.009  $ &  $ 3.881 \pm 0.342 $ &  $2.691 \pm  0.328 $ &  $2.424 \pm 0.407 $  &  $1.59  \pm 0.23  $ & $0.08$ \\
$ 15.2 $ &  $ -23.0 \pm  27.6 $ & $211.8 \pm 28.6 $  &  $0.229 \pm 0.013  $ &  $ 3.922 \pm 0.480 $ &  $3.133 \pm  0.458 $ &  $2.622 \pm 0.596 $  &  $1.48  \pm 0.31  $ & $0.02$ \\	
\noalign{\smallskip}
\hline
\noalign{\smallskip}
{\bf IC~5309} \\
\noalign{\smallskip}
\hline
\noalign{\smallskip}
$-6.5 $ &  $  74.1 \pm 38.4 $ & $107.2  \pm 27.9 $  &  $0.125 \pm 0.010  $ &  $ 2.225 \pm 0.363$ &  $1.811 \pm 0.431 $ &  $1.315 \pm 0.508 $  &  $ 0.09 \pm0.32  $ & $ 2.72 $ \\
$-3.3 $ &  $  39.1 \pm 39.5 $ & $111.1  \pm 27.1 $  &  $0.130 \pm 0.010  $ &  $ 1.884 \pm 0.362$ &  $2.233 \pm 0.432 $ &  $2.012 \pm 0.529 $  &  $-1.63 \pm0.33  $ & $ 3.99 $ \\
$-1.2 $ &  $  21.0 \pm 33.5 $ & $110.0  \pm 18.5 $  &  $0.124 \pm 0.006  $ &  $ 2.053 \pm 0.242$ &  $2.038 \pm 0.288 $ &  $1.984 \pm 0.347 $  &  $-0.46 \pm0.21  $ & $ 2.99 $ \\	
$ 0.4 $ &  $   7.9 \pm 34.0 $ & $106.6  \pm 14.1 $  &  $0.135 \pm 0.004  $ &  $ 2.445 \pm 0.164$ &  $2.061 \pm 0.197 $ &  $1.971 \pm 0.237 $  &  $ 0.45 \pm0.14  $ & $ 2.23 $ \\
$ 2.2 $ &  $  -7.1 \pm 38.7 $ & $106.9  \pm 21.3 $  &  $0.126 \pm 0.007  $ &  $ 2.262 \pm 0.274$ &  $1.996 \pm 0.324 $ &  $1.913 \pm 0.388 $  &  $-0.26 \pm0.24  $ & $ 2.74 $ \\
$ 5.0 $ &  $ -53.6 \pm 43.0 $ & $124.6  \pm 27.1 $  &  $0.121 \pm 0.010  $ &  $ 1.851 \pm 0.363$ &  $1.975 \pm 0.433 $ &  $1.271 \pm 0.539 $  &  $-0.93 \pm0.32  $ & $ 3.62 $ \\
$ 9.6 $ &  $ -81.2 \pm 46.3 $ & $116.8  \pm 38.8 $  &  $0.115 \pm 0.018  $ &  $ 2.185 \pm 0.692$ &  $2.169 \pm 0.819 $ &  $1.402 \pm 1.013 $  &  $-0.05 \pm0.61  $ & $ 2.72 $ \\
\noalign{\bigskip}
\hline
\label{tab:val_globtot}
\end{tabular}
\end{table*}
\begin{table*}
\contcaption{}
\begin{tabular}{rrrrrrrrr}
\noalign{\smallskip}
\hline
\multicolumn{1}{c}{(1)} &
\multicolumn{1}{c}{(2)} &
\multicolumn{1}{c}{(3)} &
\multicolumn{1}{c}{(4)} &
\multicolumn{1}{c}{(5)} &
\multicolumn{1}{c}{(6)} &
\multicolumn{1}{c}{(7)} &
\multicolumn{1}{c}{(8)} &
\multicolumn{1}{c}{(9)} \\
\noalign{\smallskip}
\hline
\noalign{\smallskip}
{\bf NGC~1292} \\
\noalign{\smallskip}
\hline
\noalign{\smallskip}
$-11.2 $ &  $ -41.4 \pm 24.5 $ &$ ...~~~~~~~~~ $ &  $0.085 \pm 0.011  $ &  $1.836  \pm 0.456$ &  $ 1.888 \pm  0.541 $ &  $1.568 \pm 0.635 $  &  $-0.79  \pm  0.40 $ & $ 3.70$ \\
$-5.4  $ &  $ -28.7 \pm 25.3 $ &$ ...~~~~~~~~~ $ &  $0.097 \pm 0.011  $ &  $1.824  \pm 0.464$ &  $ 1.519 \pm  0.560 $ &  $1.408 \pm 0.653 $  &  $-0.53  \pm  0.42 $ & $ 3.16$ \\
$-1.6  $ &  $   0.1 \pm 23.3 $ &$ ...~~~~~~~~~ $ &  $0.101 \pm 0.011  $ &  $1.734  \pm 0.447$ &  $ 2.019 \pm  0.540 $ &  $1.313 \pm 0.655 $  &  $-0.80  \pm  0.41 $ & $ 3.74$ \\
$ 1.2  $ &  $  10.3 \pm 23.9 $ &$ ...~~~~~~~~~ $ &  $0.103 \pm 0.011  $ &  $1.607  \pm 0.461$ &  $ 1.853 \pm  0.550 $ &  $1.416 \pm 0.673 $  &  $-0.29  \pm  0.41 $ & $ 3.01$ \\
$ 4.5  $ &  $  29.0 \pm 23.6 $ &$ ...~~~~~~~~~ $ &  $0.101 \pm 0.011  $ &  $2.009  \pm 0.460$ &  $ 1.911 \pm  0.533 $ &  $1.617 \pm 0.664 $  &  $ 0.09  \pm  0.40 $ & $ 2.82$ \\
$ 9.8  $ &  $  30.4 \pm 25.7 $ &$ ...~~~~~~~~~ $ &  $0.095 \pm 0.012  $ &  $1.922  \pm 0.478$ &  $ 1.712 \pm  0.554 $ &  $1.482 \pm 0.689 $  &  $-1.45  \pm  0.43 $ & $ 3.57$ \\
\noalign{\smallskip}
\hline
\noalign{\smallskip}
{\bf NGC~1351} \\
\noalign{\smallskip}
\hline
\noalign{\smallskip}
$-23.7 $ &  $   86.0 \pm 45.4 $ & $160.9 \pm 43.1 $  &  $0.183 \pm 0.012  $ &  $3.111  \pm 0.416$ &  $1.754 \pm 0.490 $ &  $ 2.012 \pm 0.626  $  &  $ 1.79 \pm 0.35  $ &  0.00 \\
$-14.9 $ &  $   93.9 \pm 24.0 $ & $166.8 \pm 15.3 $  &  $0.201 \pm 0.008  $ &  $3.531  \pm 0.273$ &  $2.374 \pm 0.326 $ &  $ 2.043 \pm 0.410  $  &  $ 1.73 \pm 0.22  $ &  0.00 \\
$-10.5 $ &  $   92.0 \pm 19.5 $ & $158.5 \pm 12.8 $  &  $0.202 \pm 0.005  $ &  $3.707  \pm 0.168$ &  $2.498 \pm 0.200 $ &  $ 1.892 \pm 0.251  $  &  $ 1.60 \pm 0.14  $ &  0.00 \\
$ -7.7 $ &  $   98.2 \pm 13.2 $ & $193.2 \pm 15.6 $  &  $0.214 \pm 0.007  $ &  $4.034  \pm 0.262$ &  $2.581 \pm 0.315 $ &  $ 2.485 \pm 0.402  $  &  $ 1.68 \pm 0.21  $ &  0.00 \\	
$ -5.9 $ &  $   77.6 \pm  7.4 $ & $177.5 \pm  9.2 $  &  $0.231 \pm 0.004  $ &  $4.182  \pm 0.149$ &  $2.377 \pm 0.174 $ &  $ 2.228 \pm 0.226  $  &  $ 1.85 \pm 0.12  $ &  0.00 \\
$ -4.5 $ &  $   75.9 \pm  7.7 $ & $184.9 \pm  9.2 $  &  $0.241 \pm 0.004  $ &  $3.987  \pm 0.163$ &  $2.454 \pm 0.190 $ &  $ 2.479 \pm 0.248  $  &  $ 1.80 \pm 0.13  $ &  0.00 \\
$ -3.4 $ &  $   62.3 \pm  6.4 $ & $188.3 \pm  9.0 $  &  $0.254 \pm 0.004  $ &  $4.396  \pm 0.155$ &  $2.736 \pm 0.184 $ &  $ 2.580 \pm 0.236  $  &  $ 1.63 \pm 0.13  $ &  0.00\\
$ -2.6 $ &  $   61.9 \pm  7.3 $ & $187.0 \pm  9.9 $  &  $0.260 \pm 0.005  $ &  $4.164  \pm 0.166$ &  $2.848 \pm 0.197 $ &  $ 2.828 \pm 0.253  $  &  $ 1.61 \pm 0.14  $ &  0.00\\
$ -2.0 $ &  $   57.2 \pm  8.7 $ & $195.5 \pm 11.1 $  &  $0.276 \pm 0.005  $ &  $4.560  \pm 0.185$ &  $3.000 \pm 0.219 $ &  $ 2.435 \pm 0.284  $  &  $ 1.63 \pm 0.15  $ &  0.00\\
$ -1.5 $ &  $   59.3 \pm 11.1 $ & $208.2 \pm 12.6 $  &  $0.286 \pm 0.006  $ &  $4.879  \pm 0.207$ &  $3.196 \pm 0.245 $ &  $ 2.955 \pm 0.320  $  &  $ 1.73 \pm 0.17  $ &  0.00\\	
$ -1.2 $ &  $   51.5 \pm  9.8 $ & $163.6 \pm  6.6 $  &  $0.293 \pm 0.002  $ &  $4.679  \pm 0.068$ &  $2.913 \pm 0.080 $ &  $ 3.095 \pm 0.101  $  &  $ 1.55 \pm 0.05  $ &  0.00\\
$ -0.9 $ &  $   41.9 \pm  9.8 $ & $211.0 \pm 11.4 $  &  $0.292 \pm 0.005  $ &  $4.958  \pm 0.182$ &  $2.879 \pm 0.219 $ &  $ 2.865 \pm 0.283  $  &  $ 1.73 \pm 0.15  $ &  0.00\\
$ -0.6 $ &  $   42.2 \pm 10.5 $ & $212.3 \pm 12.0 $  &  $0.300 \pm 0.005  $ &  $5.114  \pm 0.193$ &  $3.209 \pm 0.231 $ &  $ 2.850 \pm 0.301  $  &  $ 1.72 \pm 0.15  $ &  0.00 \\
$ -0.3 $ &  $   19.0 \pm  8.9 $ & $209.3 \pm 10.6 $  &  $0.302 \pm 0.005  $ &  $5.197  \pm 0.167$ &  $3.084 \pm 0.200 $ &  $ 2.890 \pm 0.250  $  &  $ 1.45 \pm 0.13  $ &  0.00 \\
$  0.0 $ &  $   -2.7 \pm 10.7 $ & $222.3 \pm 10.6 $  &  $0.297 \pm 0.005  $ &  $5.001  \pm 0.203$ &  $3.409 \pm 0.243 $ &  $ 2.942 \pm 0.310  $  &  $ 1.55 \pm 0.16  $ &  0.00 \\
$  0.3 $ &  $  -26.5 \pm 10.8 $ & $211.5 \pm 12.5 $  &  $0.297 \pm 0.005  $ &  $4.775  \pm 0.201$ &  $2.827 \pm 0.234 $ &  $ 2.745 \pm 0.313  $  &  $ 1.50 \pm 0.16  $ &  0.00 \\	
$  0.6 $ &  $  -24.9 \pm  8.5 $ & $198.7 \pm 13.3 $  &  $0.280 \pm 0.005  $ &  $4.790  \pm 0.191$ &  $2.933 \pm 0.222 $ &  $ 2.799 \pm 0.293  $  &  $ 1.69 \pm 0.15  $ &  0.00 \\
$  0.9 $ &  $  -34.9 \pm 14.6 $ & $176.1 \pm 16.2 $  &  $0.282 \pm 0.005  $ &  $4.671  \pm 0.189$ &  $2.679 \pm 0.224 $ &  $ 2.576 \pm 0.285  $  &  $ 1.46 \pm 0.16  $ &  0.00 \\
$  1.4 $ &  $  -56.2 \pm  9.5 $ & $201.6 \pm 12.4 $  &  $0.274 \pm 0.005  $ &  $4.632  \pm 0.182$ &  $2.868 \pm 0.218 $ &  $ 2.670 \pm 0.280  $  &  $ 1.69 \pm 0.15  $ &  0.00 \\
$  2.0 $ &  $  -63.3 \pm  8.1 $ & $190.2 \pm 11.4 $  &  $0.259 \pm 0.005  $ &  $4.227  \pm 0.190$ &  $2.745 \pm 0.227 $ &  $ 2.722 \pm 0.289  $  &  $ 1.60 \pm 0.15  $ &  0.00 \\
$  2.8 $ &  $  -77.6 \pm  8.5 $ & $194.9 \pm 13.8 $  &  $0.252 \pm 0.005  $ &  $4.311  \pm 0.195$ &  $2.857 \pm 0.234 $ &  $ 2.449 \pm 0.285  $  &  $ 1.66 \pm 0.16  $ &  0.00 \\
$  3.8 $ &  $  -89.4 \pm  9.2 $ & $185.8 \pm 11.4 $  &  $0.238 \pm 0.005  $ &  $4.142  \pm 0.194$ &  $3.026 \pm 0.232 $ &  $ 2.354 \pm 0.286  $  &  $ 1.59 \pm 0.16  $ &  0.00 \\	
$  5.2 $ &  $ -102.9 \pm  9.6 $ & $169.3 \pm  8.4 $  &  $0.224 \pm 0.004  $ &  $4.022  \pm 0.152$ &  $2.652 \pm 0.181 $ &  $ 2.619 \pm 0.220  $  &  $ 1.73 \pm 0.12  $ &  0.00 \\
$  7.1 $ &  $ -106.1 \pm 10.5 $ & $166.1 \pm  9.6 $  &  $0.210 \pm 0.004  $ &  $3.770  \pm 0.161$ &  $2.597 \pm 0.193 $ &  $ 2.454 \pm 0.233  $  &  $ 1.61 \pm 0.13  $ &  0.00 \\
$  9.9 $ &  $  -98.5 \pm 13.0 $ & $169.3 \pm 12.2 $  &  $0.200 \pm 0.006  $ &  $3.351  \pm 0.203$ &  $2.299 \pm 0.244 $ &  $ 1.898 \pm 0.297  $  &  $ 1.42 \pm 0.17  $ &  0.00 \\
$ 14.4 $ &  $ -105.7 \pm 25.1 $ & $157.2 \pm 14.6 $  &  $0.193 \pm 0.006  $ &  $3.275  \pm 0.209$ &  $2.098 \pm 0.251 $ &  $ 1.860 \pm 0.302  $  &  $ 1.46 \pm 0.17  $ &  0.00 \\
$ 23.6 $ &  $ -111.0 \pm 34.7 $ & $145.8 \pm 16.1 $  &  $0.179 \pm 0.007  $ &  $3.151  \pm 0.258$ &  $2.142 \pm 0.307 $ &  $ 2.078 \pm 0.377  $  &  $ 0.96 \pm 0.22  $ &  0.00 \\
\noalign{\smallskip}
\hline
\noalign{\smallskip}
{\bf NGC~1366} \\
\noalign{\smallskip}
\hline
\noalign{\smallskip}
$ -27.6 $ &  $ 50.1 \pm 41.3  $ & $141.6 \pm 29.4 $  &  $0.175 \pm 0.014  $ &  $2.588  \pm 0.492$ &  $2.320 \pm 0.566 $ &  $1.487 \pm  0.728 $  &  $1.50  \pm 0.42 $ & 0.00 \\
$ -16.5 $ &  $  6.4 \pm 17.6  $ & $198.4 \pm 22.2 $  &  $0.183 \pm 0.011  $ &  $2.862  \pm 0.407$ &  $2.380 \pm 0.490 $ &  $2.354 \pm  0.627 $  &  $1.69  \pm 0.33 $ & 0.00 \\
$  -9.5 $ &  $ -7.0 \pm 10.0  $ & $186.9 \pm 15.0 $  &  $0.200 \pm 0.007  $ &  $3.338  \pm 0.267$ &  $2.657 \pm 0.319 $ &  $2.248 \pm  0.394 $  &  $1.81  \pm 0.22 $ & 0.00 \\
$  -5.3 $ &  $-10.4 \pm  7.7  $ & $170.8 \pm 11.4 $  &  $0.204 \pm 0.006  $ &  $3.332  \pm 0.214$ &  $2.852 \pm 0.254 $ &  $2.677 \pm  0.309 $  &  $1.81  \pm 0.17 $ & 0.00 \\	
$  -3.0 $ &  $-18.1 \pm 10.5  $ & $168.2 \pm  9.2 $  &  $0.216 \pm 0.005  $ &  $3.754  \pm 0.191$ &  $2.924 \pm 0.227 $ &  $2.647 \pm  0.276 $  &  $1.68  \pm 0.15 $ & 0.00 \\
$  -1.8 $ &  $-20.7 \pm  7.4  $ & $167.6 \pm  9.1 $  &  $0.235 \pm 0.005  $ &  $3.875  \pm 0.185$ &  $3.076 \pm 0.219 $ &  $2.721 \pm  0.267 $  &  $1.89  \pm 0.15 $ & 0.00 \\
$  -1.0 $ &  $-17.7 \pm  5.8  $ & $167.0 \pm  8.1 $  &  $0.249 \pm 0.004  $ &  $4.135  \pm 0.162$ &  $3.184 \pm 0.191 $ &  $2.674 \pm  0.232 $  &  $2.02  \pm 0.13 $ & 0.00 \\
$  -0.6 $ &  $ -7.5 \pm  9.0  $ & $188.3 \pm 12.5 $  &  $0.258 \pm 0.006  $ &  $3.979  \pm 0.214$ &  $3.325 \pm 0.254 $ &  $2.971 \pm  0.313 $  &  $1.87  \pm 0.17 $ & 0.00 \\
$  -0.3 $ &  $ -6.1 \pm  7.7  $ & $186.6 \pm 11.1 $  &  $0.267 \pm 0.005  $ &  $4.402  \pm 0.195$ &  $3.404 \pm 0.231 $ &  $3.129 \pm  0.284 $  &  $1.65  \pm 0.16 $ & 0.00 \\
$  -0.0 $ &  $  4.5 \pm  8.0  $ & $195.7 \pm 11.3 $  &  $0.264 \pm 0.005  $ &  $4.326  \pm 0.193$ &  $3.265 \pm 0.229 $ &  $3.070 \pm  0.293 $  &  $1.99  \pm 0.15 $ & 0.00 \\	
$   0.3 $ &  $  9.4 \pm  9.0  $ & $202.7 \pm 11.9 $  &  $0.261 \pm 0.006  $ &  $4.205  \pm 0.208$ &  $3.532 \pm 0.246 $ &  $3.143 \pm  0.318 $  &  $1.90  \pm 0.16 $ & 0.00 \\
$   0.6 $ &  $ 14.3 \pm  4.2  $ & $156.6 \pm  6.6 $  &  $0.263 \pm 0.002  $ &  $4.218  \pm 0.054$ &  $3.107 \pm 0.063 $ &  $2.937 \pm  0.079 $  &  $1.71  \pm 0.04 $ & 0.00 \\
$   0.9 $ &  $ 22.3 \pm  8.3  $ & $195.9 \pm 11.9 $  &  $0.257 \pm 0.005  $ &  $4.255  \pm 0.202$ &  $3.207 \pm 0.237 $ &  $3.077 \pm  0.307 $  &  $1.71  \pm 0.16 $ & 0.00 \\
$   1.3 $ &  $ 24.3 \pm  6.7  $ & $179.7 \pm 10.8 $  &  $0.249 \pm 0.005  $ &  $4.087  \pm 0.189$ &  $3.208 \pm 0.221 $ &  $2.950 \pm  0.283 $  &  $1.82  \pm 0.15 $ & 0.00 \\
$   2.1 $ &  $ 15.6 \pm  9.3  $ & $171.1 \pm 11.7 $  &  $0.238 \pm 0.006  $ &  $3.814  \pm 0.217$ &  $2.940 \pm 0.257 $ &  $2.758 \pm  0.324 $  &  $1.79  \pm 0.18 $ & 0.00 \\
$   3.3 $ &  $ 13.3 \pm 12.4  $ & $170.7 \pm 14.0 $  &  $0.221 \pm 0.006  $ &  $3.689  \pm 0.231$ &  $2.870 \pm 0.274 $ &  $2.879 \pm  0.343 $  &  $1.50  \pm 0.19 $ & 0.00 \\	
$   5.6 $ &  $ -5.4 \pm  8.3  $ & $164.9 \pm 10.5 $  &  $0.207 \pm 0.006  $ &  $3.572  \pm 0.230$ &  $2.602 \pm 0.274 $ &  $2.393 \pm  0.332 $  &  $1.41  \pm 0.19 $ & 0.00 \\
$   9.8 $ &  $ -2.4 \pm 11.6  $ & $181.6 \pm 22.1 $  &  $0.199 \pm 0.008  $ &  $3.475  \pm 0.294$ &  $2.345 \pm 0.352 $ &  $2.401 \pm  0.434 $  &  $1.69  \pm 0.24 $ & 0.00 \\
$  16.8 $ &  $ -3.9 \pm  7.9  $ & $177.2 \pm 13.2 $  &  $0.192 \pm 0.006  $ &  $3.306  \pm 0.222$ &  $2.462 \pm 0.265 $ &  $2.136 \pm  0.328 $  &  $1.68  \pm 0.18 $ & 0.00 \\
$  27.9 $ &  $-60.8 \pm 48.5  $ & $173.8 \pm 35.8 $  &  $0.180 \pm 0.015  $ &  $3.244  \pm 0.535$ &  $1.961 \pm 0.639 $ &  $2.288 \pm  0.810 $  &  $1.09  \pm 0.45 $ & 0.00 \\
\noalign{\bigskip}
\hline
\end{tabular}
\end{table*}
\begin{table*}
\contcaption{}
\begin{tabular}{rrrrrrrrr}
\noalign{\smallskip}
\hline
\multicolumn{1}{c}{(1)} &
\multicolumn{1}{c}{(2)} &
\multicolumn{1}{c}{(3)} &
\multicolumn{1}{c}{(4)} &
\multicolumn{1}{c}{(5)} &
\multicolumn{1}{c}{(6)} &
\multicolumn{1}{c}{(7)} &
\multicolumn{1}{c}{(8)} &
\multicolumn{1}{c}{(9)} \\
\noalign{\smallskip}
\hline
\noalign{\smallskip}
{\bf NGC~1425} \\
\noalign{\smallskip}
\hline
\noalign{\smallskip}
$-18.9 $ &  $80.8  \pm 37.5 $ & $119.6 \pm 17.2 $  &  $ 0.162 \pm  0.008  $ &  $2.864  \pm  0.338$ &  $2.108 \pm 0.403 $ &  $1.332 \pm 0.495 $  &  $ 2.55 \pm 0.28 $ & 0.00  \\
$-15.4 $ &  $79.1  \pm 36.4 $ & $125.6 \pm 17.5 $  &  $ 0.170 \pm  0.008  $ &  $2.809  \pm  0.318$ &  $2.410 \pm 0.378 $ &  $2.233 \pm 0.461 $  &  $ 2.00 \pm 0.27 $ & 0.00  \\
$-10.8 $ &  $61.8  \pm 39.1 $ & $123.0 \pm 12.7 $  &  $ 0.169 \pm  0.005  $ &  $2.923  \pm  0.201$ &  $2.151 \pm 0.240 $ &  $2.376 \pm 0.289 $  &  $ 2.20 \pm 0.17 $ & 0.00  \\	
$ -7.4 $ &  $56.8  \pm 24.2 $ & $123.5 \pm 11.8 $  &  $ 0.175 \pm  0.004  $ &  $3.248  \pm  0.154$ &  $2.392 \pm 0.185 $ &  $2.299 \pm 0.224 $  &  $ 1.98 \pm 0.13 $ & 0.00  \\
$ -5.1 $ &  $46.1  \pm 29.8 $ & $124.5 \pm 12.7 $  &  $ 0.191 \pm  0.004  $ &  $3.298  \pm  0.175$ &  $2.451 \pm 0.205 $ &  $2.089 \pm 0.249 $  &  $ 1.98 \pm 0.15 $ & 0.00  \\
$ -3.5 $ &  $38.5  \pm 29.8 $ & $126.0 \pm 13.0 $  &  $ 0.196 \pm  0.004  $ &  $3.242  \pm  0.176$ &  $2.530 \pm 0.210 $ &  $2.278 \pm 0.255 $  &  $ 1.51 \pm 0.15 $ & 0.00  \\
$ -2.4 $ &  $34.3  \pm 19.6 $ & $132.8 \pm 12.0 $  &  $ 0.204 \pm  0.004  $ &  $3.376  \pm  0.146$ &  $2.840 \pm 0.174 $ &  $2.297 \pm 0.213 $  &  $ 1.72 \pm 0.13 $ & 0.00  \\
$ -1.6 $ &  $33.4  \pm 23.6 $ & $134.3 \pm 12.3 $  &  $ 0.212 \pm  0.004  $ &  $3.516  \pm  0.163$ &  $2.571 \pm 0.195 $ &  $2.364 \pm 0.239 $  &  $ 1.73 \pm 0.14 $ & 0.00  \\
$ -1.0 $ &  $29.6  \pm 22.4 $ & $129.7 \pm 11.2 $  &  $ 0.224 \pm  0.004  $ &  $3.461  \pm  0.144$ &  $2.796 \pm 0.174 $ &  $2.300 \pm 0.210 $  &  $ 1.87 \pm 0.12 $ & 0.00  \\	
$ -0.6 $ &  $20.8  \pm 29.9 $ & $123.7 \pm 13.1 $  &  $ 0.227 \pm  0.004  $ &  $3.534  \pm  0.173$ &  $2.853 \pm 0.209 $ &  $2.456 \pm 0.250 $  &  $ 1.74 \pm 0.15 $ & 0.00  \\
$ -0.2 $ &  $18.1  \pm 24.8 $ & $128.9 \pm 13.3 $  &  $ 0.228 \pm  0.004  $ &  $3.671  \pm  0.163$ &  $2.977 \pm 0.198 $ &  $2.709 \pm 0.245 $  &  $ 1.81 \pm 0.14 $ & 0.00  \\
$  0.0 $ &  $11.7  \pm 25.7 $ & $128.3 \pm 12.9 $  &  $ 0.229 \pm  0.004  $ &  $3.763  \pm  0.171$ &  $3.005 \pm 0.203 $ &  $2.543 \pm 0.252 $  &  $ 1.64 \pm 0.15 $ & 0.00  \\
$  0.3 $ &  $2.1   \pm 21.7 $ & $124.0 \pm 12.0 $  &  $ 0.224 \pm  0.004  $ &  $3.636  \pm  0.153$ &  $2.961 \pm 0.182 $ &  $2.583 \pm 0.225 $  &  $ 1.79 \pm 0.13 $ & 0.00  \\
$  0.6 $ &  $1.9   \pm 26.9 $ & $119.8 \pm 10.9 $  &  $ 0.218 \pm  0.004  $ &  $3.702  \pm  0.142$ &  $2.931 \pm 0.167 $ &  $2.536 \pm 0.206 $  &  $ 1.77 \pm 0.12 $ & 0.00  \\
$  1.0 $ &  $24.1  \pm 23.9 $ & $118.9 \pm 10.9 $  &  $ 0.212 \pm  0.004  $ &  $3.625  \pm  0.148$ &  $2.857 \pm 0.176 $ &  $2.461 \pm 0.215 $  &  $ 1.60 \pm 0.12 $ & 0.00  \\	
$  1.7 $ &  $37.8  \pm 31.2 $ & $121.6 \pm 12.7 $  &  $ 0.205 \pm  0.004  $ &  $3.529  \pm  0.173$ &  $2.556 \pm 0.207 $ &  $2.199 \pm 0.245 $  &  $ 1.67 \pm 0.15 $ & 0.00  \\
$  2.5 $ &  $43.1  \pm 12.3 $ & $131.9 \pm 13.9 $  &  $ 0.199 \pm  0.004  $ &  $3.390  \pm  0.180$ &  $2.361 \pm 0.216 $ &  $2.335 \pm 0.263 $  &  $ 1.74 \pm 0.15 $ & 0.00  \\
$  3.7 $ &  $54.8  \pm 24.2 $ & $131.8 \pm 14.1 $  &  $ 0.195 \pm  0.004  $ &  $3.243  \pm  0.180$ &  $2.647 \pm 0.209 $ &  $2.208 \pm 0.256 $  &  $ 1.67 \pm 0.15 $ & 0.00  \\
$  5.6 $ &  $67.1  \pm 36.2 $ & $117.2 \pm 11.9 $  &  $ 0.184 \pm  0.004  $ &  $3.041  \pm  0.152$ &  $2.437 \pm 0.180 $ &  $2.091 \pm 0.218 $  &  $ 1.75 \pm 0.13 $ & 0.00  \\
$  8.1 $ &  $76.2  \pm 43.3 $ & $110.8 \pm 19.9 $  &  $ 0.187 \pm  0.007  $ &  $3.150  \pm  0.274$ &  $2.382 \pm 0.331 $ &  $2.028 \pm 0.393 $  &  $ 1.87 \pm 0.24 $ & 0.00  \\
$ 11.5 $ &  $99.5  \pm 54.4 $ & $ 98.7 \pm 28.4 $  &  $ 0.172 \pm  0.012  $ &  $2.807  \pm  0.471$ &  $2.471 \pm 0.555 $ &  $2.064 \pm 0.677 $  &  $ 2.05 \pm 0.40 $ & 0.00  \\
$ 16.1 $ &  $108.0 \pm 38.3 $ & $107.6 \pm 22.1 $  &  $ 0.163 \pm  0.006  $ &  $2.748  \pm  0.241$ &  $2.433 \pm 0.282 $ &  $1.981 \pm 0.346 $  &  $ 2.08 \pm 0.20 $ & 0.00  \\
\noalign{\smallskip}
\hline
\noalign{\smallskip}
{\bf NGC~7515} \\
\noalign{\smallskip}
\hline
\noalign{\smallskip}
$ -14.188 $ &  $ 129.0  \pm 51.4 $ & $169.0 \pm 61.9 $  &  $0.126 \pm 0.023  $ &  $2.325  \pm 0.917$ &  $2.284 \pm 1.013 $ &  $2.231 \pm 1.270 $  &  $-1.61  \pm 0.74 $ & $5.08$ \\
$  -8.137 $ &  $  58.3  \pm 42.8 $ & $164.6 \pm 33.6 $  &  $0.171 \pm 0.015  $ &  $3.182  \pm 0.596$ &  $2.275 \pm 0.639 $ &  $2.124 \pm 0.827 $  &  $ 0.35  \pm 0.47 $ & $1.67$ \\
$  -3.742 $ &  $  33.0  \pm 23.3 $ & $179.7 \pm 20.7 $  &  $0.189 \pm 0.008  $ &  $2.982  \pm 0.311$ &  $2.784 \pm 0.339 $ &  $2.541 \pm 0.435 $  &  $ 1.79  \pm 0.23 $ & $0.26$ \\	
$  -1.821 $ &  $  32.5  \pm 26.8 $ & $145.0 \pm 15.8 $  &  $0.205 \pm 0.005  $ &  $3.459  \pm 0.201$ &  $2.842 \pm 0.218 $ &  $2.497 \pm 0.274 $  &  $ 1.80  \pm 0.15 $ & $0.32$ \\
$  -0.933 $ &  $  -0.1  \pm 22.8 $ & $155.8 \pm 15.4 $  &  $0.214 \pm 0.005  $ &  $3.649  \pm 0.215$ &  $2.778 \pm 0.237 $ &  $2.714 \pm 0.285 $  &  $ 2.12  \pm 0.16 $ & $0.21$ \\
$  -0.314 $ &  $   2.4  \pm 21.8 $ & $166.8 \pm 13.7 $  &  $0.220 \pm 0.005  $ &  $3.591  \pm 0.201$ &  $2.954 \pm 0.221 $ &  $2.883 \pm 0.269 $  &  $ 1.99  \pm 0.15 $ & $0.00$ \\
$   0.292 $ &  $  -4.0  \pm 21.9 $ & $150.8 \pm 14.9 $  &  $0.216 \pm 0.005  $ &  $3.676  \pm 0.202$ &  $2.849 \pm 0.222 $ &  $2.494 \pm 0.268 $  &  $ 2.03  \pm 0.15 $ & $0.07$ \\
$   0.909 $ &  $ -21.1  \pm 29.0 $ & $158.8 \pm 17.7 $  &  $0.207 \pm 0.007  $ &  $3.634  \pm 0.287$ &  $2.872 \pm 0.315 $ &  $2.804 \pm 0.393 $  &  $ 1.98  \pm 0.22 $ & $0.10$ \\
$   1.794 $ &  $ -31.9  \pm 23.8 $ & $172.2 \pm 17.4 $  &  $0.188 \pm 0.007  $ &  $3.324  \pm 0.260$ &  $2.754 \pm 0.286 $ &  $2.346 \pm 0.363 $  &  $ 1.92  \pm 0.20 $ & $0.02$ \\	
$   3.719 $ &  $ -26.5  \pm 31.5 $ & $156.0 \pm 18.3 $  &  $0.192 \pm 0.008  $ &  $3.279  \pm 0.323$ &  $2.741 \pm 0.355 $ &  $2.381 \pm 0.445 $  &  $ 1.79  \pm 0.25 $ & $0.00$ \\
$   7.926 $ &  $ -68.7  \pm 37.8 $ & $178.3 \pm 35.0 $  &  $0.178 \pm 0.015  $ &  $3.335  \pm 0.602$ &  $2.489 \pm 0.657 $ &  $2.348 \pm 0.843 $  &  $ 1.48  \pm 0.46 $ & $0.86$ \\
$  14.060 $ &  $-102.9  \pm 57.4 $ & $165.8 \pm 62.6 $  &  $0.136 \pm 0.024  $ &  $2.045  \pm 0.960$ &  $2.076 \pm 1.060 $ &  $1.193 \pm 1.301 $  &  $ 0.91  \pm 0.75 $ & $2.06$ \\
\noalign{\smallskip}
\hline
\noalign{\smallskip}
{\bf NGC~7531} \\
\noalign{\smallskip}
\hline
\noalign{\smallskip}
$-17.3 $ &  $   97.6  \pm  35.5 $ & $130.5 \pm  22.2 $  &  $ 0.204 \pm 0.009  $ &  $3.792  \pm 0.355 $ &  $1.497 \pm 0.401 $ &  $1.764 \pm 0.477 $  &  $ 1.83  \pm 0.28 $ & $0.04$ \\
$-11.9 $ &  $   58.8 \pm  31.5 $ & $127.7 \pm  15.6 $  &  $ 0.182 \pm 0.006  $ &  $3.193  \pm 0.198 $ &  $2.589 \pm 0.215 $ &  $2.448 \pm 0.272 $  &  $  2.01  \pm 0.15 $ & $0.03$ \\
$ -5.5 $ &  $   35.5 \pm  19.4 $ & $132.2 \pm  13.0 $  &  $ 0.200 \pm 0.003  $ &  $3.272  \pm 0.198 $ &  $2.643 \pm 0.218 $ &  $2.416 \pm 0.272 $  &  $  1.64  \pm 0.15 $ & $0.00$ \\
$ -2.5 $ &  $   20.6 \pm   8.7 $ & $133.4 \pm   8.5 $  &  $ 0.211 \pm 0.004  $ &  $3.485  \pm 0.159 $ &  $2.725 \pm 0.177 $ &  $2.481 \pm 0.219 $  &  $  1.79  \pm 0.12 $ & $0.00$ \\
$ -0.8 $ &  $   28.2 \pm   8.3 $ & $135.1 \pm   8.0 $  &  $ 0.195 \pm 0.004  $ &  $3.033  \pm 0.143 $ &  $2.841 \pm 0.159 $ &  $3.077 \pm 0.196 $  &  $  1.94  \pm 0.11 $ & $0.02$ \\
$  0.1 $ &  $   20.0 \pm  16.1 $ & $131.7 \pm   9.7 $  &  $ 0.196 \pm 0.004  $ &  $3.193  \pm 0.132 $ &  $2.686 \pm 0.147 $ &  $2.700 \pm 0.181 $  &  $  2.26  \pm 0.10 $ & $0.03$ \\
$  1.2 $ &  $   -1.5 \pm  16.0 $ & $131.6 \pm  12.2 $  &  $ 0.205 \pm 0.004  $ &  $3.239  \pm 0.148 $ &  $2.625 \pm 0.165 $ &  $2.576 \pm 0.197 $  &  $  1.89  \pm 0.12 $ & $0.01$ \\
$  3.0 $ &  $  -28.9 \pm  17.1 $ & $135.8 \pm  10.9 $  &  $ 0.204 \pm 0.004  $ &  $3.391  \pm 0.159 $ &  $2.723 \pm 0.177 $ &  $2.516 \pm 0.220 $  &  $  1.45  \pm 0.13 $ & $0.01$ \\
$  6.3 $ &  $  -53.3 \pm  26.2 $ & $127.3 \pm  10.7 $  &  $ 0.205 \pm 0.004  $ &  $3.170  \pm 0.197 $ &  $2.397 \pm 0.213 $ &  $2.384 \pm 0.270 $  &  $  1.80  \pm 0.15 $ & $0.01$ \\
$ 12.2 $ &  $  -76.7 \pm  23.4 $ & $114.1 \pm   9.0 $  &  $ 0.183 \pm 0.002  $ &  $2.918  \pm 0.197 $ &  $2.444 \pm 0.217 $ &  $1.937 \pm 0.269 $  &  $  1.79  \pm 0.15 $ & $0.35$ \\
$ 16.9 $ &  $ -100.8 \pm  42.3 $ & $116.3 \pm  28.2 $  &  $ 0.175 \pm 0.011  $ &  $2.947  \pm 0.444 $ &  $2.116 \pm 0.497 $ &  $1.504 \pm 0.590 $  &  $ -1.37  \pm 0.38 $ & $3.84$ \\
\noalign{\smallskip}
\hline
\noalign{\smallskip}
{\bf NGC~7557} \\
\noalign{\smallskip}
\hline
\noalign{\smallskip}
$-6.5 $ &  $-24.9 \pm 47.0  $ & $ 98.1 \pm  32.1 $  &  $0.176 \pm 0.012  $ &  $2.791  \pm0.473 $ &  $2.701 \pm 0.556 $ &  $2.335 \pm 0.673 $  &  $1.81  \pm 0.41 $ & 0.00 \\
$-2.2 $ &  $ -5.0 \pm 20.6  $ & $104.9 \pm  19.4 $  &  $0.177 \pm 0.004  $ &  $3.106  \pm0.162 $ &  $2.172 \pm 0.192 $ &  $2.010 \pm 0.232 $  &  $2.35  \pm 0.14 $ & 0.00 \\
$-0.4 $ &  $ 11.8 \pm 20.7  $ & $103.6 \pm  15.0 $  &  $0.179 \pm 0.004  $ &  $2.954  \pm0.157 $ &  $2.675 \pm 0.187 $ &  $2.349 \pm 0.226 $  &  $2.73  \pm 0.13 $ & 0.00 \\
$ 0.3 $ &  $ 19.1 \pm 17.1  $ & $105.6 \pm  15.0 $  &  $0.172 \pm 0.004  $ &  $2.863  \pm0.157 $ &  $2.830 \pm 0.187 $ &  $2.458 \pm 0.226 $  &  $2.88  \pm 0.13 $ & 0.00 \\
$ 1.7 $ &  $  4.3 \pm 26.1  $ & $104.0 \pm  19.8 $  &  $0.176 \pm 0.004  $ &  $3.239  \pm0.163 $ &  $2.433 \pm 0.193 $ &  $2.395 \pm 0.232 $  &  $2.29  \pm 0.14 $ & 0.00 \\
$ 5.7 $ &  $ -5.3 \pm 55.5  $ & $118.4 \pm  37.5 $  &  $0.161 \pm 0.013  $ &  $2.724  \pm0.539 $ &  $2.364 \pm 0.633 $ &  $1.820 \pm 0.777 $  &  $2.29  \pm 0.46 $ & 0.00 \\
\noalign{\bigskip}
\hline
\end{tabular}
\end{table*}
\begin{table*}
\contcaption{}
\begin{tabular}{rrrrrrrrr}
\noalign{\smallskip}
\hline
\multicolumn{1}{c}{(1)} &
\multicolumn{1}{c}{(2)} &
\multicolumn{1}{c}{(3)} &
\multicolumn{1}{c}{(4)} &
\multicolumn{1}{c}{(5)} &
\multicolumn{1}{c}{(6)} &
\multicolumn{1}{c}{(7)} &
\multicolumn{1}{c}{(8)} &
\multicolumn{1}{c}{(9)} \\
\noalign{\smallskip}
\hline
\noalign{\smallskip}
{\bf NGC~7631} \\
\noalign{\smallskip}
\hline
\noalign{\smallskip}
$-9.226 $ &  $-143.3 \pm 47.2 $ & $113.1 \pm 24.6 $  &  $0.163 \pm 0.010  $ &  $2.509  \pm 0.334$ &  $2.596 \pm 0.389 $ &  $1.902 \pm 0.461 $  &  $1.06  \pm 0.28 $ &$0.37 $ \\
$-5.616 $ &  $ -53.1 \pm 20.7 $ & $139.5 \pm 14.8 $  &  $0.180 \pm 0.009  $ &  $3.522  \pm 0.304$ &  $2.792 \pm 0.356 $ &  $2.373 \pm 0.442 $  &  $1.67  \pm 0.25 $ &$0.31 $ \\
$-2.481 $ &  $ -32.1 \pm 23.0 $ & $138.6 \pm 15.0 $  &  $0.175 \pm 0.008  $ &  $3.177  \pm 0.286$ &  $2.539 \pm 0.337 $ &  $2.124 \pm 0.417 $  &  $2.12  \pm 0.24 $ &$0.14 $ \\	
$-0.837 $ &  $ -24.2 \pm 23.0 $ & $167.7 \pm 15.0 $  &  $0.164 \pm 0.008  $ &  $2.868  \pm 0.286$ &  $2.820 \pm 0.337 $ &  $2.363 \pm 0.417 $  &  $0.48  \pm 0.24 $ &$1.48 $ \\
$ 1.959 $ &  $  57.0 \pm 23.3 $ & $142.4 \pm 16.7 $  &  $0.176 \pm 0.008  $ &  $3.561  \pm 0.269$ &  $2.291 \pm 0.318 $ &  $2.294 \pm 0.393 $  &  $0.78  \pm 0.23 $ &$1.11 $ \\
$ 5.142 $ &  $  74.6 \pm 22.2 $ & $159.1 \pm 12.8 $  &  $0.186 \pm 0.010  $ &  $2.929  \pm 0.336$ &  $3.139 \pm 0.394 $ &  $2.547 \pm 0.481 $  &  $1.34  \pm 0.28 $ &$0.25 $ \\
$ 8.875 $ &  $ 121.2 \pm 37.1 $ & $155.7 \pm 29.7 $  &  $0.190 \pm 0.020  $ &  $3.672  \pm 0.692$ &  $3.832 \pm 0.822 $ &  $2.705 \pm 1.046 $  &  $1.88  \pm 0.60 $ &$0.28 $ \\
\noalign{\smallskip}
\hline
\noalign{\smallskip}
{\bf NGC~7643} \\
\noalign{\smallskip}
\hline
\noalign{\smallskip}
$-14.3 $ &  $-129.4  \pm 63.1 $ & $130.1 \pm 65.6 $  &  $0.165 \pm 0.019  $ &  $ 2.902 \pm 0.792$ &  $3.195 \pm 0.938 $ &  $3.822 \pm 1.093 $  &  $-0.71  \pm 0.71 $ &$2.74$  \\
$ -9.0 $ &  $ -86.5  \pm 47.0 $ & $101.6 \pm 34.4 $  &  $0.175 \pm 0.008  $ &  $ 2.953 \pm 0.315$ &  $2.810 \pm 0.375 $ &  $2.243 \pm 0.452 $  &  $ 1.40  \pm 0.27 $ &$1.29$  \\
$ -2.9 $ &  $ -27.0  \pm 37.4 $ & $116.5 \pm 16.9 $  &  $0.183 \pm 0.006  $ &  $ 3.334 \pm 0.233$ &  $2.200 \pm 0.273 $ &  $2.157 \pm 0.328 $  &  $ 1.56  \pm 0.20 $ &$1.09$  \\	
$ -0.7 $ &  $  -4.7  \pm 28.0 $ & $111.3 \pm 13.0 $  &  $0.145 \pm 0.005  $ &  $ 2.759 \pm 0.195$ &  $2.201 \pm 0.231 $ &  $2.150 \pm 0.278 $  &  $ 0.21  \pm 0.17 $ &$3.02$  \\
$  0.2 $ &  $  17.0  \pm 28.0 $ & $116.9 \pm 13.0 $  &  $0.141 \pm 0.005  $ &  $ 2.711 \pm 0.195$ &  $2.294 \pm 0.231 $ &  $2.062 \pm 0.278 $  &  $ 0.21  \pm 0.17 $ &$2.67$  \\
$  1.4 $ &  $  25.6  \pm 23.2 $ & $117.1 \pm 10.0 $  &  $0.168 \pm 0.004  $ &  $ 2.931 \pm 0.158$ &  $2.262 \pm 0.190 $ &  $2.135 \pm 0.229 $  &  $ 0.86  \pm 0.13 $ &$1.75$  \\
$  5.1 $ &  $  56.8  \pm 44.4 $ & $121.6 \pm 24.7 $  &  $0.182 \pm 0.008  $ &  $ 3.096 \pm 0.311$ &  $2.516 \pm 0.374 $ &  $2.195 \pm 0.459 $  &  $ 1.13  \pm 0.27 $ &$1.39$  \\
$ 11.6 $ &  $ 148.2  \pm 56.2 $ & $130.9 \pm 66.5 $  &  $0.161 \pm 0.020  $ &  $ 2.728 \pm 0.803$ &  $2.465 \pm 0.972 $ &  $2.206 \pm 1.196 $  &  $ 1.51  \pm 0.70 $ &$1.07$  \\
\noalign{\smallskip}
\hline
\end{tabular}
\end{table*}
%%%%%%%%%%%%%%%%%%%%%%%%%%%%%%%%%%%%%%%%%%%%%%%%%%%%%%%%%%%%%%%%%%%%%%%%%%%%%%%%

\bsp

\label{lastpage}

\end{document}